\documentclass[letter,11pt]{article}
\usepackage{jheppub}
\usepackage{physics}
\usepackage{tcolorbox}
\usepackage{tikz}
\usetikzlibrary{decorations.pathmorphing}
\usepackage{amsmath}
\usepackage{graphicx}
\usepackage{subcaption}
\usepackage{slashed}
\usepackage{float}
\author[a]{Lorenzo De Ros}
\author[b]{Javier Reig Navarro}
\affiliation[a]{PRISMA$^{++}$ Cluster of Excellence \& Mainz Institute for Theoretical Physics, \\ FB 08 - Physics,
Mathematics and Computer Science,\\
Johannes Gutenberg-Universit{\"a}t Mainz,
55099 Mainz, Germany}
\affiliation[b]{Theoretical Particle Physics Laboratory (LPTP), Institute of Physics, EPFL, Lausanne, Switzerland}

\emailAdd{lderos@uni-mainz.de,javier.reignavarro@epfl.ch}

\tikzset{
  fastscalar/.style={
    dashed
  },
  fastboson/.style={
    decorate,
    decoration={
      snake,
      amplitude=0.8mm,
      segment length=2.2mm
    }
  }
}

\newcommand{\contact}[1][1]{%
\vcenter{\hbox{%
\begin{tikzpicture}[
scale=#1,
line width=0.8pt,
line cap=round,
line join=round
]

  \coordinate (C) at (0,0);

  \coordinate (LiU) at (-0.70, 0.50);
  \coordinate (LiD) at (-0.70,-0.50);
  \coordinate (RiU) at ( 0.70, 0.50);
  \coordinate (RiD) at ( 0.70,-0.50);

  \draw (LiU) -- (C);
  \draw (LiD) -- (C);
  \draw (C) -- (RiU);
  \draw (C) -- (RiD);

  \fill (C) circle[radius=3pt];

\end{tikzpicture}%
}}%
}

\newcommand{\doublecontact}[1][1]{%
  \vcenter{\hbox{%
    \begin{tikzpicture}[
      scale=#1,
      line width=0.8pt,
      line cap=round,
      line join=round
    ]

      \coordinate (L) at (-0.40,0);
      \coordinate (R) at ( 0.40,0);

   \coordinate (LiU) at (-1.25, 0.55);
  \coordinate (LiD) at (-1.25,-0.55);
  \coordinate (RiU) at ( 1.25, 0.55);
  \coordinate (RiD) at ( 1.25,-0.55);

      \draw (LiU) -- (L);
      \draw (LiD) -- (L);
      \draw (R) -- (RiU);
      \draw (R) -- (RiD);

      \draw (0,0) circle[radius=0.40];

      \fill (L) circle[radius=3pt];
      \fill (R) circle[radius=3pt];

    \end{tikzpicture}%
  }}%
}

\newcommand{\triplecontact}[1][1]{%
  \vcenter{\hbox{%
    \begin{tikzpicture}[
      scale=#1,
      line width=0.8pt,
      line cap=round,
      line join=round
    ]

      \coordinate (L) at (-0.80,0);
      \coordinate (M) at ( 0.00,0);
      \coordinate (R) at ( 0.80,0);

      \coordinate (LiU) at (-1.65, 0.65);
      \coordinate (LiD) at (-1.65,-0.65);
      \coordinate (RiU) at ( 1.65, 0.65);
      \coordinate (RiD) at ( 1.65,-0.65);

      \draw (LiU) -- (L);
      \draw (LiD) -- (L);
      \draw (R) -- (RiU);
      \draw (R) -- (RiD);

      \draw (-0.40,0) circle[radius=0.40];
      \draw ( 0.40,0) circle[radius=0.40];

      \fill (L) circle[radius=3pt];
      \fill (M) circle[radius=3pt];
      \fill (R) circle[radius=3pt];

    \end{tikzpicture}%
  }}%
}

\newcommand{\propagatorrep}[1][1]{%
\vcenter{\hbox{%
\begin{tikzpicture}[
scale=#1,
line width=0.8pt,
line cap=round,
line join=round
]

  \coordinate (LU) at (-0.40, 0.075);
  \coordinate (LD) at (-0.40,-0.075);
  \coordinate (RU) at ( 0.40, 0.075);
  \coordinate (RD) at ( 0.40,-0.075);

  \draw (LU) -- (RU);
  \draw (LD) -- (RD);

\end{tikzpicture}%
}}%
}

\newcommand{\singlebubble}[1][1]{%
\vcenter{\hbox{%
\begin{tikzpicture}[
scale=#1,
line width=0.8pt,
line cap=round,
line join=round
]

  \coordinate (L) at (-0.85,0);
  \coordinate (R) at ( 0.85,0);

  \coordinate (LU) at (-0.85, 0.075);
  \coordinate (LD) at (-0.85,-0.075);
  \coordinate (RU) at ( 0.85, 0.075);
  \coordinate (RD) at ( 0.85,-0.075);

  \coordinate (LiU) at (-1.30, 0.075);
  \coordinate (LiD) at (-1.30,-0.075);
  \coordinate (RiU) at ( 1.30, 0.075);
  \coordinate (RiD) at ( 1.30,-0.075);

  \draw (LiU) -- (LU);
  \draw (LiD) -- (LD);
  \draw (RU) -- (RiU);
  \draw (RD) -- (RiD);

  \draw[fastscalar] (LU) -- (RU);
  \draw[fastscalar] (LD) -- (RD);

  \draw[fastboson]
    (L) to[out=90,in=90,looseness=1.30] (R);

  \draw[fill=white] (L) circle[radius=0.10];
  \draw[fill=white] (R) circle[radius=0.10];

\end{tikzpicture}%
}}%
}

\newcommand{\cutbubble}[1][1]{%
  \hbox{%
    \begin{tikzpicture}[
      scale=#1,
      baseline={(0,0)},
      line width=0.8pt,
      line cap=round,
      line join=round
    ]

      \coordinate (V) at (0,0);

      \coordinate (VU) at (0, 0.075);
      \coordinate (VD) at (0,-0.075);

      \coordinate (LiU) at (-0.5, 0.075);
      \coordinate (LiD) at (-0.5,-0.075);

      \coordinate (RsU) at (0.58, 0.075);
      \coordinate (RsD) at (0.58,-0.075);

      \coordinate (Rb) at (0.58,0.64);

      \draw (LiU) -- (VU);
      \draw (LiD) -- (VD);

      \draw[fastscalar] (VU) -- (RsU);
      \draw[fastscalar] (VD) -- (RsD);

      \draw[fastboson]
        (V) to[out=90,in=180,looseness=1.10] (Rb);

      \draw[fill=white] (V) circle[radius=0.10];

    \end{tikzpicture}%
  }%
}

\newcommand{\cutbubblelabels}[1][1]{%
  \hbox{%
    \begin{tikzpicture}[
      scale=#1,
      baseline={(0,0)},
      line width=0.8pt,
      line cap=round,
      line join=round
    ]

      \coordinate (V) at (0,0);

      \coordinate (VU) at (0, 0.075);
      \coordinate (VD) at (0,-0.075);

      \coordinate (LiU) at (-0.75, 0.075);
      \coordinate (LiD) at (-0.75,-0.075);

      \coordinate (RsU) at (0.65, 0.075);
      \coordinate (RsD) at (0.65,-0.075);

      \coordinate (Rb) at (0.65,0.64);

      \draw (LiU) -- (VU);
      \draw (LiD) -- (VD);

      \draw[fastscalar] (VU) -- (RsU);
      \draw[fastscalar] (VD) -- (RsD);

      \draw[fastboson]
        (V) to[out=90,in=180,looseness=1.10] (Rb);

      \draw[fill=white] (V) circle[radius=0.10];

      \node[left] at (-0.75,0) {$s$};
      \node[right] at (0.65,0) {$b$};
      \node[above right] at (0.65,0.64) {$\phi$};
      
    \end{tikzpicture}%
  }%
}

\newcommand{\doublebubble}[1][1]{%
  \vcenter{\hbox{%
    \begin{tikzpicture}[
      scale=#1,
      line width=0.8pt,
      line cap=round,
      line join=round
    ]

      \coordinate (L)  at (-1.55,0);
      \coordinate (CL) at (-0.55,0);
      \coordinate (CR) at ( 0.55,0);
      \coordinate (R)  at ( 1.55,0);

      \coordinate (LU)  at (-1.55, 0.075);
      \coordinate (LD)  at (-1.55,-0.075);
      \coordinate (CLU) at (-0.55, 0.075);
      \coordinate (CLD) at (-0.55,-0.075);
      \coordinate (CRU) at ( 0.55, 0.075);
      \coordinate (CRD) at ( 0.55,-0.075);
      \coordinate (RU)  at ( 1.55, 0.075);
      \coordinate (RD)  at ( 1.55,-0.075);

      \coordinate (LiU) at (-2.00, 0.075);
      \coordinate (LiD) at (-2.00,-0.075);
      \coordinate (RiU) at ( 2.00, 0.075);
      \coordinate (RiD) at ( 2.00,-0.075);

      \draw (LiU) -- (LU);
      \draw (LiD) -- (LD);
      \draw (RU) -- (RiU);
      \draw (RD) -- (RiD);

      \draw[fastscalar] (LU)  -- (CLU);
      \draw[fastscalar] (LD)  -- (CLD);
      \draw[fastscalar] (CRU) -- (RU);
      \draw[fastscalar] (CRD) -- (RD);

      \draw (CLU) -- (CRU);
      \draw (CLD) -- (CRD);

      \draw[fastboson]
        (L) to[out=90,in=90,looseness=1.65] (CL);

      \draw[fastboson]
        (CR) to[out=90,in=90,looseness=1.65] (R);

      \draw[fill=white] (L)  circle[radius=0.10];
      \draw[fill=white] (CL) circle[radius=0.10];
      \draw[fill=white] (CR) circle[radius=0.10];
      \draw[fill=white] (R)  circle[radius=0.10];

    \end{tikzpicture}%
  }}%
}

\newcommand{\cutdoublebubble}[1][1]{%
  \vcenter{\hbox{%
    \begin{tikzpicture}[
      scale=#1,
      line width=0.8pt,
      line cap=round,
      line join=round
    ]

      \coordinate (L)  at (-1.55,0);
      \coordinate (CL) at (-0.55,0);
      \coordinate (CR) at ( 0.55,0);

      \coordinate (LU)  at (-1.55, 0.075);
      \coordinate (LD)  at (-1.55,-0.075);
      \coordinate (CLU) at (-0.55, 0.075);
      \coordinate (CLD) at (-0.55,-0.075);
      \coordinate (CRU) at ( 0.55, 0.075);
      \coordinate (CRD) at ( 0.55,-0.075);

      \coordinate (LiU) at (-2.00, 0.075);
      \coordinate (LiD) at (-2.00,-0.075);

      \coordinate (CutU) at (1.12, 0.075);
      \coordinate (CutD) at (1.12,-0.075);

      \coordinate (CutB) at (1.12,0.64);

      \draw (LiU) -- (LU);
      \draw (LiD) -- (LD);

      \draw[fastscalar] (LU) -- (CLU);
      \draw[fastscalar] (LD) -- (CLD);

      \draw (CLU) -- (CRU);
      \draw (CLD) -- (CRD);

      \draw[fastboson]
        (L) to[out=90,in=90,looseness=1.65] (CL);

      \draw[fastscalar] (CRU) -- (CutU);
      \draw[fastscalar] (CRD) -- (CutD);

      \draw[fastboson]
        (CR) to[out=90,in=180,looseness=1.10] (CutB);

      \draw[fill=white] (L)  circle[radius=0.10];
      \draw[fill=white] (CL) circle[radius=0.10];
      \draw[fill=white] (CR) circle[radius=0.10];

    \end{tikzpicture}%
  }}%
}

\newcommand{\crosseddoublebubble}[1][1]{%
  \vcenter{\hbox{%
    \begin{tikzpicture}[
      scale=#1,
      line width=0.8pt,
      line cap=round,
      line join=round
    ]

      \coordinate (L)  at (-1.55,0);
      \coordinate (CL) at (-0.55,0);
      \coordinate (CR) at ( 0.55,0);
      \coordinate (R)  at ( 1.55,0);

      \coordinate (LU)  at (-1.55, 0.075);
      \coordinate (LD)  at (-1.55,-0.075);
      \coordinate (CLU) at (-0.55, 0.075);
      \coordinate (CLD) at (-0.55,-0.075);
      \coordinate (CRU) at ( 0.55, 0.075);
      \coordinate (CRD) at ( 0.55,-0.075);
      \coordinate (RU)  at ( 1.55, 0.075);
      \coordinate (RD)  at ( 1.55,-0.075);

      \coordinate (LiU) at (-2.00, 0.075);
      \coordinate (LiD) at (-2.00,-0.075);
      \coordinate (RiU) at ( 2.00, 0.075);
      \coordinate (RiD) at ( 2.00,-0.075);

      \draw (LiU) -- (LU);
      \draw (LiD) -- (LD);
      \draw (RU) -- (RiU);
      \draw (RD) -- (RiD);

      \draw[fastscalar] (LU)  -- (CLU);
      \draw[fastscalar] (LD)  -- (CLD);
      \draw[fastscalar] (CRU) -- (RU);
      \draw[fastscalar] (CRD) -- (RD);

      \draw (CLU) -- (CRU);
      \draw (CLD) -- (CRD);

      \draw[fastboson]
        (L) to[out=-90,in=-90,looseness=1.30] (CR);

      \draw[fastboson]
        (CL) to[out=90,in=90,looseness=1.30] (R);

      \draw[fill=white] (L)  circle[radius=0.10];
      \draw[fill=white] (CL) circle[radius=0.10];
      \draw[fill=white] (CR) circle[radius=0.10];
      \draw[fill=white] (R)  circle[radius=0.10];

    \end{tikzpicture}%
  }}%
}

\newcommand{\crosseddoublebubblecuta}[1][1]{%
  \vcenter{\hbox{%
    \begin{tikzpicture}[
      scale=#1,
      line width=0.8pt,
      line cap=round,
      line join=round
    ]

      \coordinate (L)  at (-1.55,0);
      \coordinate (CL) at (-0.55,0);
      \coordinate (CR) at ( 0.55,0);
      \coordinate (R)  at ( 1.55,0);

      \coordinate (LU)  at (-1.55, 0.075);
      \coordinate (LD)  at (-1.55,-0.075);
      \coordinate (CLU) at (-0.55, 0.075);
      \coordinate (CLD) at (-0.55,-0.075);
      \coordinate (CRU) at ( 0.55, 0.075);
      \coordinate (CRD) at ( 0.55,-0.075);
      \coordinate (RU)  at ( 1.55, 0.075);
      \coordinate (RD)  at ( 1.55,-0.075);

      \coordinate (LiU) at (-2.00, 0.075);
      \coordinate (LiD) at (-2.00,-0.075);
      \coordinate (RiU) at ( 2.00, 0.075);
      \coordinate (RiD) at ( 2.00,-0.075);

      \draw (LiU) -- (LU);
      \draw (LiD) -- (LD);
      \draw (RU) -- (RiU);
      \draw (RD) -- (RiD);

      \draw[fastscalar] (LU)  -- (CLU);
      \draw[fastscalar] (LD)  -- (CLD);
      \draw[fastscalar] (CRU) -- (RU);
      \draw[fastscalar] (CRD) -- (RD);

      \draw (CLU) -- (CRU);
      \draw (CLD) -- (CRD);

      \draw[fastboson]
        (L) to[out=-90,in=-90,looseness=1.30] (CR);

      \draw[fastboson]
        (CL) to[out=90,in=90,looseness=1.30] (R);

      \draw[fill=white] (L)  circle[radius=0.10];
      \draw[fill=white] (CL) circle[radius=0.10];
      \draw[fill=white] (CR) circle[radius=0.10];
      \draw[fill=white] (R)  circle[radius=0.10];

      \draw[densely dashed, line width=0.8pt]
        (0,-1.10) -- (0,1.10);

    \end{tikzpicture}%
  }}%
}

\newcommand{\crosseddoublebubblecutb}[1][1]{%
  \vcenter{\hbox{%
    \begin{tikzpicture}[
      scale=#1,
      line width=0.8pt,
      line cap=round,
      line join=round
    ]

      \coordinate (L)  at (-1.55,0);
      \coordinate (CL) at (-0.55,0);
      \coordinate (CR) at ( 0.55,0);
      \coordinate (R)  at ( 1.55,0);

      \coordinate (LU)  at (-1.55, 0.075);
      \coordinate (LD)  at (-1.55,-0.075);
      \coordinate (CLU) at (-0.55, 0.075);
      \coordinate (CLD) at (-0.55,-0.075);
      \coordinate (CRU) at ( 0.55, 0.075);
      \coordinate (CRD) at ( 0.55,-0.075);
      \coordinate (RU)  at ( 1.55, 0.075);
      \coordinate (RD)  at ( 1.55,-0.075);

      \coordinate (LiU) at (-2.00, 0.075);
      \coordinate (LiD) at (-2.00,-0.075);
      \coordinate (RiU) at ( 2.00, 0.075);
      \coordinate (RiD) at ( 2.00,-0.075);

      \draw (LiU) -- (LU);
      \draw (LiD) -- (LD);
      \draw (RU) -- (RiU);
      \draw (RD) -- (RiD);

      \draw[fastscalar] (LU)  -- (CLU);
      \draw[fastscalar] (LD)  -- (CLD);
      \draw[fastscalar] (CRU) -- (RU);
      \draw[fastscalar] (CRD) -- (RD);

      \draw (CLU) -- (CRU);
      \draw (CLD) -- (CRD);

      \draw[fastboson]
        (L) to[out=-90,in=-90,looseness=1.30] (CR);

      \draw[fastboson]
        (CL) to[out=90,in=90,looseness=1.30] (R);

      \draw[fill=white] (L)  circle[radius=0.10];
      \draw[fill=white] (CL) circle[radius=0.10];
      \draw[fill=white] (CR) circle[radius=0.10];
      \draw[fill=white] (R)  circle[radius=0.10];

      \draw[densely dashed, line width=0.8pt]
        (-1.05,-1.10) -- (-1.05,1.10);

    \end{tikzpicture}%
  }}%
}

\newcommand{\cuttriplebubble}[1][1]{%
  \vcenter{\hbox{%
    \begin{tikzpicture}[
      scale=#1,
      line width=0.8pt,
      line cap=round,
      line join=round
    ]

      \coordinate (L1) at (-2.55,0);
      \coordinate (R1) at (-1.55,0);

      \coordinate (L2) at (-0.55,0);
      \coordinate (R2) at ( 0.45,0);

      \coordinate (L3) at ( 1.45,0);

      \coordinate (L1U) at (-2.55, 0.075);
      \coordinate (L1D) at (-2.55,-0.075);
      \coordinate (R1U) at (-1.55, 0.075);
      \coordinate (R1D) at (-1.55,-0.075);

      \coordinate (L2U) at (-0.55, 0.075);
      \coordinate (L2D) at (-0.55,-0.075);
      \coordinate (R2U) at ( 0.45, 0.075);
      \coordinate (R2D) at ( 0.45,-0.075);

      \coordinate (L3U) at ( 1.45, 0.075);
      \coordinate (L3D) at ( 1.45,-0.075);

      \coordinate (LiU) at (-3.00, 0.075);
      \coordinate (LiD) at (-3.00,-0.075);

      \coordinate (CutU) at (2.00, 0.075);
      \coordinate (CutD) at (2.00,-0.075);
      \coordinate (CutB) at (2.00,0.64);

      \draw (LiU) -- (L1U);
      \draw (LiD) -- (L1D);

      \draw[fastscalar] (L1U) -- (R1U);
      \draw[fastscalar] (L1D) -- (R1D);

      \draw (R1U) -- (L2U);
      \draw (R1D) -- (L2D);

      \draw[fastscalar] (L2U) -- (R2U);
      \draw[fastscalar] (L2D) -- (R2D);

      \draw (R2U) -- (L3U);
      \draw (R2D) -- (L3D);

      \draw[fastscalar] (L3U) -- (CutU);
      \draw[fastscalar] (L3D) -- (CutD);

      \draw[fastboson]
        (L1) to[out=90,in=90,looseness=1.65] (R1);

      \draw[fastboson]
        (L2) to[out=90,in=90,looseness=1.65] (R2);

      \draw[fastboson]
        (L3) to[out=90,in=180,looseness=1.10] (CutB);

      \draw[fill=white] (L1) circle[radius=0.10];
      \draw[fill=white] (R1) circle[radius=0.10];
      \draw[fill=white] (L2) circle[radius=0.10];
      \draw[fill=white] (R2) circle[radius=0.10];
      \draw[fill=white] (L3) circle[radius=0.10];

    \end{tikzpicture}%
  }}%
}

\preprint{MITP-26-046}

\title{A Unitary Fixed Point Away from Threshold:\\
Momentum-Surface EFT for Strong Mixing}

\abstract{Radiative capture in a non-relativistic system can become non-perturbative even when the underlying transition interaction is weak. We demonstrate this mechanism in a simple quantum-mechanical two-channel model with a systematic power counting, where strong mixing between scattering and bound states develops in a narrow region around a finite momentum $k_0$, leading to a perturbative violation of unitarity. In this regime, the enhanced contributions factorize, reducing the dynamics to a free scattering state interacting through a strong local coupling. Under RG evolution in the factorization scale, this local interaction approaches a unitary fixed point, analogous to that of systems with an anomalously large scattering length. This motivates an EFT organized around the finite-momentum surface $|\mathbf{k}|=k_0$, describing fermions at unitarity with a linear dispersion relation. The resummation resulting from the EFT restores unitarity and provides a systematic expansion for both elastic scattering and radiative capture.}

\begin{document}
\maketitle
\flushbottom

\newpage

%%%%%%%%%%%%%%%%%%%%%%%%%%%%%%%%%%%%%%%%%%%%%%%%%%%%%%%%%
\section{Introduction}
\label{sec:Introduction}
%%%%%%%%%%%%%%%%%%%%%%%%%%%%%%%%%%%%%%%%%%%%%%%%%%%%%%%%%

Non-Relativistic (NR) systems can exhibit strongly coupled dynamics even when the underlying interactions are short-ranged. A canonical example is two-body scattering with an unnaturally large scattering length, where non-perturbative effects give rise to universal behavior near the unitary limit~\cite{Kaplan:1998we,Kaplan:1998tg,Braaten:2004rn,Nishida:2007pj}.

Radiative capture provides a different setting in which non-perturbative dynamics can emerge, particularly when the scattering and bound states experience different potentials. An example occurs in a Non-Relativistic Effective Field Theory (NREFT) with a ${\rm SU}(N)$ gauge symmetry~\cite{Bodwin:1994jh,Brambilla:1999xf,Beneke:2013jia}: a scattering state in the adjoint representation feels a repulsive Coulomb potential, while a singlet quarkonium-like state feels an attractive one:
\begin{equation}
    \label{NRQCD_potentials}
    V_{\mathbf{1}}(r)=-C_F\frac{\alpha}{r},
    \qquad
    V_{\mathbf{adj}}(r)=\frac{1}{2N}\frac{\alpha}{r},
\end{equation}
with chromoelectric dipole gluon emission connecting the two colour configurations. The same basic structure appears in neutron--proton capture into the deuteron~\cite{Savage:1998vh}, where the incoming ${}^1S_0$ and bound ${}^3S_1$ states experience different strong potentials and an M1 photon connects the two channels. Interestingly, perturbative radiative capture cross sections exceeding the unitarity bound were first observed in related multi-channel dark sectors~\cite{Oncala:2019yvj}, and subsequently in non-relativistic ${\rm SU}(N)$ theories~\cite{Binder:2023ckj,Beneke:2024nxh}, providing the main motivation for the present work. In Ref.~\cite{Flores:2024UnitarityNR,Flores:2026NRScattering,Flores:2026yay}, this problem in the case of a Coulombic potential was approached by introducing a resummed 2PI kernel together with regulated wavefunctions. Here, we take a different route and first study the underlying mechanism in a simple two-channel toy model, chosen so that the onset and breakdown of perturbation theory can be analyzed within a well-defined power counting.

\begin{figure}[t]
    \centering
    \includegraphics[width=1.0\linewidth]{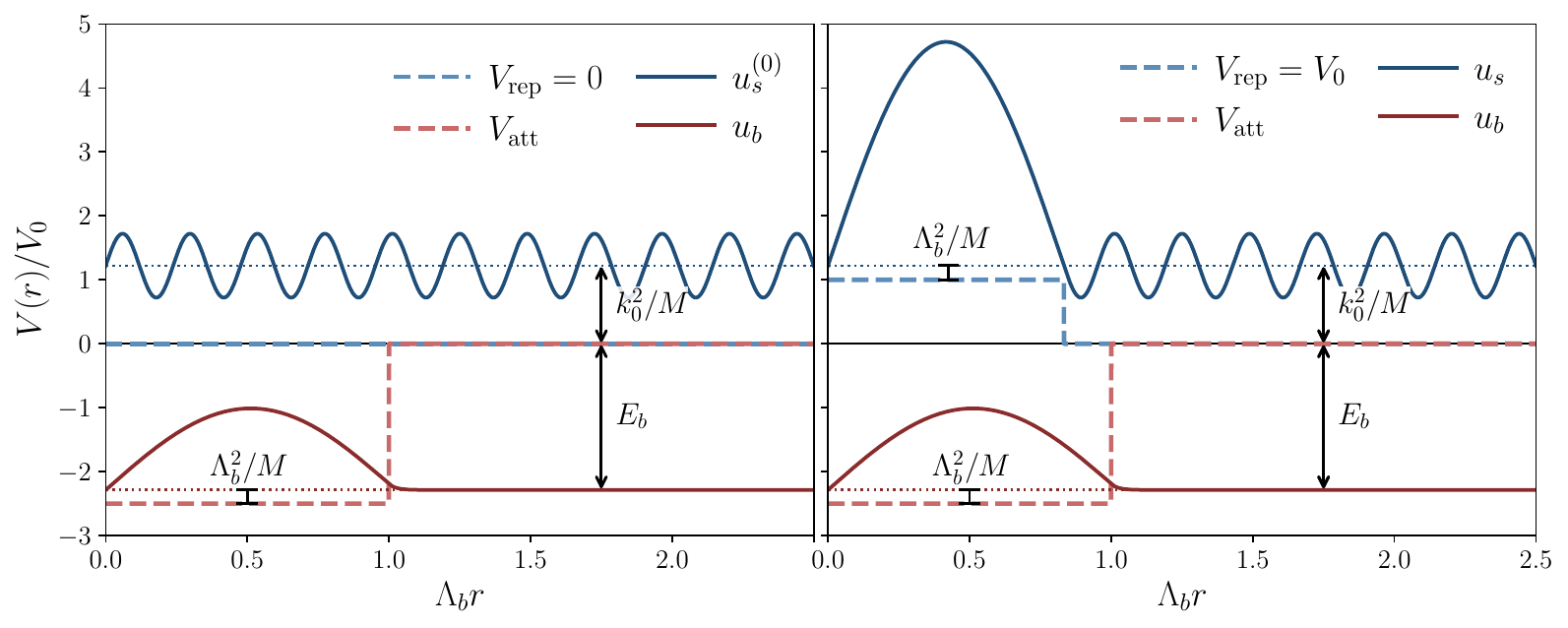}
    \caption{Schematic potentials and wavefunctions for the repulsive (blue) and attractive (red) channels. Dashed and solid curves denote the channel potentials and wavefunctions, respectively. In the left panel, the free scattering state $u_s^{(0)}$ oscillates rapidly over the support of the bound state $u_b$, leading to cancellations in the transition matrix element. In the right panel, a repulsive barrier reduces the local momentum of the scattering state $u_s$, so that the two wavefunctions vary on comparable length scales and the transition matrix element approaches its geometric size.}
    \label{fig:1}
\end{figure}

Schematically, we consider radiative capture from a scattering state $\lvert s\rangle$ into a bound state $\lvert b\rangle$ when the two states belong to different channels. The two channels experience different diagonal potentials, which we take to be repulsive and attractive and denote by $V_{\rm rep}$ and $V_{\rm att}$, respectively, and are connected by a weak off-diagonal interaction $H_{\rm off}$ describing the emission of radiation in a multipole expansion
\begin{equation}
    H=
    \begin{pmatrix}
        H_{\rm rep} & H_{\rm off}\\
        H^\dagger_{\rm off} & H_{\rm att}
    \end{pmatrix}.
\end{equation}
To make contact with the usual perturbative power counting, we first consider the transition from a free incoming state with momentum $k$.
Denoting by $\Lambda_b$ the inverse size of the bound state, for
\begin{equation}
    k\gg\Lambda_b,
\end{equation}
a free incoming plane wave $\lvert \mathbf{k} \rangle$ oscillates rapidly over the support of the bound state, as illustrated in the left panel of Fig.~\ref{fig:1}. The resulting cancellations suppress the matrix element
\begin{equation}
    \langle b \lvert H_{\rm off} \rvert \mathbf{k} \rangle \sim \frac{\Lambda_b^{3/2}}{k^3}.
\end{equation}
A repulsive barrier in the scattering channel can remove this suppression. In fact, the scattering state $\lvert s(\mathbf{k})\rangle$, an exact continuum eigenstate of $H_{\rm rep}$, can have a local momentum of order $\Lambda_b$ over the region where the bound-state wavefunction has support. Therefore, the two wavefunctions vary on comparable length scales and the oscillatory cancellations disappear, as shown in the right panel of Fig.~\ref{fig:1}. Therefore, the reduced matrix element then reaches its \emph{geometric} size
\begin{equation}
    \langle b \rvert H_{\rm off} \rvert s(\mathbf{k}) \rangle
    \sim \frac{1}{\Lambda_b^{3/2}},
\end{equation}
even though the asymptotic momentum remains $k\gg\Lambda_b$. For a barrier of height $V_0$ and width of order $\Lambda_b^{-1}$, this occurs in a restricted momentum region around
\begin{equation}
    \frac{k_0^2}{M} \sim V_0,
\end{equation}
where $M$ denotes the mass of each constituent.
This enhancement dramatically changes the scaling of the capture cross section\footnote{These momentum scalings assume monopole emission, as in the toy model studied in this work. See Sec.~\ref{sec:Multipole_Expansion}.}:
\begin{equation}
    \sigma_{\rm cap}^{\rm free}\sim\frac{\Lambda_b^3}{k^5},
    \qquad \longrightarrow \qquad
    \sigma_{\rm cap}^{\rm geom}\sim\frac{k_0}{\Lambda_b^3}.
\end{equation}
Therefore, for $k\gg\Lambda_b$, the geometric estimate exceeds the unitarity bound
\begin{equation}
    \sigma_{\rm cap} \lesssim \frac{1}{k^2}.
\end{equation}
This apparent violation of unitarity signals the breakdown of the expansion in the off-diagonal interaction.
This interpretation is consistent with the observation in non-relativistic ${\rm SU}(N)$ theories~\cite{Beneke:2024nxh} that the violation occurs when the classical turning points of the scattering and bound states coincide, so that both states have small local kinetic energy in the same spatial region.

The two-channel Hamiltonian also makes the origin of this breakdown transparent. Perturbation theory treats $H_{\rm off}$ as an insertion between eigenstates of the repulsive and attractive channel Hamiltonians, controlled by its matrix element between the scattering and bound states. Once this matrix element reaches its geometric size, the corrections generated by $H_{\rm off}$ are no longer suppressed by the small coupling alone. The eigenstates of the diagonal Hamiltonian are then a poor perturbative basis, and the repeated iteration of transitions between the two channels must be resummed. This \emph{strong mixing} is fundamentally different from the level crossing encountered at a Feshbach resonance~\cite{Chin:2010Feshbach}. In fact, the diagonal scattering- and bound-state energies remain separated, while the transition matrix element between them becomes non-perturbative.

The loss of perturbativity is manifest in elastic scattering. Insertions of $H_{\rm off}$ generate virtual transitions through the attractive channel. Near the momentum $k_0$, at which the transition matrix element to the bound state is maximal, the corresponding corrections are enhanced by the same matrix element that controls radiative capture. As a result, diagrams that are reducible with respect to repulsive-channel propagation are no longer suppressed. These enhanced diagrams factorize: the transition matrix element can be evaluated at $k_0$, leaving a free scattering propagator multiplied by a local interaction. In particular, the contribution generated by transitions through the bound-state channel is matched onto a local two-body interaction
\begin{equation}
\label{diagram:introduction}
\Sigma^{(1)}(k_0)
=
\raisebox{2.2ex}{$\singlebubble[1]$}
\quad \longrightarrow \quad
\frac{4\pi a}{M} = \contact[1],
\end{equation}
characterized by an effectively large scattering length $a$. This is analogous to \emph{fermions at unitarity}~\cite{Kaplan:1998we,Kaplan:1998tg}, where a near-threshold state promotes the leading four-fermion interaction to a relevant operator that must be resummed to all orders, and whose renormalization-group flow approaches a unitary fixed point. The same happens in our case, but at a finite momentum $k_0$ instead of at threshold.

These observations motivate an effective theory for the modes in the vicinity of the momentum at which the mixing becomes strong. Following the standard treatment of low-energy excitations near a Fermi surface, we organize the theory around a narrow \emph{momentum surface} centered at $k_0$ and describe the remaining dynamics in terms of the slow residual momentum~\cite{Shankar:1993pf,Polchinski:1992ed}
\begin{equation}
\delta k=k-k_0,
\qquad
|\delta k|\ll \frac{\Lambda_{\rm EFT}^2}{k_0} = \frac{(\pi\Lambda_b)^3}{k_0^2} \ll k_0.
\end{equation}
This hierarchy reflects two distinct expansions, a kinematic one, in $\delta k/k_0$, from linearizing the dispersion around the momentum surface, and a further expansion in $(k_0 \delta k)/\Lambda_{\rm EFT}^2$, set by the analytic structure of the microscopic interaction. The latter organizes the local operators of the EFT, starting with the contact interaction of Eq.~\eqref{diagram:introduction}, and takes the form of an effective range expansion~\cite{Bethe:1949yr} defined around the finite momentum $k_0$ instead of threshold. The resulting theory describes fermions at unitarity with a linearized dispersion relation around $k_0$, can be matched explicitly to the toy model, and provides a systematic expansion for corrections to both elastic scattering and radiative capture.

The paper is organized as follows. In Sec.~\ref{sec:Toy_Model}, we introduce the toy model with spherical potentials and monopole radiation. In Sec.~\ref{sec:Elastic_Scattering}, we compute the induced elastic scattering amplitude and show how the iteration of self-energy insertions leads to the breakdown of perturbation theory. In Sec.~\ref{sec:Renormalization_Group_and_Resummation}, we perform the required resummation, show how unitarity is restored, and discuss the RG flow of the resulting relevant interaction. In Sec.~\ref{sec:Momentum-Surface_EFT}, we introduce the momentum-surface EFT that describes the strong-mixing regime and match it to the underlying toy model. Finally, in Sec.~\ref{sec:Conclusions}, we summarize our conclusions. In Appendix~\ref{sec:Toy_Model_Conventions}, we report the conventions and the details regarding the toy model. In Appendix~\ref{sec:Power_Counting_Subleading_Corrections}, we show the suppression of the 2PI contributions and the corrections to factorization. In Appendix~\ref{sec:momentum_surface_construction}, we provide a constructive derivation of the EFT Lagrangian, as well as its Feynman rules.

\newpage
%%%%%%%%%%%%%%%%%%%%%%%%%%%%%%%%%%%%%%%%%%%%%%%%%%%%%%%%%
\section{Toy Model}
\label{sec:Toy_Model}
%%%%%%%%%%%%%%%%%%%%%%%%%%%%%%%%%%%%%%%%%%%%%%%%%%%%%%%%%

%%%%%%%%%%%%%%%%%%%%%%%%%%%%%%%%%%%%%%%%%%%%%%%%%%%%%%%%%
\subsection{Setup}
\label{sec:Setup}
%%%%%%%%%%%%%%%%%%%%%%%%%%%%%%%%%%%%%%%%%%%%%%%%%%%%%%%%%

The aim of this section is to construct a minimal toy model that captures the unitarity-violation effect of interest while admitting a clear and systematic power counting. To this end, we introduce two species of heavy fermions, $\xi$ and $\eta$, with a common mass $M$. These fields describe the NR particles of the model, which interact through finite-range, nonlocal potentials characterized by a range $\Lambda_b^{-1}$. The corresponding NR Lagrangian is
\begin{align}
    \label{L_kin}
    \mathcal{L}_{\rm kin} & =
    \xi^\dagger(\mathbf{x},t)
    \left(
        i\partial_t+\frac{\nabla^2}{2M}
    \right)
    \xi(\mathbf{x},t)
    +
    \eta^\dagger(\mathbf{x},t)
    \left(
        i\partial_t+\frac{\nabla^2}{2M}
    \right)
    \eta(\mathbf{x},t)
    \nonumber\\
    &-
    \frac{1}{2}\int \dd^3\mathbf{r}\,
    V_{\xi \xi}(r)\,
    \left[\xi^\dagger\xi\right]
    (\mathbf{x}+\mathbf{r}/2,t)\,
    \left[\xi^\dagger\xi\right]
    (\mathbf{x}-\mathbf{r}/2,t)
    \nonumber\\
    &-
    \int \dd^3\mathbf{r}\,
    V_{\xi \eta}(r)\,
    \left[\xi^\dagger\xi\right]
    (\mathbf{x}+\mathbf{r}/2,t)\,
    \left[\eta^\dagger\eta\right]
    (\mathbf{x}-\mathbf{r}/2,t)
    \nonumber\\
    &-
    \frac{1}{2}\int \dd^3\mathbf{r}\,
    V_{\eta \eta}(r)\,
    \left[\eta^\dagger\eta\right]
    (\mathbf{x}+\mathbf{r}/2,t)\,
    \left[\eta^\dagger\eta\right]
    (\mathbf{x}-\mathbf{r}/2,t),
\end{align}
where $\xi(\mathbf{x},t)$ and $\eta(\mathbf{x},t)$ are Pauli spinors.
As sketched in Fig.~\ref{fig:1}, we take the interaction between particles of the same species to be repulsive and that between particles of different species to be attractive. To simplify the discussion, we represent the potentials as a spherical barrier and a spherical well, of equal range $\Lambda_b^{-1}$ and equal and opposite heights
\begin{equation}
    \label{potentials}
    V_{\xi \xi}(r) = V_{\eta \eta}(r) = + V_0 \, \theta(1-\Lambda_b r), \qquad
    V_{\xi \eta}(r) = -V_0 \, \theta(1-\Lambda_b r).
\end{equation}
With this choice, a $(\xi,\eta)$ pair can form a bound state in the attractive spherical well. By contrast, the repulsive interactions in the $(\xi,\xi)$ and $(\eta,\eta)$ channels do not support bound states.

We further introduce a complex massless scalar field $\phi$, whose emission mediates transitions between the two heavy species,
\begin{equation}
    \xi \longleftrightarrow \eta + \phi.
\end{equation}
The scalar kinetic term and its interaction with the two fermion species are described by
\begin{equation}
    \label{L_phi}
    \mathcal{L}_\phi=\partial_\mu\phi^*(\mathbf{x},t)
    \partial^\mu\phi(\mathbf{x},t)
    -
    \left[
        g\,\xi^\dagger(\mathbf{x},t)
        \eta(\mathbf{x},t)
        \phi(\mathbf{x},t)
        +\mathrm{h.c.}
    \right],
\end{equation}
where the coupling is assumed to be perturbative $g \ll 1$. We do not consider interactions between the scalar and two fermions of the same species\footnote{A useful schematic physical picture is to regard the two species as carrying opposite charges, $\pm 1$, under a $\mathrm{U}(1)$ symmetry, in analogy with the sign structure of the Coulomb potential in an abelian gauge theory. Couplings between particles from the same species can be forbidden by assigning charge $+2$ to the scalar $\phi$, as in~\cite{Oncala:2019yvj}.}.
This interaction induces the radiative capture processes
\begin{equation}
    \xi + \xi \longrightarrow (\xi \eta) + \phi, \qquad
    \eta + \eta \longrightarrow (\xi \eta) + \phi^*.
\end{equation}
We focus in particular on capture into the $s$-wave ground state,
\begin{equation}
    \label{ground-state}
    (\xi \eta)_{(n=1,\ell=0)},
\end{equation}
although the analysis can be generalized to arbitrary bound states and partial waves.
Of the two processes above, we explicitly consider the capture of a $\xi+\xi$ scattering pair, as the $\eta+\eta$ case is completely analogous. For notational convenience, we denote the states involved in the process by
\begin{equation}
    s= \xi + \xi, \qquad b=(\xi \eta)_{(n=1,\ell=0)}.
\end{equation}
In the center-of-mass frame of the incoming pair, the transition is therefore
\begin{equation}
    s(\mathbf{k}) \longrightarrow b(\mathbf{P}_b,E_b) + \phi(\mathbf{q}),
\end{equation}
where $\mathbf{k}$ is the incoming relative momentum, $\mathbf{P}_b$ the total momentum of the bound state, $E_b<0$ its binding energy, and $\mathbf{q}$ the momentum of the emitted scalar. Energy and momentum conservation imply\footnote{More precisely, including the recoil of the bound state, energy conservation reads
\begin{equation*}
q+\frac{q^2}{4M}
=
\frac{k^2}{M}+|E_b|,
\end{equation*}
where the term $q^2/(4M)$ is subleading in the NR expansion.}
\begin{equation}
    \mathbf{P}_b=-\mathbf{q},
    \qquad
    q= \frac{k^2}{M} + |E_b|,
\end{equation}
where $k=|\mathbf{k}|$ and $q = |\mathbf{q}|$.
As motivated in Sec.~\ref{sec:Introduction}, this process has the desired property that the incoming scattering state and the outgoing bound state experience different potentials. In particular, it reproduces the qualitative structure of NRQCD quarkonium capture: the incoming channel is repulsive, as in the colour-octet configuration, whereas the outgoing channel is attractive, as in the colour-singlet configuration given in Eq.~\eqref{NRQCD_potentials}.

We focus on the regime in which the strong mixing effect arises, schematically illustrated in Fig.~\ref{fig:1}, and defined by the following assumptions. First, we require the ground state of the potential well to be deeply bound, which corresponds to the regime
\begin{equation}
\label{deep_well}
\frac{M V_0}{\Lambda_b^2} \gg 1.
\end{equation}
This condition ensures that the kinetic energy of the bound state is smaller than its potential energy, therefore it implies
\begin{equation}
    |E_b| \sim V_0.
\end{equation}
We further take the energy of the scattering state to be of the same order as the height of the repulsive barrier, and hence also as the energy of the radiated scalar
\begin{equation}
\label{barrier_height}
\frac{k^2}{M} \sim V_0 \sim q.
\end{equation}
Together, Eqs.~\eqref{deep_well} and \eqref{barrier_height} imply
\begin{equation}
\label{fast_momentum}
    k \gg \Lambda_b.
\end{equation}
This hierarchy means that the incoming state resolves the finite range of the potential, hence the size of the bound state, and the repulsive barrier therefore cannot be treated as a contact interaction. We highlight that this regime differs from the one studied in~\cite{Kaplan:1998tg,Kaplan:1998we}, where $k\ll \Lambda_b$ and the finite range potential can be treated as a local interaction.

%%%%%%%%%%%%%%%%%%%%%%%%%%%%%%%%%%%%%%%%%%%%%%%%%%%%%%%%%
\subsection{Multipole Expansion}
\label{sec:Multipole_Expansion}
%%%%%%%%%%%%%%%%%%%%%%%%%%%%%%%%%%%%%%%%%%%%%%%%%%%%%%%%%

To identify the leading radiative transition, we perform a multipole expansion (MPE) of the scalar field around the center-of-mass coordinate $\mathbf{R}$ of the heavy pair,
\begin{align}
\phi\left(
\mathbf{R}\pm\mathbf{r}/2,t
\right)
={}&
\phi(\mathbf{R},t)
\pm
\frac{r^i}{2}
\left(\partial^i\phi\right)(\mathbf{R},t)
+
\frac{r^i r^j}{8}
\left(\partial^i \partial^j \phi\right)(\mathbf{R},t)
+\cdots,
\end{align}
where $\mathbf{r}$ denotes the relative distance.
For characteristic separations $r\sim\Lambda_b^{-1}$, the validity of the MPE requires
\begin{equation}
\label{MPE}
\frac{q}{\Lambda_b}\ll 1.
\end{equation}
Combined with Eqs.~\eqref{deep_well}, \eqref{barrier_height}, and \eqref{fast_momentum}, this condition defines the hierarchy of the parameters of the toy model
\begin{equation}
    \label{all_conditions}
    \frac{\Lambda_b}{M} \ll \frac{V_0}{\Lambda_b} \ll 1.
\end{equation}
At leading order in the MPE, the scalar field is constant over the relative separation of the heavy pair, corresponding to monopole emission. The dipole term vanishes because it is odd under $\mathbf{r}\to-\mathbf{r}$, while the relevant two-body matrix element is even. The first correction is therefore quadrupole emission, suppressed by $(q/\Lambda_b)^2$ relative to the monopole contribution.

The leading monopole operator obeys the selection rule $\Delta \ell=0$. Since the outgoing bound state $\lvert b \rangle$ has $\ell=0$, monopole capture selects the $\ell=0$ component of the incoming scattering state $\lvert s^{(+)}(\mathbf{k}) \rangle$, which we denote as $\lvert s^{(+)}(k) \rangle$ and is defined in Appendix~\ref{sec:Repulsive_Channel}. The incoming particles are identical fermions, so their spatially symmetric $s$-wave state forces the spin wavefunction to be antisymmetric. The incoming pair therefore has total spin $S=0$. Spin conservation in the monopole transition then selects the $S=0$ component of the outgoing $(\xi,\eta)$ bound state, therefore the leading transition is
\begin{equation}
    {}^1S_0\longrightarrow{}^1S_0.
\end{equation}

The two-body sector is more naturally formulated in terms of spin singlet composite fields, defined as
\begin{align}
    \label{composite_field_def}
    \Phi_{\xi\xi}(\mathbf{R},\mathbf{r},t)
    &\equiv
    \left[
        \xi\left(\mathbf{R}-\frac{\mathbf{r}}{2},t\right)
        \xi\left(\mathbf{R}+\frac{\mathbf{r}}{2},t\right)
    \right]_{S=0},
    \nonumber\\
    \Phi_{\xi\eta}(\mathbf{R},\mathbf{r},t)
    &\equiv
    \left[
        \xi\left(\mathbf{R}-\frac{\mathbf{r}}{2},t\right)
        \eta\left(\mathbf{R}+\frac{\mathbf{r}}{2},t\right)
    \right]_{S=0},
\end{align}
where the subscript $S=0$ denotes projection onto the spin-singlet configuration\footnote{For two Pauli spinors $\psi$ and $\chi$, we adopt the convention
\begin{equation*}
    [\psi\chi]_{S=0}
    \equiv
    \frac{1}{\sqrt{2}}\,
    \psi^{\mathrm{T}}i\sigma^2\chi
    =
    \frac{1}{\sqrt{2}}
    \left(
        \psi_\uparrow\chi_\downarrow
        -
        \psi_\downarrow\chi_\uparrow
    \right).
\end{equation*}
}.
In terms of these composite fields, and retaining only the leading monopole term in the MPE, the Lagrangian becomes
\begin{align}
    \mathcal{L}
    &=
    \int \dd^3\mathbf{r}\,
    \Phi_{\xi \xi}^\dagger(\mathbf{R},\mathbf{r},t)
    \left[
        i\partial_t
        +\frac{\nabla_{\mathbf{R}}^2}{4M}
        +\frac{\nabla_{\mathbf{r}}^2}{M}
        -V_{\xi \xi}(r)
    \right]
    \Phi_{\xi \xi}(\mathbf{R},\mathbf{r},t)
    \nonumber\\
    &+
    \int \dd^3\mathbf{r}\,
    \Phi_{\xi \eta}^\dagger(\mathbf{R},\mathbf{r},t)
    \left[
        i\partial_t
        +\frac{\nabla_{\mathbf{R}}^2}{4M}
        +\frac{\nabla_{\mathbf{r}}^2}{M}
        -V_{\xi \eta}(r)
    \right]
    \Phi_{\xi \eta}(\mathbf{R},\mathbf{r},t)
    \nonumber\\
    &-\left[g \, 
        \phi(\mathbf{R},t)
        \int \dd^3\mathbf{r}\,
        \Phi_{\xi \xi}^\dagger(\mathbf{R},\mathbf{r},t)
        \Phi_{\xi \eta}(\mathbf{R},\mathbf{r},t)
        +\mathrm{h.c.}
    \right]+\partial_\mu \phi^* \partial^\mu \phi.
\end{align}
Therefore, at leading order in the coupling $g \ll 1$, the capture amplitude can be expressed in terms of the matrix element as
\begin{equation}
    \label{A_capture_tree-level}
    i \mathcal{A}^{(0)}_{\rm cap} (k) = \raisebox{0.5ex}{$\cutbubblelabels[1]$} = - i g \,  \langle b | s^{(+)}(k) \rangle,
\end{equation}
where the double solid line denotes the scattering state, the double dashed line the bound state, and the wiggly line the monopole radiation. The incoming scattering state $\lvert s^{(+)}(k) \rangle$ in the repulsive channel ($\xi \xi$ pair) and the final bound state $\lvert b \rangle$ in the attractive channel ($\xi \eta$ pair) are defined in the Appendices~\ref{sec:Repulsive_Channel} and~\ref{sec:Attractive_Channel}, respectively.
The corresponding capture cross section is
\begin{equation}
    \label{sigma_capture}
    (\sigma v)_{\rm cap} = \frac{q}{2\pi} \left|\mathcal{A}_{\rm cap} (k) \right|^2,
\end{equation}
where $v=2k/M$ is the relative velocity.

%%%%%%%%%%%%%%%%%%%%%%%%%%%%%%%%%%%%%%%%%%%%%%%%%%%%%%%%%
\subsection{Matrix Element}
\label{sec:Matrix_Element}
%%%%%%%%%%%%%%%%%%%%%%%%%%%%%%%%%%%%%%%%%%%%%%%%%%%%%%%%%

As shown in Eq.~\eqref{A_capture_tree-level}, the perturbative capture rate is controlled by the $s$-wave scattering-to-bound matrix element
\begin{equation}
    \langle b \rvert s^{(+)}(k) \rangle.
\end{equation}
The size of it is determined not directly by the asymptotic scattering momentum, but by the momentum of the scattering wavefunction inside the interaction region $\Lambda_b r \lesssim 1$. It is therefore useful to compare the characteristic \emph{local} momenta of the bound and scattering states, which we denote as $k_b$ and $k_s$, respectively. The definitions and normalization conventions for the corresponding wavefunctions are collected in Appendix~\ref{sec:Repulsive_Channel} and~\ref{sec:Attractive_Channel}.

%%%%%%%%%%%%%%%%%%%%%%%%%%%%%%%%%%%%%%%%%%%%%%%%%%%%%%%%%
\subsubsection{Enhancement}
\label{sec:Enhancement}
%%%%%%%%%%%%%%%%%%%%%%%%%%%%%%%%%%%%%%%%%%%%%%%%%%%%%%%%%

In the deeply-bound limit, defined in Eq.~\eqref{deep_well}, the ground-state wavefunction (solid red lines of Fig.~\ref{fig:1}) reads
\begin{equation}
    \label{wf_bound_deep}
    \psi_b(r)
    \approx
    \frac{1}{\sqrt{4\pi}\,r} \sqrt{2 \Lambda_b}
    \begin{cases}
        \sin(\pi \Lambda_b r), & \Lambda_b r<1,\\[0.6em]
        0, & \Lambda_b r>1,
    \end{cases}
\end{equation}
which oscillates inside the well with characteristic momentum
\begin{equation}
    k_b \approx \pi\Lambda_b.
\end{equation}
For an incoming free $s$-wave scattering state $\lvert k \rangle$ (solid blue line, left panel of Fig.~\ref{fig:1}), there is no distinction between the asymptotic and local momentum, so that
\begin{equation}
    k_s^{\rm free}=k,
\end{equation}
and the free scattering wavefunction reads
\begin{equation}
    \psi_k^{\rm free}(r) = \frac{\sin (k r)}{k r}.
\end{equation}
Since $k \gg \Lambda_b$ holds in the regime of interest, the local momenta also satisfy
\begin{equation}
k_s^{\rm free} \gg k_b.
\end{equation}
Therefore, the larger local momentum suppresses the matrix element through rapid oscillations of the free scattering wavefunction. This is also evident from
\begin{align}
\langle b \lvert k \rangle
&=
\int {\rm d}^3 \mathbf{r} \,
\psi_b^*(r)\psi_k^{\rm free}(r)
\nonumber\\
&\approx
2\frac{\sqrt{2\pi\Lambda_b}}{k}
\int_0^{1/\Lambda_b}{\rm d}r \,
\sin(k_b r)\sin(kr)=
-\frac{(2\pi\Lambda_b)^{3/2}
\sin(k/\Lambda_b)}
{k(k^2-\pi^2\Lambda_b^2)}.
\end{align}
Consequently, for $k \gg k_b$, we obtain the momentum scaling
\begin{equation}
    \label{matrix_element_free_scaling}
    \langle b \rvert k \rangle
    \sim
    \frac{(\pi \Lambda_b)^{3/2}}{k^3},
\end{equation}
up to an oscillatory envelope.

For the exact incoming scattering state $\lvert s^{(+)}(k) \rangle$ (solid blue line, right panel of Fig.~\ref{fig:1}), however, the momentum inside the interaction region differs from its asymptotic value. Above the barrier, the local momentum is
\begin{equation}
k_s(k)=\sqrt{k^2-MV_0},
\end{equation}
and the scattering wavefunction reads
\begin{equation}
    \label{wf_scatt_full}
    \psi^{(\pm)}_k(r)
    =
    \frac{e^{\pm i\delta(k)}}{kr}
    \begin{cases}
        N_s(k)\sin\left[k_s(k) \, r\right], & \Lambda_b r<1,\\[0.6em]
        \sin\!\left[kr+\delta(k)\right], & \Lambda_b r>1,
    \end{cases}
\end{equation}
where $N_s(k)$ is a normalization constant and $\delta(k)$ denotes the phase-shift, defined in Eqs.~\eqref{N_s_def} and~\eqref{delta_def}.
Although the asymptotic momentum satisfies $k\gg \Lambda_b$, the barrier can reduce the local momentum to the scale of the bound-state momentum. In particular, the matrix element is maximal when the scattering and the bound-state wavefunctions oscillate with comparable momenta inside the interaction region
\begin{equation}
    k_s(k_0) \approx k_b,
\end{equation}
which implicitly defines the asymptotic momentum
\begin{equation}
\label{k_0_def}
    k_0 \approx \sqrt{MV_0+\pi^2 \Lambda_b^2}.
\end{equation}
Therefore, at $k=k_0$, we obtain the scalings of the momenta
\begin{equation}
    k_s(k_0)\sim k_b\sim\Lambda_b,
\end{equation}
while $k_0\gg\Lambda_b$, and the scattering wavefunction becomes approximately
\begin{equation}
    \psi^{(\pm)}_{k_0}(r)
    \approx
    \frac{1}{k_0 r}
    \begin{cases}
        \displaystyle
        \frac{k_0}{\pi \Lambda_b}
        \sin \left(\pi \Lambda_b r\right),
        & \Lambda_b r<1, \\[1.2em]
        \displaystyle
        \sin(k_0 r - k_0 / \Lambda_b),
        & \Lambda_b r>1.
    \end{cases}
\end{equation}
This produces the maximal matrix element
\begin{align}
    \label{matrix_element_geom_scaling}
    \langle b \rvert s^{(+)}(k_0) \rangle
    &=
    \int {\rm d}^3\mathbf{r} \,
    \psi_b^*(r)\psi_{k_0}^{(+)}(r)
    \nonumber\\
    &\approx
    \frac{2\sqrt{2}}{\sqrt{\pi\Lambda_b}}
    \int_0^{1/\Lambda_b}{\rm d}r \,
    \sin^2(\pi\Lambda_b r)= \sqrt{\frac{2}{\pi}} \frac{1}{\Lambda_b^{3/2}},
\end{align}
which is set by the \emph{geometric} size of the interaction region.

%%%%%%%%%%%%%%%%%%%%%%%%%%%%%%%%%%%%%%%%%%%%%%%%%%%%%%%%%
\subsubsection{Unitarity Violation}
\label{sec:Unitarity_Violation}
%%%%%%%%%%%%%%%%%%%%%%%%%%%%%%%%%%%%%%%%%%%%%%%%%%%%%%%%%

\begin{figure}
    \centering
    \includegraphics[width=0.85\linewidth]{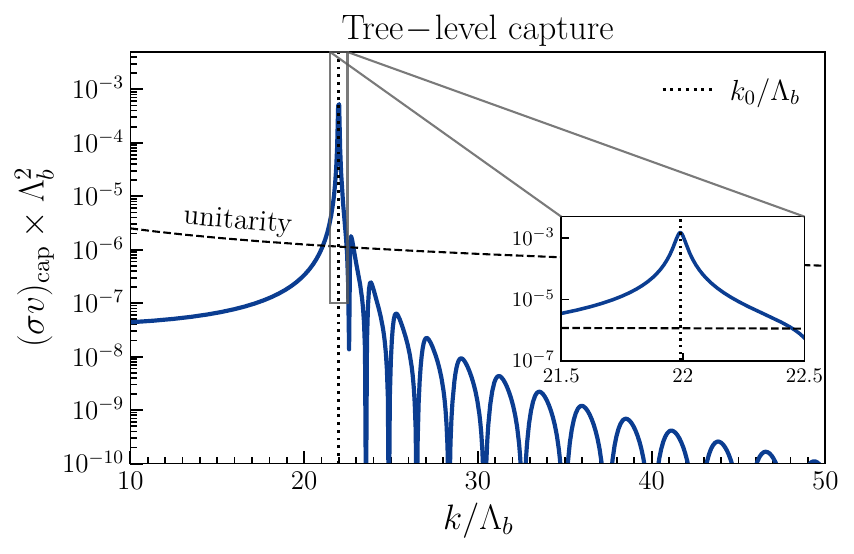}
    \caption{Tree-level cross section for the radiative capture, normalized to $\Lambda_b^2$, as a function of the incoming momentum $k/\Lambda_b$. The dashed black curve denotes the inelastic $s$-wave unitarity bound. The vertical dotted line marks the momentum $k_0$ where the strong mixing occurs. The inset zooms in the region where the rate violates the unitarity bound. For illustration, we choose $V_0/M \simeq 10^{-10}$ and $\Lambda_b/M \simeq 10^{-6}$. These values yield $MV_0/\Lambda_b^2 \simeq 10^{2} \gg 1$ and $V_0/\Lambda_b \simeq 10^{-4} \ll 1$, satisfying both the deep bound state and MPE conditions, summarized in Eq.~\eqref{all_conditions}.}
    \label{fig:2}
\end{figure}

For both cases, the perturbative capture cross section follows from Eqs.~\eqref{A_capture_tree-level} and~\eqref{sigma_capture}, giving
\begin{equation}
    \label{sigmas_scaling}
(\sigma v)_{\rm cap}^{\rm free} \approx 8\pi^2 g^2 \frac{\Lambda_b^3}{M k^4},
\qquad
(\sigma v)_{\rm cap}^{\rm geom} \approx 2 \frac{g^2}{\pi^2} \frac{k_0^2}{M \Lambda_b^3}.
\end{equation}
Unitarity imposes a bound~\cite{Griest:1989wd} on the total inelastic cross section in a fixed incoming partial wave, which for the $s$-wave reads
\begin{equation}
    \label{cap_UB}
    (\sigma v)_{\rm cap} \leq (\sigma v)_{\rm cap}^{\rm UB} \equiv \frac{2 \pi}{M k}.
\end{equation}
For the cross section of a free scattering state, as in the first of Eqs.~\eqref{sigmas_scaling}, this gives
\begin{equation}
    \label{free_unitarity}
    g^2 \left( \frac{\pi \Lambda_b}{k} \right)^3 < \mathcal{O}(1),
\end{equation}
which is always satisfied for $k\gg\Lambda_b$ and perturbative $g$. By contrast, for the exact scattering state at $k=k_0$, the unitarity condition applied to the second of Eqs.~\eqref{sigmas_scaling} leads to
\begin{equation}
    \label{geom_unitarity}
    g^2\left(\frac{k_0}{\pi \Lambda_b}\right)^3 < \mathcal{O}(1).
\end{equation}
Therefore, since $k_0\gg \pi \Lambda_b$, this condition can be strongly violated even for $g \ll 1$, provided that $g^2 k_0^3 /(\pi \Lambda_b)^3 \gg 1$.
Therefore, defining the expansion parameter
\begin{equation}
    \lambda \equiv \frac{(\pi \Lambda_b)^3}{k_0^3},
\end{equation}
the limit of strong unitarity violation can be characterized as
\begin{equation}
\label{parametrics}
    g \ll 1, \qquad \lambda \ll 1, \qquad {\rm with} \qquad \quad \frac{g^2}{\lambda} = {\rm fixed} \gg 1.
\end{equation}
Obviously, this violation signals the breakdown of the perturbative expansion in $g$, rather than a physical violation of unitarity. In Fig.~\ref{fig:2}, we show the tree-level capture cross section obtained from the exact matrix element
\begin{equation}
    \label{exact_matrix_element}
    \langle b \rvert s^{(+)}(k) \rangle =(2\pi\Lambda_b)^{3/2} e^{i\delta(k)} \frac{N_s(k)}{k} \frac{\sin\left[k_s(k)/\Lambda_b \right]}{\pi^2 \Lambda_b^2 - k_s(k)^2}.
\end{equation}
As expected from our estimates, by comparing the perturbative capture cross section with the unitarity bound, the violation occurs within a finite momentum window centered around $k_0$.

%%%%%%%%%%%%%%%%%%%%%%%%%%%%%%%%%%%%%%%%%%%%%%%%%%%%%%%%%
\subsubsection{Momentum Dependence}
\label{sec:Momentum_Dependence}
%%%%%%%%%%%%%%%%%%%%%%%%%%%%%%%%%%%%%%%%%%%%%%%%%%%%%%%%%

Beyond the strength of the unitarity violation, the momentum dependence near its maximum is an important feature of the system. Expanding the matrix element around $k=k_0$ gives
\begin{equation}
    \label{matrix_element_expansion}
    \frac{\langle b \rvert s^{(+)}(k) \rangle}{\langle b \rvert s^{(+)}(k_0) \rangle} = 1 - \frac{\pi^2}{8} \frac{(2k_0\delta k)^2}{\Lambda_{\rm EFT}^4} + \mathcal{O}\left(\frac{(2k_0 \delta k)^4}{\Lambda_{\rm EFT}^8}\right),
\end{equation}
with $\delta k= k-k_0$ and where we defined
\begin{equation}
    \label{Lambda_EFT_def}
    \Lambda_{\rm EFT}^2 \equiv \frac{(\pi \Lambda_b)^3}{k_0} = \lambda k_0^2.
\end{equation}
The scale $\Lambda_{\rm EFT}$ describes the local momentum dependence of the matrix element around its maximum.
Moreover, we highlight that since the expansion is centered at the maximum $k_0$, the leading local dependence is symmetric under deviations on either side of the maximum, $\delta k \mapsto -\delta k$.
The details on the systematics of the expansion of the location of the maximum and of the matrix element itself are described in Appendix~\ref{sec:Expansion_around_the_Maximum}.

%%%%%%%%%%%%%%%%%%%%%%%%%%%%%%%%%%%%%%%%%%%%%%%%%%%%%%%%%
\subsection{Hamiltonian Picture and Strong Mixing}
\label{sec:Hamiltonian_Picture_and_Strong_Mixing}
%%%%%%%%%%%%%%%%%%%%%%%%%%%%%%%%%%%%%%%%%%%%%%%%%%%%%%%%%

The origin of the perturbative breakdown is particularly transparent in the Hamiltonian picture. In standard quantum-mechanical perturbation theory, the mixing between two eigenstates $\lvert n\rangle$ and $\lvert m\rangle$ induced by a perturbation $H_{\rm off}$ is controlled by
\begin{equation}
    \label{weak_mixing}
    \frac{\langle n|H_{\rm off}|m\rangle}{E_n-E_m}.
\end{equation}
Perturbation theory therefore requires this ratio to remain small. In our case, the relevant mixing is that between the scattering and bound-state sectors. We separate the Hamiltonian into diagonal components $H_{\rm rep}$ and $H_{\rm att}$ associated with the two channels, $\xi \xi$ and the $\xi \eta$ pairs, 
describing the scattering and bound-state sectors, respectively, as defined in Appendices~\ref{sec:Repulsive_Channel} and~\ref{sec:Attractive_Channel}. These diagonal components are treated exactly, so that the perturbative expansion in the coupling $g$ corresponds to the expansion in the off-diagonal radiative interaction $H_{\rm off}$ that connects the two sectors
\begin{equation}
    H_{\rm off} = g \int \frac{{\rm d}^3\mathbf{q}}{(2\pi)^3\sqrt{2q}} \bigg[\lvert b, \phi(\mathbf{q}) \rangle \langle s \rvert + {\rm h.c.} \bigg].
\end{equation}
Therefore, the relevant Hamiltonian for the radiative capture process can be written as
\begin{equation}
    H = H_0 + H_{\rm off},
    \qquad
    H_0 =
    \begin{pmatrix}
        H_{\rm rep} & 0 \\
        0 & H_{\rm att} + H_\phi
    \end{pmatrix},
\end{equation}
where the free scalar Hamiltonian $H_\phi$ is defined by $H_\phi \lvert \phi(\mathbf{q}) \rangle=q\,\lvert \phi(\mathbf{q}) \rangle$ for the scalar one-particle state.
The strength of the mixing between two eigenstates of $H_0$ is controlled by the matrix element
\begin{equation}
    \label{H_int_insertion}
    \langle b,\phi(\mathbf q)|H_{\rm off}|s(\mathbf{k})\rangle
    \propto
    g\, \langle b \rvert s^{(+)}(k) \rangle,
\end{equation}
together with the energy denominators generated by propagation through
intermediate states.

The previous subsection showed that the transition matrix element is enhanced at values of the momentum close to $k_0$. The same enhancement appears in higher-order corrections involving repeated transitions between the $s$ and $b+\phi$ sectors. For instance, the next correction in $H_{\rm off}$ corresponds to the virtual process
\begin{equation}
    \label{H_int_virtual}
    s
    \longrightarrow
    b+\phi
    \longrightarrow
    s,
\end{equation}
whose expression can be written in the schematic form
\begin{align}
    \langle s(\mathbf{k})\rvert H_{\rm off}
    \frac{1}{E-H_0+i0}
    H_{\rm off}\lvert s(\mathbf{k})\rangle
    =
    \int {\rm d}\Pi_{b\phi}\,
    \frac{
        \langle s (\mathbf{k}) \rvert H_{\rm off}|b,\phi\rangle
        \langle b,\phi|H_{\rm off} \lvert s(\mathbf{k}) \rangle
    }{
        E-E_b-q+i0
    },
\end{align}
where ${\rm d}\Pi_{b\phi}$ denotes the phase-space measure of the intermediate bound-state and the monopole radiation.
The size of the perturbative corrections is therefore not set by $g$ alone, but by the enhanced matrix element of Eq.~\eqref{H_int_insertion} and the propagation of the intermediate $b+\phi$ state. Since the matrix element is enhanced for values of the momentum close to $k_0$, the validity of the perturbative expansion cannot be inferred from $g \ll 1$ alone. Moreover, at higher orders, the same insertion can be iterated
\begin{equation}
    \label{H_int_virtual_many}
    s
    \longrightarrow
    b+\phi
    \longrightarrow
    s
    \longrightarrow
    b+\phi
    \longrightarrow
    \cdots.
\end{equation}
It is therefore necessary to determine whether successive transitions between the two sectors remain perturbatively suppressed.
For this purpose, in Sec.~\ref{sec:Elastic_Scattering}, we consider the elastic scattering process in the $s$ channel, for which the higher-order corrections provide a direct measure of the convergence of the perturbative expansion.

%%%%%%%%%%%%%%%%%%%%%%%%%%%%%%%%%%%%%%%%%%%%%%%%%%%%%%%%%
\section{Elastic Scattering}
\label{sec:Elastic_Scattering}
%%%%%%%%%%%%%%%%%%%%%%%%%%%%%%%%%%%%%%%%%%%%%%%%%%%%%%%%%

The Hamiltonian picture provides an intuitive description of the mixing between the two sectors. For the explicit calculation, however, we work in the Lagrangian framework and consider elastic scattering for an incoming state in the repulsive channel. As already discussed for the capture process, the potentials in the repulsive and attractive channels are treated exactly, while the amplitude is expanded perturbatively in the off-diagonal coupling $g$.

%%%%%%%%%%%%%%%%%%%%%%%%%%%%%%%%%%%%%%%%%%%%%%%%%%%%%%%%%
\subsection{Perturbative Expansion}
\label{sec:Perturbative_Expansion}
%%%%%%%%%%%%%%%%%%%%%%%%%%%%%%%%%%%%%%%%%%%%%%%%%%%%%%%%%

In general, the $S$-matrix elements between the incoming and outgoing scattering states $\lvert s^{(\pm)}(\mathbf{k}) \rangle$ of the repulsive channel can then be written as
\begin{align}
    &S_{\rm scatt}(\mathbf{k},\mathbf{k}')
    = \langle s^{(-)}(\mathbf{k}') \rvert s^{(+)}(\mathbf{k}) \rangle
    \nonumber\\
    &
    +\frac{M k}{2\pi} \langle s^{(-)}(\mathbf{k}') \rvert
    \Big(
        [-i\hat{\Sigma}]
        +[-i\hat{\Sigma}]\,i\hat{G}_{\rm rep}\,[-i\hat{\Sigma}]
        +[-i\hat{\Sigma}]\,i\hat{G}_{\rm rep}\,[-i\hat{\Sigma}]\,
         i\hat{G}_{\rm rep}\,[-i\hat{\Sigma}]
        +\cdots
    \Big)
    \lvert s^{(+)}(\mathbf{k}) \rangle,
    \label{bubble_series}
\end{align}
where the propagator of the repulsive channel $G_{\rm rep}$ is defined in Appendix~\ref{sec:Repulsive_Channel}, together with the other conventions for the scattering states. Virtual transitions into the attractive channel are encoded in the self-energy $\Sigma$, which starts at $\Sigma=\mathcal{O}(g^2)$.

The incoming and outgoing scattering states are related by Eq.~\eqref{+to-}, therefore we substitute
\begin{equation}
    \lvert s^{(-)}(\mathbf{k}') \rangle = e^{-2i\delta(k')} \lvert s^{(+)}(\mathbf{k'}) \rangle,
\end{equation}
where $\delta(k')$ denotes the phase shift of the scattering wavefunction due to the repulsive channel only. The $S$-matrix then factorizes as
\begin{equation}
    \label{S-matrix_factorization}
    S_{\rm scatt}(\mathbf{k},\mathbf{k}') = e^{2i\delta(k')} \times \left[ 1 + \frac{Mk}{2\pi} \langle s^{(+)}(\mathbf{k}) \rvert \left([- i \hat{\Sigma}] + [- i \hat{\Sigma}] i \hat{G}_{\rm rep} [- i \hat{\Sigma}] + \cdots \right) \lvert s^{(+)}(\mathbf{k}) \rangle \right],
\end{equation}
where the first factor describes the scattering induced by the potential barrier in the repulsive channel, while the second accounts for virtual transitions into the attractive channel.
The scattering amplitude is obtained from the $S$-matrix as
\begin{equation}
    i \mathcal{A}_{\rm scatt}(\mathbf{k},\mathbf{k}') = \frac{2\pi}{M k} \left[S_{\rm scatt}(\mathbf{k},\mathbf{k}')-1\right],
\end{equation}
and its perturbative expansion reads
\begin{equation}
\label{def:pert:amplitude}
     i \mathcal{A}_{\rm scatt}(\mathbf{k},\mathbf{k}') = \sum_{n=0}^\infty i \mathcal{A}^{(n)}_{\rm scatt}(\mathbf{k},\mathbf{k}'),
\end{equation}
where $\mathcal{A}_{\rm scatt}^{(n)}=\mathcal{O}(g^{2n})$.
At $\mathcal{O}(g^0)$, the on-shell $s$-wave amplitude reads
\begin{equation}
    \label{A0_scatt}
    i \mathcal{A}^{(0)}_{\rm scatt} (k) = \propagatorrep[1]
    =
    \frac{2\pi}{M k}
    \left[e^{2i\delta(k)}-1\right],
\end{equation}
where the expressions specialized to our toy model are reported in Appendix~\ref{sec:Propagation_and_Scattering}.

We therefore compute the corrections associated with virtual transitions, starting from $A^{(1)}_{\rm scatt}$, and analyze the higher-order terms illustrated in Eq.~\eqref{H_int_virtual_many}, with the aim of identifying the parameter that controls the breakdown of the perturbative expansion near $k_0$, where the matrix element is enhanced.

%%%%%%%%%%%%%%%%%%%%%%%%%%%%%%%%%%%%%%%%%%%%%%%%%%%%%%%%%
\subsection{One-loop}
\label{sec:One-loop}
%%%%%%%%%%%%%%%%%%%%%%%%%%%%%%%%%%%%%%%%%%%%%%%%%%%%%%%%%

The leading correction from the off-diagonal interaction between the repulsive and the attractive channels arises at one-loop, i.e. at $\mathcal{O}(g^2)$.
It can be computed as a matrix element between the in and out states
\begin{align}
    &i \mathcal{A}^{(1)}_{{\rm scatt}}(E,\mathbf{k},\mathbf{k}') = \raisebox{2.2ex}{$\singlebubble[1]$} \nonumber \\
    &= \langle s^{(-)}(\mathbf{k}') \rvert \left[-i \hat{\Sigma}^{(1)}(E) \right] | s^{(+)}(\mathbf{k}) \rangle = - i e^{2i\delta(k')} \Sigma^{(1)}(E,\mathbf{k},\mathbf{k}') \nonumber\\
    &=  (-ig)^2 \mu_R^{4-d} e^{2i\delta(k')} \langle s^{(+)}(k')\rvert b\rangle \langle b\lvert s^{(+)}(k)\rangle \int \frac{\mathrm{d}^d q}{(2\pi)^d}
    \frac{i}{ E-q^0-\frac{|\mathbf{q}|^2}{4M}+|E_b|+i0 } \frac{i}{q^2+i0} \nonumber \\
    &=i g^2 \mu_R^{4-d}
    \frac{
        \Gamma(d)\Gamma(1-d)
    }{
        (4\pi)^{\frac{d-1}{2}}
        \Gamma\left(\frac{d-1}{2}\right)
    }
    \left(
        -E-|E_b|-i0
    \right)^{d-3} e^{2i \delta(k')} \langle s^{(+)}(k') \rvert b \rangle \langle b \rvert s^{(+)}(k) \rangle,
\end{align}
where $\mu_R$ denotes the renormalization scale, and $d$ is the number of space-time dimensions since we work in dimensional regularization. The matrix elements of the self-energy $\Sigma^{(1)}(E,\mathbf{k},\mathbf{k}')$ are conventionally defined between the in states $\lvert s^{(+)}\rangle $, which makes the scattering phase $e^{2i \delta (k)}$ explicit.
The two propagators inside the loop describe the radiated scalar and the bound state\footnote{The bound-state propagator should be substituted with the full spectral decomposition of the attractive channel, which includes the propagators of all bound states and all scattering states. However, the leading contribution for $k \to k_0$, arises from the ground-state propagator, as discussed in Appendix~\ref{sec:Bound-state_Propagator}.}, and the monopole emission selects the $s$-wave component of both the incoming and outgoing scattering states. The loop has only one momentum region~\cite{Smirnov:1994tg,Beneke:1997zp} contributing, which corresponds to the radiated scalar being ultrasoft $q^0 \sim |\mathbf{q}| \sim k^2/M$, therefore the recoil term $|\mathbf{q}|^2/M$ can be neglected. 

The on-shell $s$-wave component of the amplitude, after MS subtraction, becomes
\begin{align}
    \label{A1_scatt}
    i\mathcal{A}^{(1)}_{\rm scatt}(k) &= - i e^{2i \delta(k)} \Sigma^{(1)}(k) \nonumber \\
    &=i e^{2i \delta(k)}\frac{g^2}{4\pi^2}
    \left(
        \frac{k^2}{M}+|E_b|
    \right)
    \left[
        1+i\pi
        +\ln\left(
            \frac{\mu_R \sqrt{\pi e^{-\gamma_E}}}{\frac{k^2}{M}+|E_b|+i0}
        \right)
    \right]
    \left| \langle b \rvert s^{(+)}(k) \rangle \right|^2,
\end{align}
where we denote the on-shell diagonal elements of the self-energy as $\Sigma^{(1)}(k) \equiv \Sigma^{(1)}(k^2/M,k,k)$, and analogously for the amplitude.
The above expression shows that the one-loop scattering amplitude is proportional to the matrix element between the bound state and the scattering state. We highlight the analytic continuation branch cut of the logarithm generates the explicit imaginary contribution proportional to $i\pi$, corresponding to the configuration in which the intermediate bound state and radiated scalar are simultaneously on-shell.
This imaginary part corresponds to the tree-level capture rate obtained in~\eqref{A_capture_tree-level},~\eqref{sigma_capture}:
\begin{equation}
    \label{optical_theorem_one-loop}
    -2 \operatorname{Im} \Sigma^{(1)} (k) = \frac{q}{2\pi} \left|\mathcal{A}^{(0)}_{\rm cap} (k)\right|^2.
\end{equation}

%%%%%%%%%%%%%%%%%%%%%%%%%%%%%%%%%%%%%%%%%%%%%%%%%%%%%%%%%
\subsection{Loss of Perturbativity}
\label{sec:Loss_of_Perturbativity}
%%%%%%%%%%%%%%%%%%%%%%%%%%%%%%%%%%%%%%%%%%%%%%%%%%%%%%%%%

We now consider the $\mathcal{O}(g^4)$ corrections and isolate their leading behavior in the limit $k \to k_0$. We focus on the diagram containing two insertions of the one-loop self-energy $\Sigma^{(1)}$, which is two-particle reducible (2PR) with respect to the intermediate repulsive-channel propagator:
\begin{align}
    \label{2PR_off-shell}
    &i\mathcal A^{(2){\rm 2PR}}_{\rm scatt}(E,\mathbf{k}, \mathbf{k}')=  \raisebox{1.8ex}{$\doublebubble[1]$}\nonumber \\
    &= \langle s^{(-)}(\mathbf{k}') \rvert
    \left[- i \hat{\Sigma}^{(1)}(E) \right]
    i\hat{G}_{\rm rep}^{(0)}(E)
    \left[- i \hat{\Sigma}^{(1)}(E) \right]
    \lvert s^{(+)} (\mathbf{k}) \rangle
    \nonumber\\
    &= \int \frac{{\rm d}^3 \mathbf{p}}{(2\pi)^3}
    \langle s^{(-)}(\mathbf{k}') \rvert
    \left[-i \hat{\Sigma}^{(1)}(E) \right]
    \frac{i \lvert s^{(+)} (\mathbf{p}) \rangle
    \langle s^{(+)} (\mathbf{p})\rvert}
    {\frac{k^2}{M} - \frac{p^2}{M} + i0}
    \left[- i \hat{\Sigma}^{(1)}(E) \right]
    \lvert s^{(+)} (\mathbf{k}) \rangle
    \nonumber \\
    &= e^{2i \delta(k')}
    \int
    \frac{\mathrm d^3\mathbf p}{(2\pi)^3}
    \biggl[
        -i\Sigma^{(1)}
        \left(
            E,\mathbf k,\mathbf p
        \right)
    \biggr]
    \frac{i}{
        E-\frac{p^2}{M}+i0
    }
    \biggl[
        -i\Sigma^{(1)}
        \left(
            E,\mathbf p,\mathbf{k}'
        \right)
    \biggr] \nonumber \\
    &= e^{2i \delta(k')} \left[ - i \Sigma^{(1)}(E,\mathbf{k},\mathbf{k}')\right] \nonumber \\
    &\times i \frac{g^2}{4\pi^2}
    \left(
        E+|E_b|
    \right)
    \left[
        1+i\pi
        +\ln\left(
            \frac{\mu_R \sqrt{\pi e^{-\gamma_E}}}{E+|E_b|+i0}
        \right)
    \right] \int
    \frac{\mathrm d^3\mathbf p}{(2\pi)^3}
    \frac{i}{
        E-\frac{p^2}{M}+i0
    } \left| \langle b \rvert s^{(+)}(p) \rangle \right|^2.
\end{align}
Here the intermediate scattering state carries the loop momentum $\mathbf p$, and we recall that the enhancement arises from the momentum dependent matrix element $\langle b \rvert s^{(+)} (p) \rangle$, discussed in Sec.~\ref{sec:Matrix_Element}. Taking the on-shell limit, we obtain
\begin{align}
    \label{2PR_on-shell}
    &i\mathcal A^{(2){\rm 2PR}}_{\rm scatt}(k) = e^{2i \delta(k)} \left[ - i \Sigma^{(1)}(k)\right] \nonumber \\
    &\times i \frac{g^2}{4\pi^2}
    \left(
        \frac{k^2}{M}+|E_b|
    \right)
    \left[
        1+i\pi
        +\ln\left(
            \frac{\mu_R \sqrt{\pi e^{-\gamma_E}}}{\frac{k^2}{M}+|E_b|+i0}
        \right)
    \right] \int
    \frac{\mathrm d^3\mathbf p}{(2\pi)^3}
    \frac{i}{
        \frac{k^2}{M}-\frac{p^2}{M}+i0
    } \left| \langle b \rvert s^{(+)}(p) \rangle \right|^2.
\end{align}

Decomposing the loop propagator as
\begin{equation}
    \frac{1}{
        \frac{k^2}{M}-\frac{p^2}{M}+i0
    }
    =
    \mathcal{P}
    \frac{1}{
        \frac{k^2}{M}-\frac{p^2}{M}
    }
    -
    i\pi\,
    \delta\left(
        \frac{k^2}{M}-\frac{p^2}{M}
    \right),
\end{equation}
we estimate the size of the contribution from its on-shell part to the loop integral as
\begin{equation}
    \int
    \frac{{\rm d}^3 \mathbf{p}}{(2\pi)^3}
    \frac{i}{
        \frac{k^2}{M}
        -\frac{p^2}{M}
        +i0
    } \left| \langle b \rvert s^{(+)} (p)\rangle \right|^2 \sim
    \frac{Mk}{4\pi} \left| \langle b \rvert s^{(+)}(k) \rangle \right|^2.
\end{equation}
It follows that this contribution to the amplitude scales as
\begin{equation}
    \label{A2PR_scatt_estimate}
    i\mathcal A_{\rm scatt}^{(2){\rm 2PR}}(k)
    \sim
    e^{2i\delta(k)} \frac{M k}{4\pi}
    \left[\Sigma^{(1)}(k)\right]^2.
\end{equation}

Comparing the $\mathcal{O}(g^2)$ correction, obtained in Eq.~\eqref{A1_scatt}, to the $\mathcal{O}(g^4)$ 2PR amplitude, perturbativity requires their ratio to satisfy
\begin{equation}
    \label{perturbativity}
    \left|
        \frac{\mathcal A_{\rm scatt}^{(2){\rm 2PR}}(k)}
             {\mathcal A_{\rm scatt}^{(1)}(k)}
    \right|
    \sim
    \frac{Mk}{4\pi}
    \left|
        \Sigma^{(1)}(k)
    \right| \sim g^2 k^3 \left| \langle b \rvert s^{(+)} (k) \rangle \right|^2 \ll 1.
\end{equation}
Assuming the scaling of the matrix element for a free incoming scattering state, as in Eq.~\eqref{matrix_element_free_scaling}, we obtain the condition
\begin{equation}
    \label{free_perturbativity}
    \left|
        \frac{\mathcal A_{\rm scatt}^{(2){\rm 2PR}}(k)}
             {\mathcal A_{\rm scatt}^{(1)}(k)}
    \right|
    \sim g^2 k^3 \left| \langle b \rvert k \rangle \right|^2
    \sim
    g^2 \left(\frac{\pi \Lambda_b}{k}\right)^3 \ll 1.
\end{equation}
Therefore, for $g \ll 1$ and $k \gg \Lambda_b$, the perturbative expansion is well behaved.
However, near the momentum $k_0$, the matrix element reaches its geometric scaling, as in Eq.~\eqref{matrix_element_geom_scaling}, which leads to the loss of perturbativity
\begin{equation}
    \label{geom_perturbativity}
    \left|
        \frac{\mathcal A_{\rm scatt}^{(2){\rm 2PR}}(k_0)}
             {\mathcal A_{\rm scatt}^{(1)}(k_0)}
    \right|
    \sim g^2 k_0^3 \left| \langle b \rvert s^{(+)}(k_0) \rangle \right|^2
    \sim g^2 \left(\frac{k_0}{\pi \Lambda_b}\right)^3
    \sim
    \frac{g^2}{\lambda} \gg 1.
\end{equation}
Similarly, iterated insertions of the self-energy are also no longer suppressed for $g^2/\lambda \gg 1$, therefore, such contributions need to be resummed to all orders.

\begin{figure}[t]
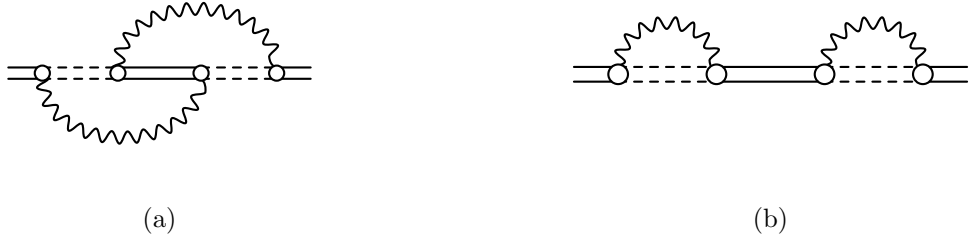

    \centering
    \begin{subfigure}[t]{0.48\linewidth}
        \centering
        \parbox[c][3cm][c]{\linewidth}{%
            \centering
            $\crosseddoublebubble[1]$
        }
        \caption{}
        \label{fig:2PI}
    \end{subfigure}
    \hfill
    \begin{subfigure}[t]{0.48\linewidth}
        \centering
        \parbox[c][3cm][c]{\linewidth}{%
            \centering
             \raisebox{6.5ex}{$\doublebubble[1.3]$}
        }
        \caption{}
        \label{fig:2PR}
    \end{subfigure}
    \caption{ Contributions to the scattering amplitude at $\mathcal{O}(g^4)$ for an incoming and an outgoing state in the repulsive channel. Diagram (a) shows the irreducible two-loop self-energy correction (2PI), while diagram (b) shows the reducible iteration of the one-loop self-energy (2PR).}
    \label{fig:two-loop-contributions}
\end{figure}
The 2PI correction at the same order $g^4$, shown in Fig.~\ref{fig:2PI}, is not enhanced in the same way. Its contribution to the scattering amplitude can be expressed as
\begin{equation}
    i \mathcal{A}^{(2){\rm 2PI}}_{\rm scatt}(\mathbf{k},\mathbf{k}') = \langle s^{(-)}(\mathbf{k}') \rvert \left[- i \hat{\Sigma}^{(2)}\right] \rvert s^{(+)}(\mathbf{k}) \rangle,
    \label{2PI:text:def}
\end{equation}
where $\Sigma^{(2)}$ represents the $\mathcal{O}(g^4)$ self-energy.
In the 2PR contribution, the intermediate repulsive-channel propagator can go on-shell while all matrix elements are evaluated sufficiently close to $k_0$. For a 2PI diagram, imposing the same condition instead restricts the available loop phase space. As shown explicitly in Appendix~\ref{sec:2PI_Diagrams}, the different cuts are suppressed by powers of $\lambda$, and for momenta close to $k_0$ the 2PI contribution therefore remains subleading
\begin{equation}
    \left|\frac{\mathcal{A}^{(2){\rm 2PI}}_{\rm scatt}(k_0)}{\mathcal{A}^{(2){\rm 2PR}}_{\rm scatt}(k_0)}\right| \sim \lambda.
\end{equation}
This distinction identifies the class of diagrams that must be resummed as the 2PR ones.
At $\mathcal{O}(g^2)$, the leading (and only) contribution comes from the one-loop self-energy $\Sigma^{(1)}$. At higher orders and close to $k_0$, the 2PR iterations of $\Sigma^{(1)}$ are no longer perturbatively suppressed, as shown above, hence they must be resummed. By constrast, 2PI corrections, i.e. higher-order corrections to the self-energy like $\Sigma^{(2)}$, remain subleading, with the first of such corrections being suppressed by $\mathcal{O}(\lambda)$ relative to the corresponding 2PR diagram. Therefore, compared to the $\mathcal{O}(g^2)$ amplitude, the 2PI contribution to the $\mathcal{O}(g
^4)$ amplitude remains perturbative
\begin{equation}
    \left|\frac{\mathcal{A}^{(2){\rm 2PI}}_{\rm scatt}(k_0)}{\mathcal{A}^{(1)}_{\rm scatt}(k_0)}\right|
    \sim
    \lambda \times \frac{g^2}{\lambda}
    \sim
    g^2.
\end{equation}
At leading order in this regime, we therefore resum the 2PR iterations of $\Sigma^{(1)}$ to all orders while treating the higher-order 2PI contributions perturbatively.

The loss of perturbativity is directly connected to the apparent violation of unitarity, discussed in Sec.~\ref{sec:Unitarity_Violation}. In fact, considering the expansion of the $S$-matrix in Eq.~\eqref{S-matrix_factorization}, unitarity imposes the $S$-matrix to satisfy
\begin{equation}
    \left|S_{\rm scatt}\right|=\left|1-i\frac{M k }{2\pi} \Sigma^{(1)}(k) + \cdots \right| \leq 1,
\end{equation}
which is strongly violated because the first self-energy correction becomes
\begin{equation}
    \frac{M k_0}{4\pi} \Sigma^{(1)}(k_0) \sim \frac{g^2}{\lambda} \gg 1.
\end{equation}
This is the same condition that makes the subsequent insertions unsuppressed, as shown in Eq.~\eqref{perturbativity}. In fact, the perturbativity conditions in Eqs.~\eqref{free_perturbativity} and~\eqref{geom_perturbativity} coincide with the unitarity conditions in Eqs.~\eqref{free_unitarity} and~\eqref{geom_unitarity}, indicating that the apparent unitarity violation is a manifestation of the breakdown of the perturbative expansion and must be cured by resumming this class of corrections.

This result has a direct interpretation in the Hamiltonian picture of
Sec.~\ref{sec:Hamiltonian_Picture_and_Strong_Mixing}. The self-energy corrections correspond to iterations of the mixing
$H_{\rm off}(E-H_0+i0)^{-1}H_{\rm off}$ between the scattering and the bound states of the two channels. Therefore, the usual perturbative criterion for weak mixing of Eq.~\eqref{weak_mixing} translates schematically into
\begin{equation}
    \left|
    \frac{\langle m|H_{\rm off}|n\rangle}{E_n-E_m}
    \right|
    \ll 1
   \qquad \longrightarrow \qquad
    \left|
    \frac{M k}{4\pi} \Sigma^{(1)}(k)
    \right|
    \ll 1.
\end{equation}

%%%%%%%%%%%%%%%%%%%%%%%%%%%%%%%%%%%%%%%%%%%%%%%%%%%%%%%%%
\section{Renormalization Group and Resummation}
\label{sec:Renormalization_Group_and_Resummation}
%%%%%%%%%%%%%%%%%%%%%%%%%%%%%%%%%%%%%%%%%%%%%%%%%%%%%%%%%

Having isolated the contributions responsible for the loss of perturbativity in the approximation of Eq.~\eqref{A2PR_scatt_estimate}, we now study the systematics of the expansion of these terms around the momentum $k_0$.

%%%%%%%%%%%%%%%%%%%%%%%%%%%%%%%%%%%%%%%%%%%%%%%%%%%%%%%%%
\subsection{Factorization}
\label{sec:Factorization}
%%%%%%%%%%%%%%%%%%%%%%%%%%%%%%%%%%%%%%%%%%%%%%%%%%%%%%%%%

The enhancement of the 2PR contribution is controlled by the region $p \to k_0$, in which the intermediate repulsive-channel state is nearly on-shell. In this region, the momentum dependence of the matrix element can be expanded around its on-shell value, while the momentum dependence of the intermediate repulsive-channel propagator must be retained explicitly. To make this separation manifest, we consider the loop integral appearing in the second $\Sigma^{(1)}$ insertion of Eq.~\eqref{2PR_on-shell}, and we extract the contribution from the on-shell matrix element
\begin{align}
    \label{add_and_subtract}
    \int
    \frac{\mathrm d^3\mathbf p}{(2\pi)^3}
    \frac{ i\left| \langle b \rvert s^{(+)}(p) \rangle \right|^2 }{
        \frac{k_0^2}{M}-\frac{p^2}{M}+i0
    }
    &=
    \left| \langle b \rvert s^{(+)}(k_0) \rangle \right|^2
    \int
    \frac{\mathrm d^3\mathbf p}{(2\pi)^3}
    \frac{i}{
        \frac{k_0^2}{M}-\frac{p^2}{M}+i0
    } \nonumber \\
    &+
    \int
    \frac{\mathrm d^3\mathbf p}{(2\pi)^3}
    \frac{i}{
        \frac{k_0^2}{M}-\frac{p^2}{M}+i0
    }
    \left[
        \left| \langle b \rvert s^{(+)}(p) \rangle \right|^2
        -
        \left| \langle b \rvert s^{(+)}(k_0) \rangle \right|^2
    \right].
\end{align}
The first term isolates the nearly on-shell contribution that is enhanced by the intermediate propagator. In the second term, instead, the difference of matrix elements vanishes for $p\to k_0$, removing this enhancement. The momentum dependence contained in this remainder is therefore not enhanced by the intermediate propagator and contributes only as a correction to the factorized result. The size of this correction can be estimated using the momentum expansion around $k_0$ derived in Eq.~\eqref{matrix_element_expansion}. In particular, defining $\delta p=p-k_0$, we have
\begin{equation}
\label{matrix:element:correction}
    \frac{
        \left|
            \langle b \rvert s^{(+)}(p) \rangle
        \right|^2
        -
        \left|
            \langle b \rvert s^{(+)}(k_0) \rangle
        \right|^2
    }{
        \left|
            \langle b \rvert s^{(+)}(k_0) \rangle
        \right|^2
    }
    = \mathcal{O} \left( \frac{(2 k_0 \delta p)^2}{\Lambda_{\rm EFT}^4} \right),
\end{equation}
where the momentum dependence of the matrix element is suppressed by powers of the scale $\Lambda_{\rm EFT}$, introduced in Eq.~\eqref{Lambda_EFT_def}. This scale represents the thickness of the \emph{momentum surface} $|\mathbf{k}| = k_0$. Moreover, it will play the role of cut-off of the EFT, which is constructed in Sec.~\ref{sec:Momentum-Surface_EFT}.
Exploiting the expansion of the matrix element, we can estimate the first correction to be
\begin{equation}
    \int \frac{{\rm d}^3 \mathbf{p}}{(2\pi)^3} \frac{i \left|\langle b \rvert s^{(+)}(p) \rangle \right|^2}{\frac{k_0^2}{M} - \frac{p^2}{M} + i0} = \left|\langle b \rvert s^{(+)}(k_0) \rangle \right|^2 \int \frac{{\rm d}^3 \mathbf{p}}{(2\pi)^3} \frac{i}{\frac{k_0^2}{M} - \frac{p^2}{M} + i0} \times \left[ 1+ \mathcal{O}\left(\lambda\right) \right],
\end{equation}
as shown explicitly in Appendix~\ref{sec:Corrections_to_Factorization}.
Therefore, the factorized expression of the 2PR amplitude at $\mathcal{O}(g^4)$ reads
\begin{equation}
    \label{A_2PR_bare}
    i \mathcal{A}^{(2){\rm 2PR}}_{\rm scatt}(k_0) = e^{2i\delta(k_0)} \left[ - i \Sigma^{(1)}(k_0) \right]^2 iI(k_0) \times \left[ 1+ \mathcal{O}\left(\lambda\right) \right],
\end{equation}
where we recognize that the loop
\begin{equation}
    \label{I_0_def}
    I(k_0) = \int \frac{{\rm d}^3\mathbf{p}}{(2\pi)^3} \frac{1}{\frac{k_0^2}{M} - \frac{p^2}{M} + i0}
\end{equation}
integrates over the propagator of a free scattering state\footnote{We highlight that this loop integral is identical to the bubble obtained in~\cite{Kaplan:1998we}, which is power divergent. In the context of a system with a large scattering length, the subtraction of the power divergence is fundamental to understanding the RG flow of the 4-fermion operator from marginal to relevant.}. Therefore, after factorization, the scattering state with $k=k_0$ is described as a free state with a large bare effective coupling, which can be identified to be
\begin{equation}
    C^{\rm bare}_0 = \Sigma^{(1)}\left(k_0\right) \left[1 + \mathcal{O}\left(g^2\right) + \mathcal{O}\left(\lambda\right) \right],
    \label{C0:bare}
\end{equation}
where the $\mathcal{O}(g^2)$ corrections correspond to the higher order 2PI corrections, which remain perturbative as discussed in Sec.~\ref{sec:Loss_of_Perturbativity}.
We remark that the original loop integral (left-hand side of Eq.~\eqref{add_and_subtract}) was UV finite because of the scaling of the matrix element at large momentum\footnote{The large momentum scaling of the matrix element $\langle b \rvert s^{(+)}(p) \rangle$ can be estimated from the matrix element of the free scattering state $\langle b \vert k \rangle$, given in Eq.~\eqref{matrix_element_free_scaling}, because at very large momentum, the effect of the potential barrier becomes negligible $\langle b \rvert s^{(+)}(p) \rangle = \mathcal{O}(p^{-3})$ for $p \to \infty$.}. However, after extracting its on-shell value, the free integral in Eq.~\eqref{I_0_def} exhibits a UV linear power divergence, which is cancelled by the equal and opposite power divergence of the correction to the factorized expression (second term in the right-hand side of Eq.~\eqref{add_and_subtract}). Therefore, such correction acts as the fixed-order counterterm of the effective coupling
\begin{equation}
    C_0^{\rm ct} = \left(C_0^{\rm bare}\right)^2 \int \frac{{\rm d}^3\mathbf{p}}{(2\pi)^3} \frac{1}{
        \frac{k_0^2}{M}-\frac{p^2}{M}+i0
    }
    \frac{\left| \langle b \rvert s^{(+)}(p) \rangle \right|^2-\left| \langle b \rvert s^{(+)}(k_0) \rangle \right|^2}{\left| \langle b \rvert s^{(+)}(k_0) \rangle \right|^2}.
\end{equation}
Following~\cite{Kaplan:1998we}, we compute the loop integral in dimensional regularization
\begin{equation}
    \label{I_0_dim_reg}
    I(k_0)
    =
    \left(\frac{\mu_F}{2}\right)^{4-d}
    \int
    \frac{{\rm d}^{d-1} \mathbf{p}}{(2\pi)^{d-1}}
    \frac{1}{
        \frac{k_0^2}{M}-\frac{p^2}{M}+i0
    }
    =
    -\,M\,
    \frac{
        \left(-k_0^2-i0\right)^{\frac{d-3}{2}}
    }{
        (4\pi)^{\frac{d-1}{2}}
    }
    \Gamma\left(\frac{3-d}{2}\right)
    \left(\frac{\mu_F}{2}\right)^{4-d},
\end{equation}
and we subtract its divergence in the power divergence subtraction (PDS) scheme, which corresponds to subtracting from Eq.~\eqref{I_0_dim_reg} the pole present in $d=3$. This requires to introduce a dependence on a \emph{factorization scale} $\mu_F$. In this scheme, the explicit expression of the loop integral reads
\begin{equation}
    I^{\rm PDS}(k_0,\mu_F) = - \frac{M}{4\pi} (\mu_F+ik_0)
\end{equation}
and the fixed-order counterterm becomes
\begin{equation}
    \label{counterterm}
    C_0^{\rm ct}(\mu_F) = - \left( C_0^{\rm bare} \right)^2 \frac{M \mu_F}{4\pi} + \mathcal{O}\left((C_0^{\rm bare})^3\right).
\end{equation}
Therefore, the renormalized coupling can be written as
\begin{align}
    \label{C_0_ren_fo}
    C_0 (\mu_F) &= C_0^{\rm bare} - C_0^{\rm ct} (\mu_F) \nonumber \\
    &= \Sigma^{(1)}(k_0) \left[1 + \frac{M \mu_F}{4\pi} \Sigma^{(1)}(k_0) + \mathcal{O} \left( \left[ \frac{M \mu_F}{4\pi} \Sigma^{(1)}(k_0) \right]^2 \right) \right] \times \left[1 + \mathcal{O}\left(g^2\right) + \mathcal{O}\left(\lambda\right) \right],
\end{align}
and substituting it in Eq.~\eqref{A_2PR_bare}, the renormalized scattering amplitude becomes
\begin{equation}
    i \mathcal{A}^{(2)}_{\rm scatt}(k_0) = e^{2i\delta(k_0)} \left[ - i C_0(\mu_F) \right]^2 iI^{\rm PDS}(k_0,\mu_F) \times \left[ 1+ \mathcal{O}(g^2) +\mathcal{O}\left( \lambda \right) \right].
\end{equation}

At fixed order, the running of the effective coupling can be obtained from Eq.~\eqref{C_0_ren_fo}.
However, this fixed-order running does not contain the full all-order information about the 2PR diagrams in the region where perturbation theory breaks down. In fact, for the natural choice $\mu_F\sim k_0$, the quantity controlling an additional factorized insertion scales as
\begin{equation}
    C_0(\mu_F)\,I^{\rm PDS}(k_0,\mu_F)
    \sim
    \frac{M k_0}{4\pi} \Sigma^{(1)}(k_0) \sim \frac{g^2}{\lambda} \gg 1,
\end{equation}
as it was shown in Sec.~\ref{sec:Loss_of_Perturbativity}, hence all the 2PR insertions are of the same order. Therefore, the consistent approach is to replace the self-energy insertions by the new factorized and $\mu_F$-dependent effective coupling $C_0$, resum these contributions to all orders, and obtain the RG running imposing the $\mu_F$ independence of the resummed scattering amplitude.

%%%%%%%%%%%%%%%%%%%%%%%%%%%%%%%%%%%%%%%%%%%%%%%%%%%%%%%%%
\subsection{Resummation and Fixed Points}
\label{sec:Resummation_and_Fixed_Points}
%%%%%%%%%%%%%%%%%%%%%%%%%%%%%%%%%%%%%%%%%%%%%%%%%%%%%%%%%

The factorization of the self-energy insertions into an effective coupling $C_0$ and the loop integral over a free scattering state allows the 2PR diagrams to take the simple form of a geometric series. Resumming this series leads to the following expression for the amplitude:
\begin{align}
    \label{A_res_series}
    i\mathcal{A}^{\rm res}_{\rm scatt}(k_0)
    &= \propagatorrep[1]+
    \raisebox{2.2ex}{$\singlebubble[1]$}
    +\raisebox{1.55ex}{$\doublebubble[1]$}
    +\cdots
    \nonumber\\
    &= \frac{2\pi}{M k_0} \left(e^{2i\delta(k_0)}-1\right)
    -i e^{2i \delta(k_0)} C_0(\mu_F)
    \sum_{n=0}^\infty
    \left[
        C_0(\mu_F)
        I^{\rm PDS}(k_0,\mu_F)
    \right]^n
    \nonumber\\
    &=\frac{2\pi}{M k_0} \left(e^{2i\delta(k_0)}-1\right)-i e^{2i\delta(k_0)}
    \frac{1}{
        \frac{1}{C_0(\mu_F)}
        +
        \frac{M}{4\pi}(\mu_F+ik_0)
    }.
\end{align}
Imposing the $\mu_F$ invariance of the resummed amplitude, we obtain the equation
\begin{equation}
\label{RGE_C_0}
    \frac{{\rm d}}{{\rm d}\mu_F}
    \left[
        \frac{1}{C_0(\mu_F)}
        +
        \frac{M\mu_F}{4\pi}
    \right]
    =
    0.
\end{equation}
The RG equation determines the $\mu_F$ dependence of $C_0$, but not its integration constant, which encodes the microscopic nature of the interaction, represented by $\Sigma^{(1)}(k_0)$. Therefore, matching the analytic part of the inverse amplitude to the microscopic theory gives
\begin{equation}
    \label{RGE_C_0_BC}
    \left[
        \frac{1}{C_0(\mu_F)}
        +
        \frac{M\mu_F}{4\pi}
    \right]_{\mu_F=k_0}
    =
    \frac{1}{\Sigma^{(1)}(k_0)} \left[1+\mathcal{O}\left(g^2\right) + \mathcal{O}(\lambda)\right].
\end{equation}
Solving the RG equation, together with the initial condition from Eq.~\eqref{RGE_C_0_BC}, we obtain the running of the effective coupling
\begin{equation}
    \label{RGE_C_0_sol}
    C_0(\mu_F)
    =
    \frac{1}{\frac{1}{\Sigma^{(1)}(k_0)}
        -
        \frac{M\mu_F}{4\pi} 
    },
\end{equation}
which leads to the expression of the scattering amplitude
\begin{equation}
    \label{A_res}
    i \mathcal{A}^{\rm res}_{\rm scatt}(k_0) = \frac{2\pi}{M k_0} \left(e^{2i\delta(k_0)} \frac{\frac{1}{\Sigma^{(1)}(k_0)}-i \frac{Mk_0}{4\pi}}{\frac{1}{\Sigma^{(1)}(k_0)}+ i \frac{Mk_0}{4\pi}} -1\right).
\end{equation}
To study the properties of the RG flow of $C_0$, it is convenient to define the dimensionless coupling
\begin{equation}
    \widehat C_0(\mu_F)
    =
    \frac{M\mu_F}{4\pi}\,
    C_0(\mu_F),
\end{equation}
whose RG equation reads
\begin{equation}
    \mu_F
    \frac{{\rm d}\widehat C_0(\mu_F)}{{\rm d}\mu_F}
    =
    \widehat C_0(\mu_F)
    \left[
        1+\widehat C_0(\mu_F)
    \right].
    \label{RG:CO}
\end{equation}
It therefore exhibits two fixed points
\begin{equation}
    \widehat C_0^* = 0,
    \qquad
    \widehat C_0^* = -1.
\end{equation}
corresponding to the non-interacting and unitary fixed points, respectively.

For the naive scaling for the matrix element, as in Eq.~\eqref{free_perturbativity}, the self-energy insertions are perturbative
\begin{equation}
    \left|
        \frac{Mk}{4\pi}\Sigma^{(1)}(k)
    \right|
    \ll1.
\end{equation}
For generic $\mu_F\sim k$, the effective coupling can then be expanded as
\begin{equation}
    \label{C_0_non-int_expansion}
    C_0(\mu_F)= \Sigma^{(1)}(k) \left[ 1 + \frac{M\mu_F}{4\pi} \Sigma^{(1)}(k) + \mathcal{O} \left( \left( \frac{M\mu_F}{4\pi} \Sigma^{(1)}(k) \right)^2 \right) \right],
\end{equation}
so that the interaction remains close to the non-interacting fixed point
$\widehat C_0^*=0$. We highlight that this expansion reproduces the
fixed-order running of $C_0$ obtained in Eq.~\eqref{C_0_ren_fo}.

Near $k_0$, instead, the matrix element reaches its geometric scaling and the self-energy insertions become non-perturbative, as in Eq.~\eqref{geom_perturbativity}. In the regime
\begin{equation}
    \left|
        \frac{Mk_0}{4\pi}\Sigma^{(1)}(k_0)
    \right|
    \gg1 ,
\end{equation}
the effective coupling can be expanded around the fixed point $\widehat C_0^*=-1$ as
\begin{equation}
    \label{C_0_unitary_expansion}
    C_0(\mu_F) = - \frac{4\pi}{M \mu_F} \left[1+\mathcal{O} \left(\frac{4\pi}{M \mu_F \Sigma^{(1)}(k_0)}\right)\right].
\end{equation}
To characterize the departure from this fixed point, we write
\begin{equation}
    \widehat C_0
    =
    -1+\delta\widehat C_0.
\end{equation}
Linearizing the RG equation around $\widehat C_0^*=-1$ gives
\begin{equation}
    \mu_F
    \frac{{\rm d}\delta\widehat C_0}{{\rm d}\mu_F}
    =
    -\delta\widehat C_0
    +
    \mathcal O\left(
        \delta\widehat C_0^2
    \right),
\end{equation}
so that $\delta\widehat C_0\propto 1/\mu_F$. The unitary fixed point therefore possesses a relevant direction. This means that an arbitrarily small departure grows as $\mu_F$ is lowered, and the fixed point is reached only for a tuned value of the microscopic coupling. 

Using the matching condition at $k_0$, Eq.~\eqref{RGE_C_0_BC}, the parameter controlling this relevant deformation can be identified with the inverse of a \emph{scattering length} 
\begin{equation}
    \label{scattering_length}
    \frac{1}{a}
    =
    \frac{4\pi}{
        M\Sigma^{(1)}(k_0)
    }
    \left[
        1
        +
        \mathcal O\left(
            g^2
        \right)
    \right],
\end{equation}
in terms of $a$,
\begin{equation}
    \widehat C_0(\mu_F)
    =
    -\frac{1}{
        1-\dfrac{1}{\mu_F a}
    }
    =
    -1
    -
    \frac{1}{\mu_F a}
    +
    \mathcal O\left(
        \frac{1}{(\mu_F a)^2}
    \right).
    \label{Dimensionless:C0}
\end{equation}
where the second equality holds for $|\mu_F a| \gg1$, namely close to the unitary fixed point. This reproduces the linearized flow with $\delta\widehat C_0=-1/(\mu_F a)$, and the fixed point is exactly recovered only for $|a|\to\infty$. We stress that $\widehat C_0^*=-1$ is a UV fixed point in the sense that, for any finite $a$, it is approached as $\mu_F\to\infty$.

Having adopted the same notation as~\cite{Kaplan:1998we} shows the analogy between this system and the problem of a bound state close to threshold considered there, where $a$ is the standard two-body scattering length. In the present case, $a$ likewise characterizes the two-body scattering dynamics, however the difference is that here this scattering parameter is defined around the finite momentum $k_0$, where the interaction becomes strong, rather than through the threshold expansion around $k=0$.
This suggests constructing an effective description directly around this
finite-momentum region. Writing
\begin{equation}
    k=k_0+\delta k,
    \qquad
    |\delta k|\ll k_0,
\end{equation}
the scale $k_0$ specifies the momentum surface, while the residual
momentum $\delta k$ describes the slow dynamics in its vicinity. This
separation naturally motivates an effective field theory organized around
the momentum surface $|\mathbf{k}|=k_0$, which will be constructed in Sec.~\ref{sec:Momentum-Surface_EFT}.

%%%%%%%%%%%%%%%%%%%%%%%%%%%%%%%%%%%%%%%%%%%%%%%%%%%%%%%%%
\subsection{Unitarity}
\label{sec:Unitarity}
%%%%%%%%%%%%%%%%%%%%%%%%%%%%%%%%%%%%%%%%%%%%%%%%%%%%%%%%%

The resummation obtained in Sec.~\ref{sec:Resummation_and_Fixed_Points} leads once again to an $S$-matrix compatible with unitarity.
Eq.~\eqref{A_res} makes explicit that the full $S$-matrix factorizes into the scattering from the potential barrier in the repulsive channel and the effect of channel mixing
\begin{equation}
    \label{S_res}
    S^{\rm res}_{\rm scatt}(k_0) =e^{2i\delta(k_0)} \times\frac{\frac{1}{\Sigma^{(1)}(k_0)}-i\frac{Mk_0}{4\pi}}{\frac{1}{\Sigma^{(1)}(k_0)}+i\frac{Mk_0}{4\pi}},
\end{equation}
as discussed in Sec.~\ref{sec:Perturbative_Expansion}. Since the contribution from the potential barrier is a phase, it does not affect the modulus of the $S$-matrix. The violation of unitarity is caused by the large self-energy insertions, as explained in Sec.~\ref{sec:Loss_of_Perturbativity}, which have now been resummed. Close to the unitary fixed point, where the inverse self-energy can be treated as a small perturbation, as in Eq.~\eqref{C_0_unitary_expansion}, unitarity is restored:
\begin{equation}
    \left|S^{\rm res}_{\rm scatt}(k_0)\right| = \left|\frac{ \frac{1}{\Sigma^{(1)}(k_0)}-i\frac{Mk_0}{4\pi}}{\frac{1}{\Sigma^{(1)}(k_0)}+i\frac{Mk_0}{4\pi}}\right| = 1+\mathcal{O} \left(\frac{4\pi}{M k_0 \Sigma^{(1)}(k_0)}\right).
\end{equation}
Equivalently, after resummation, the capture cross section is compatible with the inelastic unitarity bound, which was shown to be violated at fixed perturbative order in Sec.~\ref{sec:Unitarity_Violation}. The resummed capture cross section can be obtained by summing over all the inelastic cuts (cuts of the self-energies) of all the diagrams appearing in Eq.~\eqref{A_res_series}, which gives
\begin{align}
    \label{resummed_capture_cross_section}
    (\sigma v)_{\rm cap}^{\rm res} (k_0) &= \frac{q}{2\pi} \times \left| \raisebox{0.6ex}{$\cutbubble[1]$}+ \raisebox{2.1ex}{$\cutdoublebubble[1]$}+\raisebox{2.0ex}{$\cuttriplebubble[1]$}+\ldots\right|^2 \nonumber \\
    &=
    \frac{
        -2\,\operatorname{Im}\Sigma^{(1)}(k_0)
    }{
        \left|
            1+\frac{iM k_0}{4\pi}\Sigma^{(1)}(k_0)
        \right|^2
    }
\end{align}
This expression  explicitly satisfies the optical theorem in the form
\begin{equation}
    (\sigma v)_{\rm cap} = \frac{2\pi}{M k} \left(1-\left|S_{\rm scatt}\right|^2\right).
\end{equation}
In fact, substituting the expression for the $S$-matrix from Eq.~\eqref{S_res}, the optical theorem can be written as
\begin{equation}
    (\sigma v)_{\rm cap}^{\rm res} = 2 \operatorname{Im} \mathcal{A}^{\rm res}_{\rm scatt} - \frac{Mk}{2\pi} \left|\mathcal{A}^{\rm res}_{\rm scatt}\right|^2,
\end{equation}
where the second term on the right-hand side subtracts the contribution of the elastic cuts (cuts of the repulsive-channel propagators) from the full imaginary part of the amplitude.

%%%%%%%%%%%%%%%%%%%%%%%%%%%%%%%%%%%%%%%%%%%%%%%%%%%%%%%%%
\section{Momentum-Surface EFT}
\label{sec:Momentum-Surface_EFT}
%%%%%%%%%%%%%%%%%%%%%%%%%%%%%%%%%%%%%%%%%%%%%%%%%%%%%%%%%

In the previous section, we showed that the breakdown of perturbation
theory near momentum $k_0$ is associated with the repeated 2PR propagation of
the repulsive-channel state. At $k=k_0$, the transition matrix elements
factorize and the corresponding coefficient $C_0$ can be resummed to all
orders, restoring unitarity and giving rise to the unitary RG fixed
point discussed above.
This naturally suggests constructing an EFT directly around the momentum
surface $|\mathbf{k}|=k_0$. Expansions around a momentum surface are standard in systems with a Fermi surface~\cite{Shankar:1993pf,Polchinski:1992ed}, where the EFT is organized around the Fermi momentum and describes small residual momenta. In the present case, the relevant surface is defined by $|\mathbf{k}|=k_0$, and the residual momentum around it is
\begin{equation}
    \delta k=k-k_0,
    \qquad {\rm with} \quad
    |\delta k|\ll k_0,
\end{equation}

The momentum-surface EFT involves two expansions. First, the condition $|\delta k|/k_0\ll1$ defines the thin shell around $k_0$ and allows us to linearize the kinematics around the momentum surface. Corrections to this approximation are therefore controlled by $\delta k/ k_0$.
Second, the microscopic interaction has a much stronger momentum dependence around $k_0$. As discussed in Sec.~\ref{sec:Momentum_Dependence}, the enhanced matrix element is centered around its maximum at $k_0$, and its leading momentum dependence is even in $\delta k$. The corresponding local expansion, shown in Eq.~\eqref{matrix_element_expansion}, is controlled by the smaller scale
\begin{equation}
    \Lambda_{\rm EFT}^2
    =
    \frac{(\pi \Lambda_b)^{3}}{k_0} =\lambda k_0^2,
\end{equation}
with $\lambda \ll 1$.
The EFT therefore describes the dynamics for residual momentum satisfying the hierarchy
\begin{equation}
\label{EFT:expansion:hierarchy}
    |\delta k|
    \lesssim
    \frac{\Lambda_{\rm EFT}^2}{k_0}
    \ll
    k_0.
\end{equation}
The expansion in $(k_0\delta k)/ \Lambda_{\rm EFT}^2$ controls the momentum dependence inherited from the leading microscopic interaction, while corrections to the linearized momentum-surface description are suppressed by $\delta k/k_0$.
In the following, we construct this EFT, describe its power counting and match its Wilson coefficients to the microscopic toy model.

%%%%%%%%%%%%%%%%%%%%%%%%%%%%%%%%%%%%%%%%%%%%%%%%%%%%%%%%%
\subsection{Lagrangian}
\label{sec:Lagrangian}
%%%%%%%%%%%%%%%%%%%%%%%%%%%%%%%%%%%%%%%%%%%%%%%%%%%%%%%%%

To expand the momentum around $k_0$, we write $k=k_0+\ell$ with $|\ell|\ll k_0$, where $\ell$ denotes the residual momentum. For power-counting purposes, we will denote by $\delta k$ the characteristic residual momentum of the physical process, so that $\ell\sim\delta k$. Removing the common reference energy $k_0^2/M$, the residual dispersion relation is
\begin{equation}
\frac{k^2-k_0^2}{M} = \frac{2k_0\ell+\ell^2}{M} = 2\frac{k_0 \ell}{M}\left[1+\mathcal{O}\left(\frac{\delta k}{k_0}\right)\right].
\end{equation}
The expansion of the kinetic term is therefore controlled by $\delta k/k_0$. The interactions are expanded independently in local operators around the same momentum surface, with their natural momentum dependence organized in powers of $k_0\delta k$ and controlled by the smaller scale $\Lambda_{\rm EFT}^2$, as described by Eq.~\eqref{EFT:expansion:hierarchy}.

To construct the theory explicitly, we project the two-particle $\xi\xi$ sector onto the $s$-wave and separate the rapid momentum and energy dependence associated with $k_0$ from the slow residual dynamics. The remaining modes describing the radial motion are captured by the reduced radial field $\chi(\mathbf{R},\ell,t)$, which is defined as
\begin{equation}
    r \int \frac{ {\rm d}\Omega_{\hat{\mathbf{r}}}}{\sqrt{4\pi}} \Phi_{\xi\xi}(\mathbf{R}, \mathbf{r},t) = e^{-i\frac{k_0^2}{M}t}
    \left[
        e^{ik_0r}\chi(\mathbf{R},r,t)
        -
        e^{-ik_0r}\chi(\mathbf{R},-r,t)
    \right].
\end{equation}
Near the momentum surface, the center-of-mass motion factorizes from the relative dynamics, so that the low-energy expansion is entirely organized in the residual radial momentum. The details of this construction, together with the normalization of the field and the corresponding Feynman rules, are given in Appendix~\ref{sec:momentum_surface_construction}.

Keeping the leading terms in $\delta k/k_0$, the momentum-surface EFT can be written in position space as
\begin{align}
\label{EFT_Lagrangian}
    \mathcal{L}_{\rm EFT}
    ={}&
    \int {\rm d}r\,
    \chi^\dagger(\mathbf{R},r,t)
    \left(
        i\partial_t
        +\frac{\nabla_{\mathbf R}^2}{4M}
        +2i\frac{k_0}{M}\partial_r
    \right)
    \chi(\mathbf{R},r,t)
    \nonumber\\
    &-
    \frac{k_0^2}{\pi}
    \int {\rm d}r\,
    \chi^\dagger(\mathbf{R},r,t)
    \delta(r)
    \Bigg[
        C_0
        -2i C_1 k_0
        \overleftrightarrow{\partial_r}
        \nonumber\\
    &\hspace{4.5cm}
        -2C_2^{(1)}k_0^2
        \left(
            \overrightarrow{\partial_r^2}
            +
            \overleftarrow{\partial_r^2}
        \right)
        +4C_2^{(2)}k_0^2
        \overleftarrow{\partial_r}\,
        \overrightarrow{\partial_r}
        +\cdots
    \Bigg]
    \chi(\mathbf{R},r,t).
\end{align}
The coefficients $C_{n}$ denote the Wilson coefficients of the local operators in the EFT expansion. At a given order in $\delta k$, several derivative structures may appear, as illustrated by $C_2^{(1)}$ and $C_2^{(2)}$ above. The equivalent momentum-space form of the EFT is given in Eq.~\eqref{Momentum:shell:EFT}.

As explained at the beginning of this section, the linearized dispersion relation requires $|\delta k|/k_0\ll1$, while the derivative expansion of the interactions, which is organized in powers of $k_0\delta k$, is controlled by the smaller scale $\Lambda_{\rm EFT}^2$, giving rise to the hierarchy of scales in Eq.~\eqref{EFT:expansion:hierarchy}. The latter condition is therefore the more restrictive one, and sets the actual range of validity of the momentum-surface EFT.

%%%%%%%%%%%%%%%%%%%%%%%%%%%%%%%%%%%%%%%%%%%%%%%%%%%%%%%%%
\subsection{Power Counting}
\label{sec:Power_Counting}
%%%%%%%%%%%%%%%%%%%%%%%%%%%%%%%%%%%%%%%%%%%%%%%%%%%%%%%%%

Denoting $\delta k$ the characteristic residual momentum scale of the process as in the previous section
\begin{equation}
    \ell\sim\delta k,
    \qquad
    \partial_r\sim\delta k,
    \qquad
    \partial_t\sim\frac{k_0\delta k}{M}.
\end{equation}
Introducing a characteristic center-of-mass momentum $P$, the kinetic action fixes the scaling of the momentum-surface field to
\begin{equation}
    \chi(\mathbf R,r,t)
    \sim
    P^{3/2}(\delta k)^{1/2},
\end{equation}
so that the kinetic lagrangian scales as
\begin{equation}
    \mathcal L_{\rm kin}
    \sim
    P^3\frac{k_0\delta k}{M}.
\end{equation}
For an interaction contributing at order $(\delta k)^n$, the radial $s$-wave measure gives an overall factor $k_0^2$, while each additional pair of derivatives around the momentum surface gives an extra factor of order $k_0\delta k$. The corresponding interaction therefore scales as
\begin{equation}
    \mathcal L_n
    \sim
    P^3 \delta k \, k_0^2 C_{n}\,
    (k_0\delta k)^n.
    \label{Lagrangian:scaling}
\end{equation}

As shown in the previous section, for $k \to k_0$ the leading interaction $C_0$ lies close to the unitary fixed point, as in Eq.~\eqref{C_0_unitary_expansion}. Therefore, at $\mu_F \sim k_0$, it scales as
\begin{equation}
    C_0
    \sim
    \frac{1}{M k_0}.
\end{equation}
Using the scaling above for the interaction with $n=0$, we then find
\begin{equation}
    \label{L_0_sim_L_kin}
    \mathcal L_0
    \sim
    P^3 k_0^2 C_0\,\delta k
    \sim
    P^3\frac{k_0\delta k}{M}
    \sim
    \mathcal L_{\rm kin},
\end{equation}
showing explicitly that the leading interaction contributes at the same order as the kinetic term and must therefore be treated non-perturbatively.
We remark that the Lagrangian of Eq.~\eqref{EFT_Lagrangian}, together with the power counting of Eq.~\eqref{L_0_sim_L_kin} represents an example of a theory of \emph{fermions at unitarity}~\cite{Ho_2004, Braaten:2004rn,Nishida:2006eu,Nishida:2010tm}, with a linear dispersion relation~\cite{Shankar:1993pf,Polchinski:1992ed}.

Moreover, in the considered microscopic model, the expansion in the residual momentum around $k_0$ breaks down at $k_0 \delta k\sim \Lambda_{\rm EFT}^2$, therefore the momentum-surface EFT should be organized as an expansion in the residual momentum over this breakdown scale. In particular, the interaction terms must satisfy
\begin{equation}
    \frac{\mathcal L_{n+1}}{\mathcal L_n}
    \sim
    \frac{k_0 \delta k}{\Lambda_{\rm EFT}^2}, \qquad
    \frac{\mathcal{L}_n}{\mathcal{L}_0} \sim \frac{\Lambda_{\rm EFT}^2}{k_0^2} \left(\frac{k_0 \delta k}{\Lambda_{\rm EFT}^2}\right)^n,
\end{equation}
for $n>0$. Therefore, the higher-derivative operators are perturbative for $k_0 \delta k\ll\Lambda_{\rm EFT}^2$.
Using the scaling in Eq.~\eqref{Lagrangian:scaling}, the Wilson coefficients with $n>0$ scale as
\begin{equation}
    \frac{C_{n+1}}{C_{n}}
    \sim
    \frac{1}{\Lambda_{\rm EFT}^2}, \qquad C_n \sim \frac{1}{M k_0^3 \Lambda_{\rm EFT}^{2n-2}},
    \label{power:counting:hierarchy}
\end{equation}
in the absence of additional structure. Our choice of performing the EFT expansion around the maximum of the matrix element introduces an additional symmetry in the microscopic theory, under which the amplitude is symmetric up to subleading corrections for $\delta k \mapsto -\delta k$, verified explicitly in Appendix~\ref{sec:Expansion_around_the_Maximum} and shown in Eq.~\eqref{matrix_element_expansion}. 
Consequently, only the even terms in $\delta k$ are controlled by the microscopic scale $\Lambda_{\rm EFT}$. By contrast, the odd terms in $\delta k$ arise from corrections to the linearized momentum-surface description and are therefore suppressed by powers of $\delta k/k_0$. This implies that the coefficients $C_n$ with odd $n$ are additionally suppressed by $\Lambda_{\rm EFT}^2/k_0^2\sim\lambda$ relative to the coefficients with even $n$. The power counting introduced above will be verified explicitly by matching to the microscopic toy model. 

In this discussion, we have not yet distinguished between bare $C^{\rm bare}_n$ and renormalized $C_n(\mu_F)$ Wilson coefficients with respect to the factorization scale. The Lagrangian of Eq.~\eqref{EFT_Lagrangian} is to be understood as written in terms of the bare coefficients. However, the scalings of the coefficients of Eq.~\eqref{power:counting:hierarchy} only arise after renormalization with respect to the factorization scale $\mu_F$.

%%%%%%%%%%%%%%%%%%%%%%%%%%%%%%%%%%%%%%%%%%%%%%%%%%%%%%%%%
\subsection{Matching to the Toy Model}
\label{sec:Matching_to_the_Toy_Model}
%%%%%%%%%%%%%%%%%%%%%%%%%%%%%%%%%%%%%%%%%%%%%%%%%%%%%%%%%

In Sec.~\ref{sec:Resummation_and_Fixed_Points}, the matching was performed exactly at $k=k_0$. After factorizing the enhanced 2PR contribution, the bare interaction was identified at leading order with the one-loop self-energy, up to the higher-order corrections indicated in Eq.~\eqref{C0:bare}. We now extend this matching to momenta in the vicinity of $k_0$, which leads to the definition of the momentum-dependent interaction
\begin{align}
    \label{Cbare_k}
    C^{\rm bare}(k) = \Sigma^{(1)}(k) \left[1 + \mathcal{O}(g^2) + \mathcal{O}(\lambda) \right],
\end{align}
where $\mathcal{O}(g^2)$ denotes the higher-order corrections to the self-energy (discussed in Appendix~\ref{sec:2PI_Diagrams}), while $\mathcal{O}(\lambda)$ denotes the corrections to the factorization of the 2PR diagrams (discussed in Appendix~\ref{sec:Corrections_to_Factorization}). An explicit expression for these terms is given in Eq.~\eqref{C:bare:app}.
For $(k_0 \delta k) \ll \Lambda_{\rm EFT}^2$, the expansion of $C^{\rm bare}$ can be identified with the EFT expansion in terms of the bare Wilson coefficients
\begin{equation}
    C^{\rm bare}(k)
    =
    \sum_{n=0}^{\infty}
    C^{\rm bare}_n
    (2k_0\delta k)^n.
    \label{EFT_tree_matching}
\end{equation}

It is convenient to study the $\mu_F$ dependence of the renormalized Wilson coefficients after having resummed the amplitude to all orders, as already discussed in Sec.~\ref{sec:Factorization} and in Sec.~\ref{sec:Resummation_and_Fixed_Points}.
Therefore, in analogy with Eq.~\eqref{A_res_series}, the expression of the resummed amplitude can be extended for $\delta k \ll \Lambda_{\rm EFT}^2/k_0 $ to
\begin{equation}
    \label{A_res_general_k}
    i\mathcal A_{\rm scatt}^{\rm res}(k)
    =
    \frac{2\pi}{Mk}
    \left(e^{2i\delta(k)}-1\right)
    -
    i e^{2i\delta(k)}
    \frac{1}{
        \dfrac{1}{C(\mu_F,k)}
        +\frac{M}{4\pi} (\mu_F+ik)
    },
\end{equation}
where $C(\mu_F,k)$ represents the renormalized momentum-dependent interaction. Its explicit expression can be obtained by requiring the resummed amplitude to be $\mu_F$-independent
\begin{equation}
    \frac{{\rm d}}{{\rm d}\mu_F}
    \left[
        \frac{1}{C(\mu_F,k)}
        +
        \frac{M\mu_F}{4\pi}
    \right]
    =
    0,
    \label{RG:C}
\end{equation}
together with the initial condition of the RG equation\footnote{We remark that the RG flow of the interaction is understood with respect to the factorization scale $\mu_F$. This scale is distinct from the renormalization scale $\mu_R^{\rm EFT}$ of the momentum-surface EFT.}
\begin{equation}
    \left[\frac{1}{C(\mu_F,k)} + \frac{M \mu_F}{4\pi}\right]_{\mu_F=k} = \frac{1}{C^{\rm bare}(k)} = \frac{1}{\Sigma^{(1)}(k)}\left[1+\mathcal{O}(g^2)+\mathcal{O}(\lambda)\right],
    \label{matching_general_k}
\end{equation}
which leads to
\begin{equation}
    \label{C_explicit}
    C(\mu_F,k) = \frac{1}{\frac{1}{\Sigma^{(1)}(k)} - \frac{M \mu_F}{4\pi}}.
\end{equation}

The expression of the resummed amplitude obtained in Eq.~\eqref{A_res_general_k} suggests to define an \emph{effective range expansion}~\cite{Bethe:1949yr} to capture its momentum dependence. Therefore, the analytic part of its denominator can be expanded around the momentum surface as
\begin{equation}
    i\mathcal A_{\rm scatt}^{\rm res}(k)
    =
    \frac{2\pi}{Mk}
    \left(e^{2i\delta(k)}-1\right)
    -
    i e^{2i\delta(k)} \frac{4\pi}{M}
    \frac{1}{
        \frac{1}{a}
    -
    \frac{\Lambda_{\rm EFT}}{2}
    \sum_{n=1}^{\infty}
    \left(\Lambda_{\rm EFT} r_n\right)
    \left[
        \frac{2k_0\delta k}{\Lambda_{\rm EFT}^2}
    \right]^n+ik
    },
\end{equation}
where the scattering length $a$ has already been identified in Eq.~\eqref{scattering_length} and the parameters $r_n$ are the equivalent of the effective range parameters. We highlight that, differently from the standard scenario, the effective range expansion we introduce is defined around a finite momentum surface, rather than around threshold. The parameters can be identified from the expansion of the one-loop self-energy\footnote{In general, the self-energy depends on the renormalization scale $\mu_R$ of the microscopic model, as shown in Eq.~\eqref{A1_scatt}. To match its expansion with the EFT expansion, we set it to the value $\mu_R \sim k^2/M + |E_b|$, which makes the logarithm of Eq.~\eqref{A1_scatt} $\mathcal{O}(1)$.} from Eq.~\eqref{A1_scatt}
\begin{align}
    \label{shell_effective_range_expansion}
    \frac{4\pi}{M \Sigma^{(1)}(k)}
    &=
    \frac{1}{a}
    -
    \frac{\Lambda_{\rm EFT}}{2}
    \sum_{n=1}^{\infty}
    \left(\Lambda_{\rm EFT} r_n\right)
    \left[
        \frac{2k_0\delta k}{\Lambda_{\rm EFT}^2}
    \right]^n
    \nonumber\\
    &=
    \frac{1}{a}
    -\frac{r_1}{2}(2k_0\delta k)
    -\frac{r_2}{2\Lambda_{\rm EFT}^2}(2k_0\delta k)^2
    -\frac{r_3}{2\Lambda_{\rm EFT}^4}(2k_0\delta k)^3
    -\cdots.
\end{align}
For the considered toy model, their explicit expression, as well as their general scaling, reads as follows
\begin{align}
    a&= - \frac{g^2}{4\pi} (1+i\pi) \frac{k_0}{\Lambda_{\rm EFT}^2} = - \frac{1+i\pi}{4\pi} \frac{g^2}{\lambda} \frac{1}{k_0},
    \label{matching_a}
    \\[4pt]
    r_1
    &= \mathcal{O}\left( \frac{\Lambda_{\rm EFT}^2}{g^2k_0^3} \right) = \mathcal{O} \left( \frac{\lambda}{g^2} \frac{1}{k_0} \right),
    \label{matching_r1}
    \\[4pt]
    r_2
    &= \frac{2\pi^3}{g^2(1+i\pi)} \frac{1}{k_0},
    \label{matching_r2}
    \\[4pt]
    r_{2n+1}
    &= \mathcal{O}\left( \frac{\Lambda_{\rm EFT}^2}{g^2k_0^3} \right) = \mathcal{O} \left( \frac{\lambda}{g^2} \frac{1}{k_0} \right),
    \label{matching_r2np1}
    \\[4pt]
    r_{2n}
    &= \mathcal{O}\left( \frac{1}{g^2 k_0} \right).
    \label{matching_r2n}
\end{align}
Such scalings show that the regime of strong unitarity violation ($g^2/\lambda ={\rm fixed} \gg 1$, as in Eq.~\eqref{parametrics}) can be understood as the condition of a large scattering length
\begin{equation}
    k_0 |a| \sim \frac{g^2}{\lambda} \gg 1.
    \label{k0a:eq}
\end{equation}
Moreover, since the $r_{2n}$ and the $r_{2n+1}$ parameters have $n$-independent scaling, we have $r_{2n+2}/r_{2n}=\mathcal{O}(1)$ and $r_{2n+1}/r_{2n-1}=\mathcal{O}(1)$, therefore the effective range expansion is well defined.
As already mentioned in Sec.~\ref{sec:Power_Counting}, the parameters $r_n$ for odd $n$ are further suppressed by an extra factor of $\lambda$, which is a consequence of our choice of performing the EFT expansion around the maximum of the matrix element $k_0$, described in Sec.~\ref{sec:Expansion_around_the_Maximum}. We highlight that all the parameters are complex as they arise from the expansion of the self-energy, which has on-shell cuts.

The renormalized Wilson coefficients can be obtained by expanding the interaction
\begin{equation}
    C(\mu_F,k) = \sum_{n=0}^\infty C_n(\mu_F) (2k_0\delta k)^n,
\end{equation}
and by comparing order by order with the expansion of Eq.~\eqref{C_explicit}.
The first three Wilson coefficients can be expressed in terms of the effective range parameters in the usual form
\begin{align}
    C_0(\mu_F)
    &=
    \frac{4\pi}{M}
    \frac{1}{
        \dfrac{1}{a}-\mu_F
    },
    \label{C0_matching}
    \\[4pt]
    C_1(\mu_F)
    &=
    \frac{4\pi}{M}
    \frac{r_1/2}{
        \left(
            \dfrac{1}{a}-\mu_F
        \right)^2
    },
    \label{C1_matching}
    \\[4pt]
    C_2(\mu_F)
    &=\frac{4\pi}{M}
    \left(\frac{1}{\frac{1}{a}-\mu_F}\right)^3 \left[\frac{r_1^2}{4} + \frac{1}{2} \frac{r_2}{\Lambda_{\rm EFT}^2} \left(\frac{1}{a} - \mu_F\right)\right].
    \label{C2_matching}
\end{align}
After renormalization, we can now study the scaling of the Wilson coefficients. For $\mu_F\sim k_0$, and more generally for $\mu_F \gg 1/a$, the toy model sits close to the unitary fixed point\footnote{For $\mu_F\lesssim1/|a|$, the flow of Eq.~\eqref{Dimensionless:C0} drives the coupling towards the non-interacting fixed point. However, this EFT is valid only in a thin shell around $k_0$, so extrapolating the flow to $\mu_F\lesssim1/|a|$ lies beyond its domain of validity. Such scales correspond to momenta far from the surface, where the mixing is weak. There, the appropriate description is fixed-order perturbation theory in the original degrees of freedom.} of Eq.~\eqref{Dimensionless:C0}. Moreover, Eq.~\eqref{C2_matching} for $C_2$ contains two terms, one proportional to $r_1^2$ and one proportional to $r_2$. Using the scalings of Eqs.~\eqref{matching_r1} and~\eqref{matching_r2}, one can explicitly verify that the $r_2$ term always dominates $C_2$ within the regime $\mu_F\gg1/a$. Lastly, taking this into account and using Eqs.~\eqref{matching_r2np1} and~\eqref{matching_r2n}, we can explicitly verify the power counting
\begin{equation}
\label{C_scalings_general}
C_n
\sim \frac{r_n}{M\mu_F^2\Lambda_{\rm EFT}^{2n-2}} \sim
\frac{1}{M g^2 k_0 \mu_F^2 \Lambda_{\rm EFT}^{2n-2}}\times
\begin{cases}
1, & n\ \text{even},\\[2mm]
\lambda, & n\ \text{odd}.
\end{cases}
\end{equation}
for $n \geq 1$. For $\mu_F \sim k_0$, this reproduces the scaling derived in Sec.~\ref{sec:Power_Counting}, and in particular in Eq.~\eqref{power:counting:hierarchy}. The odd operators carry the additional suppression $\Lambda_{\rm EFT}^2/k_0^2\sim \lambda$.

%%%%%%%%%%%%%%%%%%%%%%%%%%%%%%%%%%%%%%%%%%%%%%%%%%%%%%%%%
\subsection{Factorization Scale and Fixed-Points in the EFT}
\label{sec:Factorization_Scale_in_the_EFT}
%%%%%%%%%%%%%%%%%%%%%%%%%%%%%%%%%%%%%%%%%%%%%%%%%%%%%%%%%
We now study how the factorization-scale dependence is inherited by the local operators of the momentum-surface EFT. Substituting the expansion of the momentum-dependent interaction of Eq.~\eqref{EFT_tree_matching} into the RG equation of Eq.~\eqref{RG:C}, we obtain
\begin{equation}
    \frac{{\rm d}C_n}{{\rm d}\mu_F}
    =
    \frac{M}{4\pi}
    \sum_{m=0}^{n} C_m C_{n-m} .
    \label{local_C_RGE}
\end{equation}
The local expansion is organized in powers of the residual momentum $\delta k$ around the fixed surface $k=k_0$, so that $k^2-k_0^2\simeq2k_0\delta k$ counts as one power of the expansion. Since each term contains a factor $(2k_0)^n$ and $[C_n]=1/(M p^{2n+1})$, the natural dimensionless coefficient associated with the local operator basis is
\begin{equation}
    \widehat C_n
    \equiv
    \frac{M}{4\pi}
    (2k_0)^n\mu_F^{n+1}C_n,
\end{equation}
whose RG equation becomes
\begin{equation}
    \beta_{n}
    \equiv
    \mu_F\frac{{\rm d}\widehat C_{n}}{{\rm d}\mu_F}
    =
    (n+1)\widehat C_{n}
    +
    \sum_{m=0}^{n}
    \widehat C_{m}\widehat C_{n-m}.
    \label{hn_beta_main}
\end{equation}
Note that, as expected, $\beta_0$ is the same  RG equation as in Eq.~\eqref{RG:CO}. At $\widehat C_0^*=-1$, the condition $\beta_1=0$ is satisfied for
arbitrary $\widehat C_1$. Writing $\widehat C_1^*=c$, the remaining
fixed-point equations give
\begin{equation}
    \widehat C_n^\star=-(-c)^n,
    \qquad n\geq1.
    \label{fixed_point_family}
\end{equation}
The interacting solution therefore forms a one-parameter family of
fixed points\footnote{The $\widehat C_1$ direction is tangent to this
family and is therefore marginal.}. The matching derived in the previous section makes this structure
explicit. In particular,
\begin{align}
    \widehat C_1
    &=
    k_0 r_1\,\widehat C_0^2
    \equiv
    c\,\widehat C_0^2,
    \qquad
    c\equiv k_0r_1
    \sim
    \frac{\lambda}{g^2}
    \ll1,
    \label{eq:C1hat}
    \\
    \widehat C_n
    &\supset
    (k_0r_1)^n\widehat C_0^{\,n+1}
    =
    c^n\widehat C_0^{\,n+1}.
    \label{eq:Cnhat_fixed}
\end{align}
Therefore, as $\widehat C_0\to-1$, the $\mu_F$-independent contributions
approach the one-parameter fixed-point family given in
Eq.~\eqref{fixed_point_family}. Since in our toy model $c\ll1$
(Eq.~\eqref{parametrics}), the fixed point relevant for our setup lies
close to the $c=0$ member of this family. At leading order in $c$, we
therefore take
\begin{equation}
    \widehat C_0^\star=-1,
    \qquad
    \widehat C_{n>0}^\star=0.
    \label{interacting_fixed_point_main}
\end{equation}
This shows that the unitary fixed point identified in Sec.~\ref{sec:Resummation_and_Fixed_Points} from the running of $C_0$ is indeed a fixed point of the full momentum-dependent interaction.

Linearizing around the fixed point in Eq.~\eqref{interacting_fixed_point_main},
with $\widehat C_n=\widehat C_n^\star+\delta\widehat C_n$, gives
\begin{equation}
    \mu_F
    \frac{{\rm d}}{{\rm d}\mu_F}
    \delta\widehat C_n
    =
    (n-1)\delta\widehat C_n
    +\mathcal O(\delta\widehat C^2).
    \label{linearized_Cn}
\end{equation}
With respect to this $\mu_F$ flow, for which $\widehat C_0^\star=-1$ is the UV fixed point of Sec.~\ref{sec:Resummation_and_Fixed_Points}, a perturbation is relevant if it grows as $\mu_F$ is lowered towards the IR, irrelevant if it is driven to zero there, and marginal if its linearized scaling vanishes. Therefore, we see that $C_0$ is relevant, $C_1$ is marginal and the $C_{n\geq2}$ directions are irrelevant.

This behavior can also be seen directly from the matched Wilson
coefficients. For example for the $C_2$ coefficient of Eq.~\eqref{C2_matching},
\begin{equation}
    \widehat C_2
    =
    k_0^2r_1^2\widehat C_0^3
    +
    \frac{1}{\lambda}\mu_F r_2\widehat C_0^2 \sim \frac{1}{\lambda}\mu_F r_2\widehat C_0^2,
    \label{eq:C2hat}
\end{equation}
where the first term belongs to the fixed-point family, while the second
is the deformation away from it. More generally, 
\begin{equation}
    \delta\widehat C_n \sim \frac{r_n \mu_F^{n-1}}{\lambda^{n-1}  k_0^{n-2}} \widehat C_0^2
    \sim (k_0 r_n)\left(\frac{k_0\mu_F}{\Lambda_{\rm EFT}^2}\right)^{n-1},
    \qquad n\geq1,
    \label{eq:deltaCnhat}
\end{equation}
The explicit $\mu_F^{\,n-1}$ dependence agrees with Eq.~\eqref{linearized_Cn}, providing a direct check of the RG classification obtained above. 

It is worth comparing this structure with the unitary fixed point found in NR systems with a large scattering length, as in Refs.~\cite{Kaplan:1998we,Birse:1998dk}. There, the EFT is expanded around zero momentum rather than around a momentum surface and the local expansion contains only $C_{2n}$ operators. The beta functions have the same recursive structure as Eq.~\eqref{hn_beta_main}, but the canonical scaling is different because each successive operator carries two powers of momentum. The fixed point again sits at $\widehat C_0^\star=-1$, but there is no marginal direction. The fixed point equations uniquely force $\widehat C_{2n>0}^\star=0$, so the threshold unitary fixed point is isolated, as opposed to the one-parameter family found in the present case. At the exact fixed point, the system with an infinite scattering length at threshold is associated with an enhanced Non-Relativistic Conformal Symmetry (NRCFT) \cite{Mehen:1999nd,Nishida:2007pj}.

This difference in fixed-point structure is also reflected in the approach to the fixed point. In Refs.~\cite{Kaplan:1998we,Birse:1998dk}, the higher Wilson coefficients are determined recursively by lower-order effective-range parameters, with new parameters entering only through subleading corrections. In our model, instead, Eqs.~\eqref{C_scalings_general} and~\eqref{eq:deltaCnhat} show that each $C_n$ is dominated by the corresponding effective-range parameter $r_n$.  Together with the marginal $C_1$ direction, this reflects a different universality class from that of Refs.~\cite{Kaplan:1998we,Birse:1998dk}.

%%%%%%%%%%%%%%%%%%%%%%%%%%%%%%%%%%%%%%%%%%%%%%%%%%%%%%%%%
\subsection{Resummation in the EFT}
\label{sec:Surface_Resummation}
%%%%%%%%%%%%%%%%%%%%%%%%%%%%%%%%%%%%%%%%%%%%%%%%%%%%%%%%%
The power counting of Sec.~\ref{sec:Power_Counting} shows that the leading interaction $C_0$ has to be resummed to all orders, while higher $C_n$ can be treated perturbatively.
At tree-level, the EFT scattering amplitude, derived in Appendix~\ref{EFT:Feyn:rules}, is
\begin{equation}
    i\mathcal A_{\rm EFT}^{\rm tree}
    =
    \contact[1]
    =
    -i
    \sum_{n=0}^{\infty}
    C^{\rm bare}_{n}(2k_0\delta k)^n .
\end{equation}
A one-loop bubble containing a $C^{\rm bare}_{n}$ and a $C^{\rm bare}_{m}$ vertex gives
\begin{align}
    i\mathcal A_{\mathrm{EFT},nm}^{\rm 1-loop}
    &=
    \doublecontact[1]
    \nonumber\\
    &=
    (-iC^{\rm bare}_{n})(-iC^{\rm bare}_{m})
    \frac{k_0^2}{\pi}
    (2k_0)^{n+m}
    \int_{-\infty}^{+\infty}
    \frac{{\rm d}\ell}{2\pi}\,
    \frac{i\ell^{n+m}}
    {E-2\frac{k_0}{M}\ell+i0},
    \label{shell_bubble}
\end{align}
where, for an on-shell external state, the residual energy is $E=2k_0\delta k/M$. Retaining the on-shell pole contribution\footnote{The remaining analytic contribution, corresponding to the principal-value part of Eq.~\eqref{shell_bubble}, renormalizes the local Wilson coefficients. This would introduce an EFT renormalization scale $\mu_R^{\rm EFT}$, whose running cannot change the power counting inherited from the factorization scale dependence. Therefore, it is possible to work in a scheme where such analytic contributions vanish.} gives
\begin{equation}
    i\mathcal A_{\rm EFT}^{\rm 1-loop}
    =
    -\frac{Mk_0}{4\pi}
    \left[
        \sum_{n=0}^{\infty}
        C^{\rm bare}_{n}(2k_0\delta k)^n
    \right]^2.
    \label{shell_on_shell_loop}
\end{equation}
Therefore, each additional bubble produces another iteration of the same interaction, and the series can be summed geometrically
\begin{align}
    i\mathcal A_{\rm scatt}^{\rm EFT}(k)
    &=\contact[1]
    +
    \doublecontact[1]
    +
    \triplecontact[1]
    +\cdots \nonumber \\
    &= \frac{
        -i\displaystyle\sum_{n=0}^{\infty}
        C^{\rm bare}_{n}(2k_0\delta k)^n
    }{
        1+i\frac{Mk_0}{4\pi}\displaystyle\sum_{n=0}^{\infty}
        C^{\rm bare}_{n}(2k_0\delta k)^n
    } = \frac{
        -i\displaystyle\sum_{n=0}^{\infty}C_n (\mu_F) (2k_0\delta k)^n
    }{
        1+\frac{M}{4\pi} (\mu_F+ik_0)\displaystyle\sum_{n=0}^{\infty}
        C_n (\mu_F) (2k_0\delta k)^n
    },
    \label{full_shell_resummation}
\end{align}
where in the last step we substituted the bare with the renormalized Wilson coefficients using Eq.~\eqref{matching_general_k}.

Keeping the $C_0$ dependence exact and expanding perturbatively in the higher-derivative interactions, namely in powers of the residual momentum, gives
\begin{align}
    \label{EFT_scatt}
    i\mathcal A_{\rm scatt}^{\rm EFT}(k)
    &=
    \frac{-i \displaystyle\sum_{n=0}^\infty
        C_n(\mu_F)(2k_0\delta k)^n}{
        1+\frac{M}{4\pi} (\mu_F + ik_0) C_0(\mu_F)
    }+\mathcal O\!\left(C_{n>0}^2\right).
\end{align}
Finally, Eq.~\eqref{full_shell_resummation} describes only the scattering generated by the off-diagonal interaction. Restoring the elastic scattering from the repulsive channel only, the scattering amplitude can be written as
\begin{equation}
    i\mathcal A_{\rm scatt}(k)
    =
    \frac{2\pi}{Mk}
    \left(
        e^{2i\delta(k)}-1
    \right)
    +
    e^{2i\delta(k)}
    i \mathcal{A}_{\rm scatt}^{\rm EFT}.
    \label{shell_EFT_full_amplitude}
\end{equation}

%%%%%%%%%%%%%%%%%%%%%%%%%%%%%%%%%%%%%%%%%%%%%%%%%%%%%%%%%
\subsection{Results and Discussion}
\label{sec:Results_and_Discussion}
%%%%%%%%%%%%%%%%%%%%%%%%%%%%%%%%%%%%%%%%%%%%%%%%%%%%%%%%%

\begin{figure}[t]
    \centering
    \includegraphics[width=1.0\textwidth]{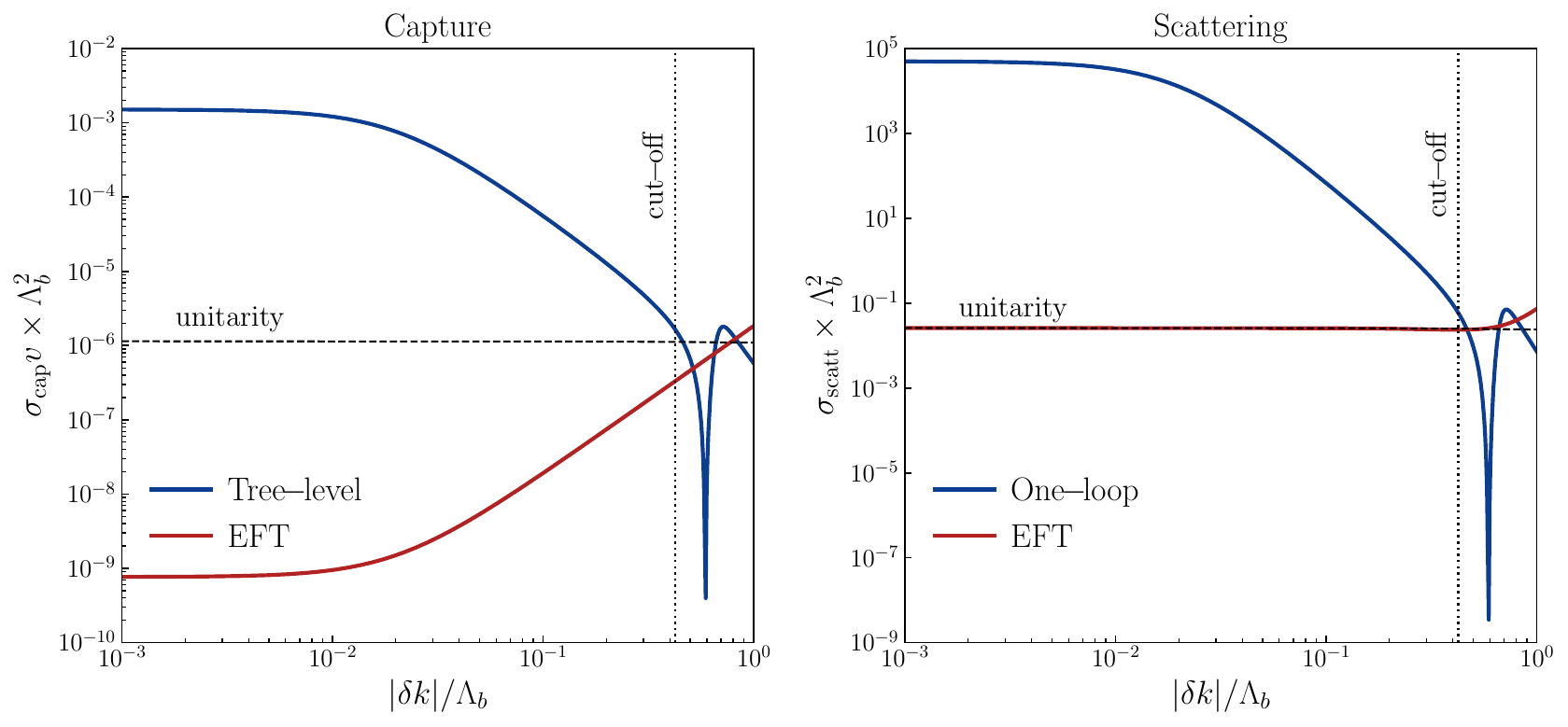}
    \caption{The left panel shows the capture cross section, obtained from the tree-level computation (blue line) and from the EFT expansion (red line) in the momentum region $\delta k \lesssim \Lambda_{\rm EFT}^2/k_0$, which corresponds to the range shown in the inset plot of Fig.~\ref{fig:2}. The right panel shows the scattering cross section from the one-loop computation (blue line) and the EFT expansion (red line). We compare the capture and scattering cross sections to the inelastic and elastic unitarity bounds (dashed black lines), respectively. The EFT cut-off (black dotted lines) is chosen as the breakdown of the effective range expansion $2r_2(k_0 \delta k)^2/\Lambda_{\rm EFT}^2=k_0$, namely at $\delta k=g \Lambda_{\rm EFT}/(2\pi^{3/2})$. The numerical values of the parameters of the toy model are the same as the ones chosen for Fig.~\ref{fig:2}.}
    \label{fig:4}
\end{figure}

Having obtained the resummed and systematically expanded expression of the amplitude for $\delta k \ll \Lambda_{\rm EFT}^2/k_0$, we now derive the scattering and capture cross sections and compare them with the unitarity bound. The elastic scattering cross section can be obtained as
\begin{equation}
    \sigma_{\rm scatt} = 4\pi \left|\frac{M}{4\pi} \mathcal{A}_{\rm scatt}\right|^2,
\end{equation}
However, since we aim to compare it with the unitarity bound
\begin{equation}
    \label{scatt_UB}
    \sigma_{\rm scatt}^{\rm UB} \equiv \frac{4\pi}{k^2},
\end{equation}
we define the scattering cross section
\begin{equation}
    \sigma^{\rm EFT}_{\rm scatt} = 4\pi \left|\frac{M}{4\pi} \mathcal{A}_{\rm scatt}^{\rm EFT} \right|^2,
\end{equation}
which is the quantity that saturates the unitarity bound. In fact, it is independent of the phase-shift from the repulsive channel and includes only the contribution from the off-diagonal interaction.
In Fig.~\ref{fig:4} (right panel), we compare the cross sections obtained from the one-loop scattering amplitude of Eq.~\eqref{A1_scatt}, to the one from the EFT expansion of Eq.~\eqref{EFT_scatt}, which resums the contributions of $C_0$ to all orders. The one-loop cross section is shown to be $3$ orders of magnitude above the unitarity bound, which implies the loss of perturbativity, while the EFT result saturates the unitarity bound.

\begin{figure}[t]
    \centering
    \includegraphics[width=0.55\textwidth]{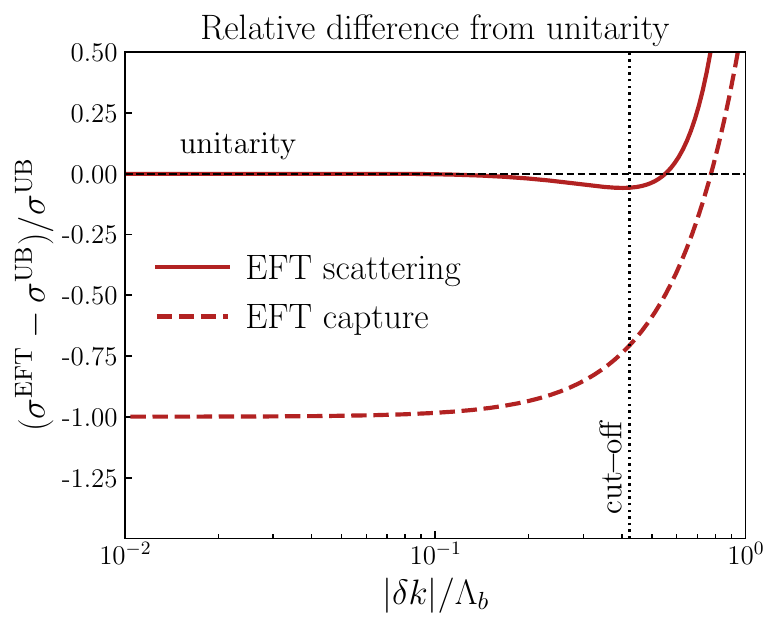}
    \caption{Relative size between the cross sections from the EFT expansion and the unitarity bounds, for the elastic (red solid line) and inelastic (red dashed line) processes. We show the momentum dependence in the region of validity of the EFT expansion $k_0\delta k<\Lambda_{\rm EFT}^2$. While the scattering cross section saturates the unitarity bound, the capture cross section is suppressed. The numerical values of the parameters of the toy model are the same as the ones chosen for Fig.~\ref{fig:2} and Fig.~\ref{fig:4}.}
    \label{fig:5}
\end{figure}

The capture cross section can be obtained from the inelastic contributions encoded in the imaginary part of the Wilson coefficients $C_n$. In the momentum-surface EFT, Eq.~\eqref{resummed_capture_cross_section} takes the form
\begin{align}
    \label{EFT_capture}
    (\sigma v)^{\rm EFT}_{\rm cap}(k)
    &=
    -\frac{
        2\,\sum_{n=0}^\infty
            \operatorname{Im} C_n(\mu_F)(2k_0\delta k)^n
    }{
        \left|
            1+\frac{M}{4\pi}(\mu_F+ik_0)C_0(\mu_F)
        \right|^2
    } + \mathcal{O}\left(C_{n>0}^2\right).
\end{align}
In Fig.~\ref{fig:4} (left panel), we compare the capture cross section from the tree-level amplitude of Eq.~\eqref{A_capture_tree-level}, to the EFT expansion of Eq.~\eqref{EFT_capture}. The tree-level capture exceeds the unitarity bound for $\delta k < \Lambda_{\rm EFT}^2/k_0$, as already discussed in Sec.~\ref{sec:Momentum_Dependence}. In fact, it corresponds to the momentum dependence of the capture cross section near its maximum, shown in the inset plot of Fig.~\ref{fig:2}. Moreover, in the EFT, the capture is suppressed. This can be seen from Eq.~\eqref{EFT_capture}, where the cross section is shown to scale as $(\sigma v)_{\rm cap} \propto \operatorname{Im} C_0/|C_0|^2$, which is suppressed in presence of a large scattering length. This can be understood by verifying that the EFT capture satisfies the optical theorem
\begin{equation}
    (\sigma v)_{\rm cap} = 2 \operatorname{Im} \mathcal{A}_{\rm scatt} - \frac{Mk}{2\pi} \left|\mathcal{A}_{\rm scatt}\right|^2 = 2 \operatorname{Im} \mathcal{A}^{\rm EFT}_{\rm scatt} - \frac{Mk}{2\pi} \left|\mathcal{A}_{\rm scatt}^{\rm EFT}\right|^2.
\end{equation}
In Fig.~\ref{fig:5}, we compare the relative size of the EFT capture and scattering cross sections to the corresponding unitarity bounds of Eqs.~\eqref{cap_UB} and~\eqref{scatt_UB}.
It shows that, as the unitarity bound is saturated by the elastic process, the inelastic process is consequently suppressed.

%%%%%%%%%%%%%%%%%%%%%%%%%%%%%%%%%%%%%%%%%%%%%%%%%%%%%%%%%
\section{Conclusions}
\label{sec:Conclusions}
%%%%%%%%%%%%%%%%%%%%%%%%%%%%%%%%%%%%%%%%%%%%%%%%%%%%%%%%%

In this work, we have studied the breakdown of perturbation theory in radiative capture between two non-relativistic channels that experience different potentials. Using a simple toy model, we developed a systematic power counting that keeps track of the multipole-expanded off-diagonal interaction, as well as of the non-perturbative scattering-to-bound matrix element. In the presence of a potential barrier, the wavefunction of a fast scattering state with a specific momentum $k_0$ can vary on the same spatial scale as the wavefunction of a deep bound state in another channel and the transition matrix element reaches its geometric size. As a consequence, the weak off-diagonal transition induces a strong mixing between the two eigenstates of the unperturbed Hamiltonian, causing a loss of perturbativity.
We identified the enhanced contributions as the 2PR diagrams, which must be resummed to all orders. By contrast, the 2PI corrections to the self-energy remain perturbative.

Near $k_0$, the enhanced 2PR contributions factorize. The self-energy insertion evaluated at $k_0$ can therefore be interpreted as a non-perturbatively large effective interaction between free scattering states. Treating this interaction as a local operator requires a factorization scale $\mu_F$, which allows us to study the RG flow of the effective interaction. We find that the local interaction associated with the self-energy at $k_0$ sits close to a unitary fixed point. The RG structure is analogous to that of a non-relativistic system with an anomalously large scattering length, where a four-fermion interaction becomes relevant and must be iterated non-perturbatively~\cite{Kaplan:1998we,Kaplan:1998tg}. The important difference is that the strong dynamics found here is not organized around the physical threshold $k=0$, but around a finite momentum surface $|\mathbf{k}|=k_0$.

The resummation also admits an interpretation in terms of the analytic structure of the scattering amplitude. The resummed amplitude develops a complex pole, as for an unstable resonance, and the geometric series is analogous to the self-energy resummation that gives a propagator a finite width. The physical interpretation is nevertheless different. Here the pole lies close to the momentum surface, at a complex residual momentum $\delta k$ whose real and imaginary parts are both small compared with $k_0$. This pole then plays the role of the shallow pole of a system with a large scattering length, but now located near the momentum surface rather than near threshold.

These observations motivate the momentum-surface EFT constructed in this work. Writing the relative momentum as a hard reference momentum $k_0$ plus a small residual momentum, the dynamics associated with $k_0$ is separated from the slow motion in its vicinity, in close analogy with effective theories organized around a Fermi surface~\cite{Shankar:1993pf,Polchinski:1992ed}. The dispersion relation becomes linear at leading order in the residual momentum, while the interactions admit a local expansion controlled by the smaller breakdown scale $\Lambda_{\rm EFT}$. The leading four-fermion contact interaction is relevant, whereas higher-derivative interactions remain perturbative within the cut-off of the EFT. Therefore, the resulting theory provides an example of fermions at unitarity with a linear dispersion relation.

The RG flow of the full momentum-dependent interaction further clarifies the structure of this theory. The interacting solution belongs to a one-parameter family of fixed points. The model considered here realizes the member with the leading contact interaction at the unitary fixed point and vanishing higher-derivative interactions. Linearizing the flow shows that the leading interaction is relevant, the first derivative interaction is marginal, and the higher interactions are irrelevant. The momentum-surface theory therefore shares with threshold unitarity the relevance of the leading contact interaction, while exhibiting a distinct fixed-point structure associated with the expansion around a finite momentum.

Finally, the resummed theory makes the restoration of unitarity manifest. In the strong-mixing region, the elastic scattering generated by the channel mixing approaches the unitarity limit, whereas radiative capture is suppressed. This follows directly from the optical theorem: once the elastic contribution saturates the unitarity bound, the inelastic contribution must become small. The large fixed-order capture cross section is therefore just a diagnostic of the breakdown of the original perturbative organization. After the enhanced 2PR contributions are resummed, both elastic scattering and capture are described consistently within a controlled EFT expansion.

A natural next step is to apply this framework to the non-relativistic ${\rm SU}(N)$ systems that originally motivated this work. In that case, the emission of the non-abelian gauge boson is described by a dipole interaction rather than by the monopole interaction used in the toy model. Moreover, the Coulombic scattering and bound-state wavefunctions have a substantially more complicated analytic structure, and the long-range nature of the interaction removes the simple notion of an interaction range that was useful in organizing the toy model. Studying the factorization and the momentum-surface power counting in this setting is left for future work.

More broadly, systems with a large scattering length are common when a bound state lies close to threshold. By contrast, to our knowledge there is currently no established physical realization of the finite-momentum mechanism discussed here, in which strong mixing and an effectively unitary interaction emerge only in a restricted momentum window away from the physical threshold. Nevertheless, the ingredients required by the toy model are sufficiently simple that it would be surprising if such a regime could not be realized in any physical system. Identifying concrete examples, in atomic, nuclear, condensed-matter, or particle-physics settings, would be particularly interesting.

\section*{Acknowledgements}

We thank Martin Beneke and Riccardo Rattazzi for comments, and helpful and insightful discussions. Moreover, we thank Tobias Binder, Mathias Garny, Stefan Lederer, Nicholas Leister, Andrea Luzio, Filippo Nardi, Marc Riembau and Michael Stadlbauer for useful discussions. The work of JRN is partially supported by the Swiss National Science Foundation under contract 200020-213104. LDR is supported by the Cluster of Excellence Precision Physics, Fundamental Interactions and Structure of Matter (PRISMA++, EXC 2118/2, Project ID 390831469), funded by the German Research Foundation (DFG).

\newpage
\appendix

%%%%%%%%%%%%%%%%%%%%%%%%%%%%%%%%%%%%%%%%%%%%%%%%%%%%%%%%%
\section{Toy Model Conventions}
\label{sec:Toy_Model_Conventions}
%%%%%%%%%%%%%%%%%%%%%%%%%%%%%%%%%%%%%%%%%%%%%%%%%%%%%%%%%

%%%%%%%%%%%%%%%%%%%%%%%%%%%%%%%%%%%%%%%%%%%%%%%%%%%%%%%%%
\subsection{Repulsive Channel}
\label{sec:Repulsive_Channel}
%%%%%%%%%%%%%%%%%%%%%%%%%%%%%%%%%%%%%%%%%%%%%%%%%%%%%%%%%

We summarize the definitions and conventions for the channel of a $\xi\xi$ pair. As defined in Eq.~\eqref{potentials}, a $\xi \xi$ pair feels a repulsive spherical potential barrier, therefore this channel does not support any bound states. In terms of the composite operator, the in/out scattering wavefunctions are defined as
\begin{equation}
    \label{scattering_wavefunction_definition}
    \langle \Omega \left| \Phi_{\xi \xi} (\mathbf{R},\mathbf{r},t)
    \right| s^{(\pm)}(\mathbf{P}=\mathbf{0},\mathbf{k}) \rangle
    =
    e^{- i \frac{k^2}{M} t}
    \psi^{(\pm)}_\mathbf{k} (\mathbf{r}).
\end{equation}
The scattering states are eigenstates of the corresponding diagonal
component of the Hamiltonian
\begin{equation}
    H_{\rm rep} \lvert s^{(\pm)}(\mathbf{k}) \rangle
    =
    \frac{k^2}{M} \lvert s^{(\pm)}(\mathbf{k}) \rangle,
\end{equation}
or, equivalently, their wavefunctions solve the Schr\"odinger equation
\begin{equation}
    \left[
        - \frac{\nabla^2_{\mathbf{r}}}{M}
        + V_0 \, \theta(1-\Lambda_b r)
        - \frac{k^2}{M}
    \right]
    \psi^{(\pm)}_{\mathbf{k}}(\mathbf{r})
    =0.
\end{equation}
The in and out scattering wavefunctions are related by
\begin{equation}
    \psi^{(-)}_{\mathbf{k}}(\mathbf{r})
    =
    \psi^{(+)*}_{-\mathbf{k}}(\mathbf{r}),
\end{equation}
and we choose the scattering states to be normalized as
\begin{equation}
    \label{scattering_normalization}
    \langle s^{(\pm)}\left(\mathbf{k}'\right)
    \rvert s^{(\pm)}(\mathbf{k}) \rangle
    =
    (2\pi)^3 \delta^{(3)}(\mathbf{k}-\mathbf{k}').
\end{equation}
We denote the $s$-wave component of the scattering state as
\begin{equation}
    \lvert s^{(\pm)}(k) \rangle = \int \frac{{\rm d}\Omega_{\hat{\mathbf{k}}}}{4\pi} \lvert s^{(\pm)}(\mathbf{k}) \rangle
\end{equation}
In the toy model with a spherical barrier potential, the scattering
wavefunctions admit simple analytic expressions, which we present below
together with the corresponding propagator, scattering amplitude, and
fine-tuned momentum limit.

%%%%%%%%%%%%%%%%%%%%%%%%%%%%%%%%%%%%%%%%%%%%%%%%%%%%%%%%%
\subsubsection{Scattering-state Wavefunction}
\label{sec:Scattering-state_Wavefunction}
%%%%%%%%%%%%%%%%%%%%%%%%%%%%%%%%%%%%%%%%%%%%%%%%%%%%%%%%%

For an $s$-wave scattering state with energy greater than the height of the potential barrier, $k^2/M>V_0$, the wavefunction reads
\begin{equation}
    \psi^{(\pm)}_k(r)
    =
    \frac{e^{\pm i\delta(k)}}{kr}
    \begin{cases}
        N_s(k)\sin(k_s r), & \Lambda_b r<1,\\[0.6em]
        \sin\!\left(kr+\delta(k)\right), & \Lambda_b r>1,
    \end{cases}
\end{equation}
therefore, the in and out states are related by a phase
\begin{equation}
    \label{+to-}
    \lvert s^{(+)}(k) \rangle = e^{2i\delta(k)} \lvert s^{(-)}(k) \rangle.
\end{equation}
We highlight that $k$ is the asymptotic momentum, while $k_s=\sqrt{k^2-MV_0}$ denotes the local momentum inside the interaction region. The normalization coefficient $N_s(k)$ is fixed by the smooth matching of the scattering solution
\begin{equation}
    \label{N_s_def}
    \frac{1}{N_s^2(k)}
    =
    \sin^2\!\left(k_s/\Lambda_b\right)
    +
    \frac{k_s^2}{k^2}
    \cos^2\!\left(k_s/\Lambda_b\right),
\end{equation}
as well as the phase-shift
\begin{equation}
    \label{delta_def}
    e^{2i\delta(k)}
    =
    e^{-2ik/\Lambda_b}
    \frac{
        k_s\cot(k_s/\Lambda_b)+ik
    }{
        k_s\cot(k_s/\Lambda_b)-ik
    }.
\end{equation}

%%%%%%%%%%%%%%%%%%%%%%%%%%%%%%%%%%%%%%%%%%%%%%%%%%%%%%%%%
\subsubsection{Propagation and Scattering}
\label{sec:Propagation_and_Scattering}
%%%%%%%%%%%%%%%%%%%%%%%%%%%%%%%%%%%%%%%%%%%%%%%%%%%%%%%%%

Because of the repulsive potential, this channel supports no bound states, therefore its scattering states provide a complete spectral resolution. In the frame of reference of the center of mass of the $\xi \xi$ pair, $\mathbf{P}=\mathbf{0}$, the propagator of the repulsive channel can be written as
\begin{equation}
    \label{scattering_propagator}
    i\hat{G}_{\rm rep}(E)
    =
    \frac{i}{E-\hat{H}_{\rm rep}+i0}
    =
    \int
    \frac{\mathrm{d}^3\mathbf{p}}{(2\pi)^3}
    \frac{
        i\lvert s^{(+)}(\mathbf{p}) \rangle
        \langle s^{(+)}(\mathbf{p}) \rvert
    }{
        E-\frac{p^2}{M} + i0
    }.
\end{equation}
The $s$-wave component of the on-shell scattering amplitude can be expressed as
\begin{equation}
    \label{A0_scatt_general}
    i\mathcal{A}^{(0)}_{\rm scatt}(k)
    =
    \frac{2\pi}{Mk}
    \left[e^{2i \delta(k)}-1\right].
\end{equation}
In our toy model, this leads to the simple expression
\begin{equation}
    \label{A0_scatt_toy_model}
    i\mathcal{A}^{(0)}_{\rm scatt}(k)
    =
    \frac{2\pi}{Mk}
    \left[
        e^{-2ik/\Lambda_b}
        \frac{
            k_s\cot(k_s/\Lambda_b)+ik
        }{
            k_s\cot(k_s/\Lambda_b)-ik
        }
        -1
    \right].
\end{equation}
We recall that, at this stage, the off-diagonal interaction between the two channels is absent, so Eq.~\eqref{A0_scatt_toy_model} contains only the exact potential scattering in the repulsive channel.

%%%%%%%%%%%%%%%%%%%%%%%%%%%%%%%%%%%%%%%%%%%%%%%%%%%%%%%%%
\subsection{Attractive Channel}
\label{sec:Attractive_Channel}
%%%%%%%%%%%%%%%%%%%%%%%%%%%%%%%%%%%%%%%%%%%%%%%%%%%%%%%%%

We summarize the definitions and conventions for the channel of a $\xi\eta$ pair. Differently from a $\xi\xi$ pair, a $\xi \eta$ pair feels an attractive spherical potential well, which can support bound states. In terms of the composite operator, the bound-state wavefunction is defined as
\begin{equation}
    \label{bound_wavefunction_definition}
    \langle \Omega \left| \Phi_{\xi \eta} (\mathbf{R},\mathbf{r},t)
    \right| b(\mathbf{P}_b=\mathbf{0},E_b) \rangle
    =
    e^{- i E_b t} \psi_b (\mathbf{r}).
\end{equation}
The bound state is an eigenstate of the corresponding diagonal component
of the Hamiltonian,
\begin{equation}
    H_{\rm att} \lvert b\rangle
    =
    E_b \lvert b\rangle,
\end{equation}
or, equivalently, its wavefunction solves the Schr\"odinger equation
\begin{equation}
    \left[
        -\frac{\nabla^2_{\mathbf{r}}}{M}
        - V_0 \, \theta(1-\Lambda_b r)
        - E_b
    \right]
    \psi_b(\mathbf{r})
    = 0.
\end{equation}
We choose the bound state to be normalized as
\begin{equation}
    \label{bound_normalization}
    \langle b \lvert b \rangle = 1.
\end{equation}
In the toy model with a spherical potential well, the bound-state wavefunction admits a simple analytic expression.

%%%%%%%%%%%%%%%%%%%%%%%%%%%%%%%%%%%%%%%%%%%%%%%%%%%%%%%%%
\subsubsection{Bound-state Wavefunction}
\label{sec:Bound-state_Wavefunction}
%%%%%%%%%%%%%%%%%%%%%%%%%%%%%%%%%%%%%%%%%%%%%%%%%%%%%%%%%

The wavefunction of an $s$-wave bound state of a spherical potential well
reads
\begin{equation}
    \label{wf_bound_full}
    \psi_b(r)
    =
    \frac{N_b}{\sqrt{4\pi}\,r}
    \begin{cases}
        \sin(k_b r), & \Lambda_b r<1,\\[0.6em]
        \sin(k_b/\Lambda_b)\,
        e^{-\gamma(r-\Lambda_b^{-1})}, & \Lambda_b r>1,
    \end{cases}
\end{equation}
where we denote
\begin{equation}
    \gamma=\sqrt{M|E_b|}.
\end{equation}
The local momentum of the bound state inside the interaction region
$\Lambda_b r < 1$ is defined as
\begin{equation}
    \frac{k_b^2}{M}=V_0-|E_b|,
\end{equation}
and its energy eigenvalue can be obtained by solving
\begin{equation}
    k_b\cot\!\left(k_b/\Lambda_b\right)=-\gamma.
\end{equation}
The normalization coefficient can be written as
\begin{equation}
    \frac{1}{N_b^2} = \frac{1}{2 \Lambda_b}+\frac{1}{2\gamma}.
\end{equation}

%%%%%%%%%%%%%%%%%%%%%%%%%%%%%%%%%%%%%%%%%%%%%%%%%%%%%%%%%
\subsubsection{Deep-Well Limit}
\label{sec:Deep-Well_Limit}
%%%%%%%%%%%%%%%%%%%%%%%%%%%%%%%%%%%%%%%%%%%%%%%%%%%%%%%%%

For the toy model described in Sec.~\ref{sec:Toy_Model}, in order for the mixing to be strong, we assume the potential well to be deep, which leads to the requirement
\begin{equation}
    \frac{MV_0}{\Lambda_b^2} \gg 1.
\end{equation}
From this condition, we obtain that the wavefunction outside the interaction region is exponentially suppressed, and for the ground state, the momentum inside the interaction region approaches $k_b \approx \pi \Lambda_b$.
Therefore, the ground-state wavefunction can be approximated as
\begin{equation}
    \psi_b(r)
    \approx
    \frac{1}{\sqrt{4\pi}\,r} \sqrt{2 \Lambda_b}
    \begin{cases}
        \sin(\pi \Lambda_b r), & \Lambda_b r<1,\\[0.6em]
        0, & \Lambda_b r>1.
    \end{cases}
\end{equation}

%%%%%%%%%%%%%%%%%%%%%%%%%%%%%%%%%%%%%%%%%%%%%%%%%%%%%%%%%
\subsubsection{Bound-state Propagator}
\label{sec:Bound-state_Propagator}
%%%%%%%%%%%%%%%%%%%%%%%%%%%%%%%%%%%%%%%%%%%%%%%%%%%%%%%%%

In the spectral decomposition of the attractive channel, there are both scattering states and bound states. Therefore, the full propagator of this channel schematically reads
\begin{equation}
    i G_{\rm att}(E,\mathbf{P}_b) = \frac{i}{E-H_{\rm att} + i0} \left[ \sum_n \lvert b_n \rangle \langle b_n \rvert + \int \frac{{\rm d}^3\mathbf{p}}{(2\pi)^3} \lvert s^{(+)}_{\rm att}(\mathbf{p}) \rangle \langle s^{(+)}_{\rm att} (\mathbf{p}) \rvert \right],
\end{equation}
where the index $n$ runs over all bound states and $\lvert s^{(+)}_{\rm att}(\mathbf{p}) \rangle$ denotes the continuum eigenstates of $H_{\rm att}$. However, the aim of this work is to study the strong mixing regime which occurs between the scattering state of the repulsive channel $\lvert s^{(+)}(k_0) \rangle$ and the ground state of the attractive channel $\lvert b \rangle$. Therefore, we approximate the spectral decomposition of the attractive channel propagator retaining only the ground-state pole
\begin{equation}
    \label{bound_propagator}
    iG_{\rm att}(E,\mathbf{P}_b)
    \approx
    \frac{
        i\lvert b \rangle \langle b \rvert
    }{
        E-\frac{P_b^2}{4M}+|E_b|+i0
    }.
\end{equation}
To understand the regime of validity of this, we can estimate the momentum $k_1$ at which the strong mixing with the first $s$-wave excited state occurs. Assuming the first excited state to also be a deep bound state, we can write the equivalent of Eq.~\eqref{k_0_def}, which reads
\begin{equation}
    \frac{k_1^2}{M} \approx V_0 + \frac{4\pi^2 \Lambda_b^2}{M}.
\end{equation}
Therefore, the difference between the momenta corresponding to the strong mixing with the ground state and the first excited state reads
\begin{equation}
    k_1 - k_0 \approx \frac{3\pi^2}{2} \frac{\Lambda_b^2}{k_0}.
\end{equation}
This momentum difference has to be compared with the width of the region of strong mixing around $k_0$ obtained in Eq.~\eqref{Lambda_EFT_def}. We can see that the strong mixing with the first excited state occurs at momenta much further from $k_0$ than the width of the region
\begin{equation}
    k_1-k_0 \gg \frac{\Lambda_{\rm EFT}^2}{k_0}=\lambda k_0,
\end{equation}
and therefore it can neglected.

%%%%%%%%%%%%%%%%%%%%%%%%%%%%%%%%%%%%%%%%%%%%%%%%%%%%%%%%%
\subsection{Matrix Element Derivation}
\label{sec:Matrix_Element_Derivation}
%%%%%%%%%%%%%%%%%%%%%%%%%%%%%%%%%%%%%%%%%%%%%%%%%%%%%%%%%

For our toy model, the exact matrix element between the scattering state of the repulsive channel and the ground state of the attractive channel can be computed analytically. In position space, it corresponds to the overlap integral between the scattering and the bound-state wavefunctions
\begin{equation}
    \langle b \rvert s^{(+)}(\mathbf{k}) \rangle = \langle b \rvert s^{(+)}(k) \rangle = \int {\rm d}^3 \mathbf{r} \, \psi_b^*(r) \psi^{(+)}_k(r),
\end{equation}
where the first equality follows from angular-momentum orthogonality, since the $\ell=0$ bound state only has a non-vanishing overlap with the $s$-wave component of the scattering state.
From the analytic expressions of the scattering-state wavefunction given in Eq.~\eqref{wf_scatt_full}, and the bound-state wavefunction given in Eq.~\eqref{wf_bound_full}, we can write the matrix element as
\begin{align}
    &\langle b \rvert s^{(+)}(k) \rangle
    =
    \int {\rm d}^3 \mathbf{r} \,
    \psi_b^*(\mathbf{r}) \psi^{(+)}_k (r)
    \nonumber \\
    &=
    \frac{\sqrt{\pi}\,e^{i\delta}N_b}{k}
    \left\{
        N_s
        \left[
            \frac{
                \sin\left(\frac{k_b-k_s}{\Lambda_b}\right)
            }{
                k_b-k_s
            }
            -
            \frac{
                \sin\left(\frac{k_b+k_s}{\Lambda_b}\right)
            }{
                k_b+k_s
            }
        \right]
    \right.
    \nonumber \\
    & \qquad \qquad \qquad \quad
    +
    \left.
        2\sin\left(\frac{k_b}{\Lambda_b}\right)
        \frac{
            \gamma\sin\left(\delta+\frac{k}{\Lambda_b}\right)
            +
            k\cos\left(\delta+\frac{k}{\Lambda_b}\right)
        }{
            \gamma^2+k^2
        }
    \right\}.
\end{align}
We recall that the momentum of the scattering state inside the interaction region $k_s(k)$, as well as the normalization $N_s(k)$ and the phase-shift $\delta(k)$, are functions of the asymptotic momentum, defined in Eq.~\eqref{N_s_def} and~\eqref{delta_def}.

%%%%%%%%%%%%%%%%%%%%%%%%%%%%%%%%%%%%%%%%%%%%%%%%%%%%%%%%%
\subsubsection{Expansion around the Maximum}
\label{sec:Expansion_around_the_Maximum}
%%%%%%%%%%%%%%%%%%%%%%%%%%%%%%%%%%%%%%%%%%%%%%%%%%%%%%%%%

In the deep-well limit, the matrix element can be simplified as
\begin{align}
    \langle b \rvert s^{(+)}(k) \rangle &= 2 \sqrt{2\pi \Lambda_b} \frac{e^{i\delta(k)}N_s}{k} \int_0^{1/\Lambda_b} {\rm d}r \, \sin(\pi \Lambda_b r) \sin(k_s r) \nonumber \\
    &=(2\pi\Lambda_b)^{3/2} e^{i\delta(k)} \frac{N_s}{k} \frac{\sin(k_s/\Lambda_b)}{\pi^2 \Lambda_b^2 - k_s^2}.
\end{align}
The momentum at which the maximum of the matrix element occurs can be computed perturbatively in powers of $\Lambda_b^2/(M V_0)$ (the deep-well limit), and it reads
\begin{equation}
    k_0^2 = M V_0+\pi ^2 \Lambda_b^2-3 \pi ^2 \frac{\Lambda_b^4}{M V_0} + \mathcal{O} \left(\frac{\Lambda_b^6}{M^2 V_0^2}\right).
\end{equation}
Therefore, expanding around the maximum in powers of $\delta k = k-k_0$, we obtain
\begin{align}
    \langle b \rvert s^{(+)}(k) \rangle =& \sqrt{\frac{2}{\pi}} \frac{e^{i\delta(k_0)}}{\Lambda_b^{3/2}} + \mathcal{O} \left( \frac{\Lambda_b}{\sqrt{M V_0}} \right) + \delta k \times \mathcal{O} \left( \frac{\sqrt{\Lambda_b}}{(M V_0)^{3/2}} \right) \nonumber \\
    &- \delta k^2 \times \frac{e^{i \delta(k_0)}}{\sqrt{2} \pi^{9/2}} \frac{(M V_0)^2}{\Lambda_b^{15/2}} \left[1+\mathcal{O} \left( \frac{\Lambda_b^2}{M V_0} \right)\right],
\end{align}
where the first order correction in $\delta k$ is suppressed, as expected.
Therefore, we can write
\begin{equation}
    \frac{\langle b \rvert s^{(+)}(k) \rangle}{\langle b \rvert s^{(+)}(k_0) \rangle} = \left[ 1 - \frac{1}{2\pi^4} \frac{k_0^4 \delta k^2}{\Lambda_b^6} \right] \times \left[ 1+\mathcal{O} \left( \frac{\Lambda_b}{k_0} \right) \right].
\end{equation}
From this expansion, we obtain the natural scale of the momentum dependence around the maximum $\Lambda_{\rm EFT}^2=(\pi \Lambda_b)^3/k_0$, which corresponds to the definition of the cut-off of the EFT description, derived in Sec.~\ref{sec:Momentum-Surface_EFT}.

%%%%%%%%%%%%%%%%%%%%%%%%%%%%%%%%%%%%%%%%%%%%%%%%%%%%%%%%%
\section{Power Counting Subleading Corrections}
\label{sec:Power_Counting_Subleading_Corrections}
%%%%%%%%%%%%%%%%%%%%%%%%%%%%%%%%%%%%%%%%%%%%%%%%%%%%%%%%%

As explained in Sec.~\ref{sec:Elastic_Scattering}, the sum of the $\mathcal{O}(g^2)$ and $\mathcal{O}(g^4)$ orders of the scattering amplitude can be written as
\begin{align}
    \mathcal{A}^{(1)}_{\rm scatt}(k) + \mathcal{A}^{(2)}_{\rm scatt}(k)
    =&- e^{2i\delta(k)} \left[
    \Sigma^{(1)}(k)
    +
    \left[\Sigma^{(1)}(k)\right]^2 I(k)
    +
   \Sigma^{(2)}(k) \right.
    \nonumber\\
    & \left.+
    \Sigma^{(1)}(k)
    \int\frac{{\rm d}^3\mathbf p}{(2\pi)^3}
    \frac{
        \Sigma^{(1)}(p)-\Sigma^{(1)}(k)
    }{
        \dfrac{k^2}{M}-\dfrac{p^2}{M}+i0
    }
     \right].
\end{align}

Here, the first term is the leading $\mathcal{O}(g^2)$ contribution computed in Eq.~\eqref{A1_scatt}, the second term is the factorized 2PR contribution introduced in Eq.~\eqref{add_and_subtract}, the third term collects the 2PI corrections introduced in Eq.~\eqref{2PI:text:def}, and the last term is the correction to the on-shell factorization of the 2PR diagram. $I(k)$ is defined in Eq.~\eqref{I_0_def}. Factoring out the leading contribution, one obtains
\begin{align}
    \mathcal{A}^{(1)}_{\rm scatt}(k) + \mathcal{A}^{(2)}_{\rm scatt}(k)
    =
   - e^{2i\delta(k)} \Sigma^{(1)}(k)
    \Bigg\{
    1
    &+
    I(k)
    \Bigg[
    \Sigma^{(1)}(k)
    +
    \frac{
       \Sigma^{(2)}(k)
    }{
        \Sigma^{(1)}(k) I(k)
    }
    \nonumber\\
    &+
    \frac{1}{I(k)}
    \int\frac{{\rm d}^3\mathbf p}{(2\pi)^3}
    \frac{
        \Sigma^{(1)}(p)-\Sigma^{(1)}(k)
    }{
        \dfrac{k^2}{M}-\dfrac{p^2}{M}+i0
    }
    \Bigg]
    \Bigg\}.
\end{align}
The quantity multiplying $I(k)$ is precisely the bare effective coupling at order $g^4$, introduced in Eq.~\eqref{C0:bare} and Eq.~\eqref{Cbare_k},
\begin{equation}
    C^{\rm bare}(k) =\Sigma^{(1)}(k) + \frac{\Sigma^{(2)}(k)}{\Sigma^{(1)}(k) I(k)} + \int \frac{1}{\frac{k^2}{M}-\frac{p^2}{M}} \frac{\Sigma^{(1)}(p)-\Sigma^{(1)}(k)}{I(k)}.
    \label{C:bare:app}
\end{equation}
The purpose of this appendix is to show explicitly that the bare effective coupling is dominated by the leading contribution $\Sigma^{(1)}(k)$ for all momenta $k$ in the vicinity of the momentum surface $k_0$, and to determine the suppression of the subleading corrections. In particular, we will obtain
\begin{equation}
    C^{\rm bare}(k)
    =
    \Sigma^{(1)}(k)
    \left[
        1
        +
        \mathcal{O}(g^2)
        +
        \mathcal{O}(\lambda)
    \right].
\end{equation}
The $\mathcal{O}(g^2)$ suppression arises from the $\mathcal{O}(g^4)$ 2PI diagrams and is derived in Sec.~\ref{sec:2PI_Diagrams}, while the $\mathcal{O}(\lambda)$ correction comes from corrections to factorization and is derived in Sec.~\ref{sec:Corrections_to_Factorization}.
%%%%%%%%%%%%%%%%%%%%%%%%%%%%%%%%%%%%%%%%%%%%%%%%%%%%%%%%%
\subsection{2PI Diagrams}
\label{sec:2PI_Diagrams}
%%%%%%%%%%%%%%%%%%%%%%%%%%%%%%%%%%%%%%%%%%%%%%%%%%%%%%%%%

As discussed in Sec.~\ref{sec:Loss_of_Perturbativity}, higher order 2PI diagrams are subleading for external momenta close to $k_0$. In 2PR diagrams, all matrix elements can be simultaneously evaluated within the momentum region where the mixing is strong. By contrast, in higher order 2PI diagrams this is possible only within a restricted region of phase space. To determine the size of this suppression, we analyze the cuts of the $\mathcal{O}(g^4)$ diagram in Fig.~\ref{fig:2PI}. The cuts expose the relevant on-shell kinematics, allowing us to power count the region of phase space in which the non-perturbative matrix elements are enhanced, following the logic of the method of regions~\cite{Smirnov:1994tg,Beneke:1997zp}. The full diagram is
\begin{align}
    i \mathcal{A}^{(2){\rm 2PI}}_{\rm scatt}
    &= e^{2i\delta(k)} (-ig)^4
    \left| \langle b \rvert s^{(+)}(k) \rangle \right|^2
    \nonumber \\
    &\quad \times
    \int \frac{{\rm d}^4 q_1}{(2\pi)^4}
    \frac{i}{\frac{k^2}{M}- q_1^0 + |E_b| + i0}
    \frac{i}{q_1^2 + i0}
    \int \frac{{\rm d}^4 q_2}{(2\pi)^4}
    \frac{i}{\frac{k^2}{M}- q_2^0 + |E_b| + i0}
    \frac{i}{q_2^2 + i0}
    \nonumber \\
    &\quad \times
    \int \frac{{\rm d}^3 \mathbf{p}}{(2\pi)^3}
    \frac{i \left| \langle b \rvert s^{(+)}(p) \rangle \right|^2}
    {\frac{k^2}{M}-q_1^0-q_2^0-\frac{p^2}{M}+i0},
\end{align}
where $q_{1,2}$ denote the 4-momenta while $k=|\mathbf{k}|$ and $p=|\mathbf{p}|$ denote the norm of the 3-momenta.

The first cut can be estimated as
\begin{align}
    &\crosseddoublebubblecuta[1]
    \sim
    e^{2i\delta(k)} g^4
    \left| \langle b \rvert s^{(+)}(k) \rangle \right|^2
    \int_0^\infty {\rm d}|\mathbf{q}_1|\,
    \frac{|\mathbf{q}_1|}
    {\frac{k^2}{M}+|E_b|-|\mathbf{q}_1|+i0}
    \nonumber \\
    &\quad \times
    \int_0^\infty {\rm d}|\mathbf{q}_2|\,
    \frac{|\mathbf{q}_2|}
    {\frac{k^2}{M}+|E_b|-|\mathbf{q}_2|+i0}
    \left[
    M\,p_\star(|\mathbf{q}_1|,|\mathbf{q}_2|)
    \left|
    \langle b \rvert
    s^{(+)}\!\left(p_\star(|\mathbf{q}_1|,|\mathbf{q}_2|)\right)
    \rangle
    \right|^2 \right],
\end{align}
where we defined
\begin{equation}
    p_\star(|\mathbf{q}_1|,|\mathbf{q}_2|)
    =
    \sqrt{k^2-M\left(|\mathbf{q}_1|+|\mathbf{q}_2|\right)}.
\end{equation}
For the diagram to receive the same strong-mixing enhancement as the corresponding 2PR contribution, the matrix element in the integrand must be evaluated inside the momentum region where it remains comparable to its peak value, i.e.\ within the intrinsic resonance width $\Lambda_{\rm EFT}^2=(\pi \Lambda_b)^3/k_0$ derived in Sec.~\ref{sec:Expansion_around_the_Maximum},
\begin{equation}
    \left|
    p_\star(|\mathbf{q}_1|,|\mathbf{q}_2|)-k_0
    \right|
    \lesssim
    \frac{\Lambda_{\rm EFT}^2}{k_0}.
\end{equation}
For external momentum $k \simeq k_0$, this implies
\begin{equation}
    |\mathbf{q}_1|+|\mathbf{q}_2|
    \lesssim
    \frac{k}{M} \frac{\Lambda_{\rm EFT}^2}{k_0},
\end{equation}
hence the available phase space of the two radiated scalars is suppressed by
\begin{equation}
    \frac{|\mathbf{q}_1||\mathbf{q}_2|}
    {(k^2/M)^2}
    \lesssim
    \frac{\Lambda_{\rm EFT}^4}{k_0^4}.
\end{equation}

The second cut can similarly be estimated as
\begin{align}
    &\crosseddoublebubblecutb[1]
    \sim
    e^{2i\delta(k)} g^4
    \left| \langle b \rvert s^{(+)}(k) \rangle \right|^2
    \left(\frac{k^2}{M}+|E_b|\right)
    \nonumber \\
    &\quad \times
    \int_0^\infty {\rm d}|\mathbf{q}_2|\,
    \frac{|\mathbf{q}_2|}
    {\frac{k^2}{M}-|\mathbf{q}_2|+|E_b|+i0}
    \int_0^\infty {\rm d}p\,
    p^2
    \frac{
    \left| \langle b \rvert s^{(+)}(p) \rangle \right|^2
    }
    {|E_b|+|\mathbf{q}_2|+\frac{p^2}{M}-i0}.
\end{align}
The propagator appearing in the inner integral has no singularity for physical momenta
\begin{equation}
    |E_b|+|\mathbf{q}_2|+\frac{p^2}{M}>0,
\end{equation}
and is therefore smooth across the momentum region in which the matrix element is strongly enhanced. As a result, the enhanced contribution to the $p$ integral is restricted to the support of the strong-mixing around the $k_0$ surface, which has width $k_0 |\delta k|<\Lambda_{\rm EFT}^2=\lambda k_0^2$, as established in Sec.~\ref{sec:Expansion_around_the_Maximum}. For $k \to k_0$, the suppression factor can be estimated as
\begin{equation}
    \int_{(1-\lambda)k_0}^{(1+\lambda)k_0}
    {\rm d}p \, p^2
    \frac{
    \left| \langle b \rvert s^{(+)}(p) \rangle \right|^2
    }
    {|E_b|+|\mathbf{q}_2|+\frac{p^2}{M}-i0}
    \sim \left| \langle b \rvert s^{(+)}(k_0) \rangle \right|^2 \frac{\lambda k_0^3}{|E_b|+|\mathbf{q}_2|+\frac{k_0^2}{M}+i0}.
\end{equation}
Since the remaining cut is identical to the second one, we conclude that the $\mathcal{O}(g^4)$ 2PI contribution is suppressed relative to the corresponding 2PR diagrams
\begin{equation}
    \left|\frac{\mathcal{A}^{(2){\rm 2PI}}(k_0,\mu_R)}{\mathcal{A}^{(2){\rm 2PR}}(k_0,\mu_R)}\right| = \mathcal{O} \left( \frac{\Lambda_{\rm EFT}^2}{k_0^2} \right) = \mathcal{O} \left(\lambda \right).
\end{equation}

%%%%%%%%%%%%%%%%%%%%%%%%%%%%%%%%%%%%%%%%%%%%%%%%%%%%%%%%%
\subsection{Corrections to Factorization}
\label{sec:Corrections_to_Factorization}
%%%%%%%%%%%%%%%%%%%%%%%%%%%%%%%%%%%%%%%%%%%%%%%%%%%%%%%%%

%%%%%%%%%%%%%%%%%%%%%%%%%%%%%%%%%%%%%%%%%%%%%%%%%%%%%%%%%
\subsubsection{Momentum Independent Corrections}
\label{sec:Momentum_Independent_Corrections}
%%%%%%%%%%%%%%%%%%%%%%%%%%%%%%%%%%%%%%%%%%%%%%%%%%%%%%%%%

In Sec.~\ref{sec:Factorization}, we assumed that the loop integral over the relative momentum of the scattering state in the repulsive channel factorizes into the product of the matrix element and the loop over the free scattering state
\begin{align}
    &\left(\frac{\mu_F}{2}\right)^{4-d} \int \frac{{\rm d}^{d-1} \mathbf{p}}{(2\pi)^{d-1}} \frac{\left|\langle b \rvert s^{(+)}(p) \rangle \right|^2}{\frac{k_0^2}{M} - \frac{p^2}{M} + i0} \nonumber \\
    &= \left|\langle b \rvert s^{(+)}(k_0) \rangle \right|^2 \left(\frac{\mu_F}{2}\right)^{4-d} \int \frac{{\rm d}^{d-1} \mathbf{p}}{(2\pi)^{d-1}} \frac{1}{\frac{k_0^2}{M} - \frac{p^2}{M} + i0} + \mathcal{O}\left(\lambda\right),
\end{align}
Therefore, the corrections to the factorization formula start at $\mathcal{O}(\Lambda_{\rm EFT}^2/k_0^2)$ and can be explicitly computed from the integral
\begin{equation}
    \left(\frac{\mu_F}{2}\right)^{4-d} \int \frac{{\rm d}^{d-1} \mathbf{p}}{(2\pi)^{d-1}} \frac{1}{\frac{k_0^2}{M} - \frac{p^2}{M} + i0} \frac{\left|\langle b \rvert s^{(+)}(p) \rangle \right|^2 - \left|\langle b \rvert s^{(+)}(k_0) \rangle \right|^2}{\left|\langle b \rvert s^{(+)}(k_0) \rangle \right|^2}.
\end{equation}

We first consider the part of the integration domain outside the momentum shell $|p-k_0|> \Lambda_{\rm EFT}^2/k_0 = \lambda k_0$. The leading correction can be obtained by retaining only the matrix element evaluated at $k_0$, which is the location of the maximum, because the matrix element evaluated at $p$ is always smaller $\left|\langle b \rvert s^{(+)}(p) \rangle \right| < \left|\langle b \rvert s^{(+)}(k_0) \rangle \right|$. Therefore, we can approximate this part of the integral in the PDS scheme as
\begin{align}
    &\left(\frac{\mu_F}{2}\right)^{4-d} \int_{|p-k_0|>\lambda k_0} \frac{{\rm d}^{d-1} \mathbf{p}}{(2\pi)^{d-1}} \frac{1}{\frac{k_0^2}{M} - \frac{p^2}{M} + i0} \frac{\left|\langle b \rvert s^{(+)}(p) \rangle \right|^2 - \left|\langle b \rvert s^{(+)}(k_0) \rangle \right|^2}{\left|\langle b \rvert s^{(+)}(k_0) \rangle \right|^2} \nonumber \\
    &\sim -\left(\frac{\mu_F}{2}\right)^{4-d} \frac{2^{2-d}}{\pi^{(d-1)/2}\Gamma\left(\frac{d-1}{2}\right)} \left[\int_0^{(1-\lambda)k_0} + \int_{(1+\lambda)k_0}^\infty \right] \frac{{\rm d}p \, p^{d-2}}{\frac{k_0^2}{M} - \frac{p^2}{M} + i0} \nonumber \\
    &= \frac{M \mu_F}{4\pi} - \frac{3}{4 \pi^2} \lambda M k_0 + \mathcal{O} \left(\lambda^2 M k_0 \right),
\end{align}
where we remark that the pole of the propagator lies outside the integration domain.
We highlight that the first term corresponds to the linear power divergence contained in the counterterm of Eq.~\eqref{counterterm}, which cancels against the divergence of the factorized term.
Then, we consider the integration domain within the momentum shell $|p-k_0|<\lambda k_0$, where we can substitute the expansion of the matrix element for $p \to k_0$ from Eq.~\eqref{matrix_element_expansion}, which gives
\begin{align}
    &\left(\frac{\mu_F}{2}\right)^{4-d} \int_{|p-k_0|<\lambda k_0} \frac{{\rm d}^{d-1} \mathbf{p}}{(2\pi)^{d-1}} \frac{1}{\frac{k_0^2}{M} - \frac{p^2}{M} + i0} \frac{\left|\langle b \rvert s^{(+)}(p) \rangle \right|^2 - \left|\langle b \rvert s^{(+)}(k_0) \rangle \right|^2}{\left|\langle b \rvert s^{(+)}(k_0) \rangle \right|^2} \nonumber \\
    &\sim \left(\frac{\mu_F}{2}\right)^{4-d} \frac{2^{2-d}}{\pi^{(d-1)/2}\Gamma\left(\frac{d-1}{2}\right)} \int_{(1-\lambda)k_0}^{(1+\lambda)k_0} {\rm d}p \frac{p^{d-2}}{\frac{k_0^2}{M} - \frac{p^2}{M} + i0} \left[- \pi^2 \frac{(k_0 \delta p)^2}{\Lambda_{\rm EFT}^4} + \mathcal{O} \left( \frac{(k_0\delta p)^3}{\Lambda_{\rm EFT}^6} \right)\right] \nonumber \\
    &= -\frac{1}{4} \lambda M k_0 + \mathcal{O} \left( \lambda^2 M k_0 \right).
\end{align}
Therefore, combining the two parts of the integration domain, we show that the first correction to factorization reads
\begin{align}
    &\left(\frac{\mu_F}{2}\right)^{4-d} \int \frac{{\rm d}^{d-1} \mathbf{p}}{(2\pi)^{d-1}} \frac{\left|\langle b \rvert s^{(+)}(p) \rangle \right|^2}{\frac{k_0^2}{M} - \frac{p^2}{M} + i0} \nonumber \\
    &= \left[ \left|\langle b \rvert s^{(+)}(k_0) \rangle \right|^2 \left(\frac{\mu_F}{2}\right)^{4-d} \int \frac{{\rm d}^{d-1} \mathbf{p}}{(2\pi)^{d-1}} \frac{1}{\frac{k_0^2}{M} - \frac{p^2}{M} + i0} \right] \times \left[1 + \mathcal{O}\left(\lambda\right) \right].
\end{align}

%%%%%%%%%%%%%%%%%%%%%%%%%%%%%%%%%%%%%%%%%%%%%%%%%%%%%%%%%
\subsubsection{Momentum Dependent Corrections}
\label{sec:Momentum_Dependent_Corrections}
%%%%%%%%%%%%%%%%%%%%%%%%%%%%%%%%%%%%%%%%%%%%%%%%%%%%%%%%%

We now show that the momentum dependence of the corrections to factorization is suppressed compared to the momentum dependence of the factorized formula, order by order in $k_0 \delta k/\Lambda_{\rm EFT}^2$. We show it explicitly for $\mathcal{O}((k_0\delta k)^2/\Lambda_{\rm EFT}^4)$, which can be generalized.
Considering the loop integral with the full $k$ momentum dependence, it can be rewritten as
\begin{align}
    &\left(\frac{\mu_F}{2}\right)^{4-d} \int \frac{{\rm d}^{d-1} \mathbf{p}}{(2\pi)^{d-1}}
    \frac{ \left| \langle b \rvert s^{(+)}(p) \rangle \right|^2 }{
        \frac{k^2}{M}-\frac{p^2}{M}+i0
    }
    = \left| \langle b \rvert s^{(+)}(k) \rangle \right|^2
    \Bigg\{ \left(\frac{\mu_F}{2}\right)^{4-d} \int \frac{{\rm d}^{d-1} \mathbf{p}}{(2\pi)^{d-1}}
    \frac{1}{
        \frac{k^2}{M}-\frac{p^2}{M}+i0
    } \nonumber \\
    &+\left(\frac{\mu_F}{2}\right)^{4-d} \int \frac{{\rm d}^{d-1} \mathbf{p}}{(2\pi)^{d-1}}
    \frac{1}{
        \frac{k^2}{M}-\frac{p^2}{M}+i0
    }
    \frac{\left|\langle b \rvert s^{(+)}(p) \rangle \right|^2 - \left|\langle b \rvert s^{(+)}(k) \rangle \right|^2}{\left|\langle b \rvert s^{(+)}(k) \rangle \right|^2} \Bigg\}.
\end{align}
Then, following the derivation of Sec.~\ref{sec:Momentum_Independent_Corrections}, we compute the relative correction separating the integration domain, starting from $|p-k_0|>\Lambda_{\rm EFT}^2/k_0 = \lambda k_0$. Since $p$ lies outside the momentum shell, while $k$ lies inside, we can neglect the matrix element evaluated at $p$ because $|\langle b \rvert s^{(+)}(p) \rangle |<|\langle b \rvert s^{(+)}(k) \rangle |$ holds.
Therefore, we obtain
\begin{align}
    &\left(\frac{\mu_F}{2}\right)^{4-d} \int_{|p-k_0|>\lambda k_0} \frac{{\rm d}^{d-1} \mathbf{p}}{(2\pi)^{d-1}}
    \frac{1}{
        \frac{k^2}{M}-\frac{p^2}{M}+i0
    }
    \frac{\left|\langle b \rvert s^{(+)}(p) \rangle \right|^2 - \left|\langle b \rvert s^{(+)}(k) \rangle \right|^2}{\left|\langle b \rvert s^{(+)}(k) \rangle \right|^2} \nonumber \\
    &\sim\frac{M \mu_F}{4\pi} - \frac{3}{4\pi^2} \lambda M k_0 + \mathcal{O} \left( \lambda^2 M k_0 \right).
\end{align}
Inside the momentum shell $|p-k_0|<\lambda k_0$, expand the integrand in $\delta p= p-k_0$ and $\delta k=k-k_0$ as
\begin{align}
    &\left(\frac{\mu_F}{2}\right)^{4-d} \int_{|p-k_0|<\lambda k_0} \frac{{\rm d}^{d-1} \mathbf{p}}{(2\pi)^{d-1}} \frac{1}{\frac{k^2}{M} - \frac{p^2}{M} + i0} \frac{\left|\langle b \rvert s^{(+)}(p) \rangle \right|^2 - \left|\langle b \rvert s^{(+)}(k) \rangle \right|^2}{\left|\langle b \rvert s^{(+)}(k) \rangle \right|^2} \nonumber \\
    &\sim \left(\frac{\mu_F}{2}\right)^{4-d} \frac{2^{2-d}}{\pi^{(d-1)/2}\Gamma\left(\frac{d-1}{2}\right)} \int_{(1-\lambda)k_0}^{(1+\lambda)k_0} {\rm d}p \frac{p^{d-2}}{\frac{2k_0}{M} (\delta k - \delta p) + i0} \nonumber \\
    & \qquad \qquad \qquad \qquad \qquad \times \left[- \pi^2 \frac{(k_0 \delta p)^2 - (k_0\delta k)^2}{\Lambda_{\rm EFT}^4} + \mathcal{O} \left( \frac{(k_0 \delta p)^3}{\Lambda_{\rm EFT}^6},\frac{(k_0 \delta k)^3}{\Lambda_{\rm EFT}^6} \right)\right] \nonumber \\
    &= -\frac{1}{4} \lambda M k_0 + \mathcal{O}\left(\lambda^2 M k_0\right) - \frac{1}{2} M k_0 \frac{k_0 \delta k}{\Lambda_{\rm EFT}^2} \left[1+\mathcal{O} \left( \lambda \right)\right] \nonumber \\
    &+ \frac{1}{4} \lambda M k_0 \frac{(k_0 \delta k)^2}{\Lambda_{\rm EFT}^4} \left[1+\mathcal{O} \left(\frac{\Lambda_{\rm EFT}}{k_0}\right)\right] + \mathcal{O}\left( \lambda M k_0 \frac{(k_0 \delta k)^3}{\Lambda_{\rm EFT}^6}\right).
\end{align}
Therefore, the $\delta k^2$ correction to factorization scales as $\lambda M k_0 (k_0 \delta k)^2/\Lambda_{\rm EFT}^4$. Since it enters Eq.~\eqref{C:bare:app} divided by the factorized loop, $I(k)\sim Mk_0$, its relative size is $\lambda (k_0 \delta k)^2/\Lambda_{\rm EFT}^4$, therefore, the factorization correction is suppressed by an additional factor $\lambda$.

Moreover, it is interesting to note the presence of an additional linear correction suppressed only by $(k_0 \delta k)/\Lambda_{\rm EFT}^2$. However, such a term represents the loop correction to the location of the maximum of the effective interaction. Indeed, the position $k_0$ was determined from the tree-level matrix element and is generically shifted by higher-loop corrections. The linear term can therefore be absorbed  into the definition of a renormalized maximum $k_0^R$, corresponding to the extremum of the interaction including the $\mathcal{O}(g^4)$ correction. Determining this shift is beyond the scope of the present work.

%%%%%%%%%%%%%%%%%%%%%%%%%%%%%%%%%%%%%%%%%%%%%%%%%%%%%%%%%
\section{Derivation of the Momentum-Surface EFT}
\label{sec:momentum_surface_construction}
%%%%%%%%%%%%%%%%%%%%%%%%%%%%%%%%%%%%%%%%%%%%%%%%%%%%%%%%%

In this appendix, we summarize the construction of the momentum-surface EFT used in Sec.~\ref{sec:Lagrangian}. We begin from the two-particle EFT for a $\xi\xi$ pair in the ${}^1S_0$ channel. Since the relevant dynamics is localized around the momentum $k_0$, the contact interactions are organized in powers of
\begin{equation}
    -\overleftrightarrow{\nabla}^{\,2}-k_0^2,
\end{equation}
where
$\overleftrightarrow{\nabla}
=(\overrightarrow{\nabla}-\overleftarrow{\nabla})/2$. 
Schematically, the EFT takes the form
\begin{align}
    \mathcal L_{\rm EFT}
    ={}&
    \xi^\dagger(\mathbf x,t)
    \left(
        i\partial_t+\frac{\nabla^2}{2M}
    \right)
    \xi(\mathbf x,t)
    \nonumber\\
    &-
    \Bigg\{
        C_0
        \left[\xi\xi\right]^\dagger_{S=0}(\mathbf x,t)
        \left[\xi\xi\right]_{S=0}(\mathbf x,t)
        \nonumber\\
    &\qquad
        +\frac{C_1}{2}
        \left(
            \left[\xi\xi\right]^\dagger_{S=0}(\mathbf x,t)
            \left[
                \xi
                \left(
                    -\overleftrightarrow{\nabla}^{\,2}-k_0^2
                \right)
                \xi
            \right]_{S=0}(\mathbf x,t)
            +{\rm h.c.}
        \right)
        +\cdots
    \Bigg\},
\end{align}
Higher-order operators contain additional powers of $-\overleftrightarrow{\nabla}^{\,2}-k_0^2$ and are accompanied by the corresponding Wilson coefficients.

At this stage, however, the interaction and kinetic terms are organized around different reference momenta. The interactions are already expanded around the surface $k=k_0$,  while the kinetic term retains its usual form around $k=0$. As discussed in Sec.~\ref{sec:Lagrangian}, we parameterize momenta close to the surface as $k=k_0+\ell$, where $\ell$ is the residual momentum and $\delta k$ denotes the characteristic residual momentum scale of the process, so that  $\ell\sim\delta k$ and $|\ell|\ll k_0$. The kinetic term can then likewise be expanded around $k_0$.

Since we are interested only in the $s$-wave channel, it is convenient to rewrite the two-particle sector in terms of the bilocal field $\Phi_{\xi\xi}(\mathbf R,\mathbf r,t)$, defined in Eq.~\eqref{composite_field_def}, where $\mathbf R$ and $\mathbf r$ denote the center-of-mass and relative coordinates, respectively. This makes the separation between center-of-mass and relative motion explicit and allows for a direct partial-wave projection. Since the interaction depends only on the relative coordinate, the center-of-mass motion factorizes and the momentum-surface construction acts only on the relative momentum. The kinetic term then becomes
\begin{equation}
    \int {\rm d}^3\mathbf r\,
    \Phi_{\xi\xi}^\dagger(\mathbf R,\mathbf r,t)
    \left(
        i\partial_t
        +\frac{\nabla_{\mathbf R}^2}{4M}
        +\frac{\nabla_{\mathbf r}^2}{M}
    \right)
    \Phi_{\xi\xi}(\mathbf R,\mathbf r,t),
\end{equation}
while the contact interactions are localized at $\mathbf r=0$ and are organized in powers of
\begin{equation}
    -\nabla_{\mathbf r}^2-k_0^2.
\end{equation}
We then project directly onto the $s$-wave component,
\begin{equation}
    \Phi_{00}(\mathbf R,r,t)
    =
    \int {\rm d}\Omega_{\hat{\mathbf r}}\,
    Y_{00}^*(\hat{\mathbf r})\,
    \Phi_{\xi\xi}(\mathbf R,\mathbf r,t),
\end{equation}
and retain only the radial mode $\Phi_{00}$. Using $ \delta^{(3)}(\mathbf r)
    =
    \frac{\delta(r)}{4\pi r^2}$,
the $s$-wave Lagrangian becomes
\begin{align}
    \mathcal L_{\rm EFT}
    ={}&
    \int {\rm d}r\,r^2\,
    \Phi_{00}^\dagger(\mathbf R,r,t)
    \left(
        i\partial_t
        +\frac{\nabla_{\mathbf R}^2}{4M}
        +\frac{1}{M r^2}
        \partial_r\!\left(r^2\partial_r\right)
    \right)
    \Phi_{00}(\mathbf R,r,t)
    \nonumber\\
    &-
    \int {\rm d}r\,
    \frac{\delta(r)}{4\pi}
    \Bigg[
        C_0\,
        \Phi_{00}^\dagger(\mathbf R,r,t)
        \Phi_{00}(\mathbf R,r,t)
        +\cdots
    \Bigg].
\end{align}

It is convenient to introduce the reduced radial field
\begin{equation}
    \varphi(\mathbf R,r,t)
    =
    r\,\Phi_{00}(\mathbf R,r,t),
\end{equation}
for which the radial kinetic operator takes the one-dimensional form
\begin{align}
    \mathcal L_{\rm EFT}
    ={}&
    \int {\rm d}r\,
    \varphi^\dagger(\mathbf R,r,t)
    \left(
        i\partial_t
        +\frac{\nabla_{\mathbf R}^2}{4M}
        +\frac{\partial_r^2}{M}
    \right)
    \varphi(\mathbf R,r,t)
    \nonumber\\
    &-
    \int {\rm d}r\,
    \frac{\delta(r)}{4\pi r^2}
    \Bigg[
        C_0\,
        \varphi^\dagger(\mathbf R,r,t)
        \varphi(\mathbf R,r,t)
        \nonumber\\
    &\hspace{2.8cm}
        +
        \frac{C_1}{2}
        \left(
            \varphi^\dagger(\mathbf R,r,t)
            \left(
                -\partial_r^2-k_0^2
            \right)
            \varphi(\mathbf R,r,t)
            +{\rm h.c.}
        \right)
        +\cdots
    \Bigg].
\end{align}
The apparent factor $1/r^2$ in the contact interaction is compensated by the regular near-origin behavior $\varphi(\mathbf R,r,t)\sim r$.

In terms of creation and annihilation operators, the reduced radial field takes the form, 
\begin{align}
    \varphi(\mathbf{R},r,t)
    &= \int \frac{{\rm d}^3 \mathbf{P}}{(2\pi)^3}\int \frac{\mathrm{d}^3\mathbf{k}}{(2\pi)^3}\,
    \left[a_{\frac{\mathbf{P}}{2}-\mathbf{k}}a_{\frac{\mathbf{P}}{2}+\mathbf{k}}\right]_{S=0}
    e^{-i\left( \frac{P^2}{4M} +\frac{k^2}{M}\right)t + i \mathbf{P} \cdot \mathbf{R}}
    \frac{\sqrt{4\pi}}{k} \sin (kr).
    \label{varphi_ops_modes}
\end{align}
The appearance of $\sin(kr)$ reflects the regular $s$-wave radial solution. Since $\varphi$ is defined on the half-line $r\geq0$, it is therefore natural to introduce the Fourier sine transform
\begin{equation}
    \varphi(\mathbf R,r,t)
    =
    \int_0^\infty
    \frac{{\rm d}k}{2\pi}\,
    2\sin(kr)\,
    \varphi(\mathbf R,k,t).
\end{equation}
Near the origin $ 2\sin(kr)\simeq 2kr$, so that the leading contact interaction becomes
\begin{align}
    &\int_0^\infty {\rm d}r\,
    \frac{\delta(r)}{4\pi r^2}
    \varphi^\dagger(\mathbf R,r,t)
    \varphi(\mathbf R,r,t)
    \nonumber\\
    &\qquad=
    \int_0^\infty
    \frac{{\rm d}k}{2\pi}
    \int_0^\infty
    \frac{{\rm d}k'}{2\pi}\,
    \frac{kk'}{\pi}\,
    \varphi^\dagger(\mathbf R,k',t)
    \varphi(\mathbf R,k,t).
\end{align}
The factor $kk'/\pi$ originates from the radial $s$-wave normalization inherited from the partial-wave projection and the definition of the reduced radial field.  The two-particle Lagrangian can therefore be written in relative-momentum space as
\begin{align}
\mathcal L_{\rm EFT}
={}&
\int_0^\infty
\frac{{\rm d}k}{2\pi}\,
\varphi^\dagger(\mathbf R,k,t)
\left(
    i\partial_t
    +\frac{\nabla_{\mathbf R}^2}{4M}
    -\frac{k^2}{M}
\right)
\varphi(\mathbf R,k,t)
\nonumber\\
&-
\int_0^\infty
\frac{{\rm d}k}{2\pi}
\int_0^\infty
\frac{{\rm d}k'}{2\pi}\,
\frac{kk'}{\pi}\,
\varphi^\dagger(\mathbf R,k',t)
\nonumber\\
&\qquad\times
\Bigg[
    C_0
    +\frac{C_1}{2}
    \left(
        k^2+(k')^2-2k_0^2
    \right)
    +\cdots
\Bigg]
\varphi(\mathbf R,k,t).
\label{radial_momentum_EFT}
\end{align}

The remaining step is to separate the rapid dependence associated with the reference momentum and energy from the residual dynamics. Using $k=k_0+\ell$, the relative kinetic energy decomposes as
\begin{equation}
    \frac{k^2}{M}
    =
    \frac{k_0^2}{M}
    +
    \frac{2k_0\ell+\ell^2}{M}.
\end{equation}
The first term is a common reference energy, while the second contains the residual dynamics. Factoring out the corresponding rapid time dependence, we write
\begin{equation}
    \varphi(\mathbf R,r,t)
    =
    e^{-i\frac{k_0^2}{M}t}
    \int_{-\infty}^{+\infty}
    \frac{{\rm d}\ell}{2\pi}\,
    2\sin\!\left[(k_0+\ell)r\right]
    \chi(\mathbf R,\ell,t).
\end{equation}
Here the residual-momentum integral has been extended to the full real line, with a regulator understood. This does not imply that the EFT is valid for arbitrarily large $|\ell|$; modes outside the momentum shell are encoded in the Wilson coefficients. Defining the residual position-space field by
\begin{equation}
    \chi(\mathbf R,r,t)
    =
    \frac{1}{i}
    \int_{-\infty}^{+\infty}
    \frac{{\rm d}\ell}{2\pi}\,
    e^{i\ell r}
    \chi(\mathbf R,\ell,t),
\end{equation}
the reduced radial field can be written as
\begin{equation}
    \varphi(\mathbf R,r,t)
    =
    e^{-i\frac{k_0^2}{M}t}
    \left[
        e^{ik_0r}\chi(\mathbf R,r,t)
        -
        e^{-ik_0r}\chi(\mathbf R,-r,t)
    \right].
    \label{varphi_chi_modes}
\end{equation}
The phases $e^{\pm ik_0r}$ encode the rapid radial dependence associated with the reference momentum, while $\chi$ describes the slow dynamics normal to the momentum surface. The relative minus sign ensures the regularity condition
\begin{equation}
    \varphi(\mathbf R,0,t)=0.
\end{equation}
Substituting $k=k_0+\ell$ into Eq.~\eqref{radial_momentum_EFT}, the kinetic term becomes
\begin{equation}
    i\partial_t
    +\frac{\nabla_{\mathbf R}^2}{4M}
    -\frac{2k_0\ell+\ell^2}{M},
\end{equation}
while the interaction terms are expanded using
\begin{equation}
    k^2-k_0^2
    =
    2k_0\ell+\ell^2,
    \qquad
    \frac{kk'}{\pi}
    =
    \frac{k_0^2}{\pi}
    \left(
        1+\frac{\ell}{k_0}
    \right)
    \left(
        1+\frac{\ell'}{k_0}
    \right).
\end{equation}
The terms proportional to $\ell^2$ and to $\ell/k_0$ in the radial measure are corrections to the kinematic expansion and are suppressed by powers of $\delta k/k_0$. These corrections are distinct from the derivative expansion of the local interaction, whose natural momentum dependence is organized in powers of $k_0 \ell \sim k_0 \delta k$ and controlled by the smaller scale $\Lambda_{\rm EFT}^2$. The scaling of individual interaction operators may be further suppressed by symmetries of the microscopic theory.

Keeping the leading terms in the kinematic expansion in $\delta k /k_0$, while retaining the local derivative expansion of the interactions, we obtain the momentum-space form of the momentum-surface EFT,
\begin{align}
\label{Momentum:shell:EFT}
\mathcal{L}_{\rm EFT}
={}&
\int_{-\infty}^{+\infty}
\frac{{\rm d}\ell}{2\pi}\,
\chi^\dagger(\mathbf R,\ell,t)
\left(
    i\partial_t
    +\frac{\nabla_{\mathbf R}^2}{4M}
    -2\frac{k_0\ell}{M}
\right)
\chi(\mathbf R,\ell,t)
\nonumber\\
&-
\frac{k_0^2}{\pi}
\int_{-\infty}^{+\infty}
\frac{{\rm d}\ell}{2\pi}
\int_{-\infty}^{+\infty}
\frac{{\rm d}\ell'}{2\pi}\,
\chi^\dagger(\mathbf R,\ell',t)
\nonumber\\
&\qquad\times
\Bigg[
    C_0
    +C_1k_0(\ell+\ell')
    +2C_2^{(1)}k_0^2
    \left(
        \ell^2+\ell'^2
    \right)
    +4C_2^{(2)}k_0^2\ell\ell'
    +\cdots
\Bigg]
\chi(\mathbf R,\ell,t).
\end{align}
Fourier transforming the residual momentum gives the equivalent position-space form,
\begin{align}
    \mathcal{L}_{\rm EFT}
    ={}&
    \int {\rm d}r\,
    \chi^\dagger(\mathbf R,r,t)
    \left(
        i\partial_t
        +\frac{\nabla_{\mathbf R}^2}{4M}
        +2i\frac{k_0}{M}\partial_r
    \right)
    \chi(\mathbf R,r,t)
    \nonumber\\
    &-
    \frac{k_0^2}{\pi}
    \int {\rm d}r\,
    \chi^\dagger(\mathbf R,r,t)
    \delta(r)
    \Bigg[
        C_0
        -2iC_1k_0\overleftrightarrow{\partial_r}
        \nonumber\\
    &\hspace{3.8cm}
        -2C_2^{(1)}k_0^2
        \left(
            \overrightarrow{\partial_r}^{\,2}
            +
            \overleftarrow{\partial_r}^{\,2}
        \right)
        +4C_2^{(2)}k_0^2
        \overleftarrow{\partial_r}
        \overrightarrow{\partial_r}
        +\cdots
    \Bigg]
    \chi(\mathbf R,r,t).
\end{align}
When several independent operators appear at the same derivative order, we denote the corresponding Wilson coefficients by $C_n^{(j)}$, where $j$ labels the different operator structures.

%%%%%%%%%%%%%%%%%%%%%%%%%%%%%%%%%%%%%%%%%%%%%%%%%%%%%%%%%
\subsection{Feynman Rules and Normalization}
%%%%%%%%%%%%%%%%%%%%%%%%%%%%%%%%%%%%%%%%%%%%%%%%%%%%%%%%%
\label{EFT:Feyn:rules}
At leading order in the kinematic expansion in $\delta k/k_0$, the residual dispersion relation is linear,
\begin{equation}
    E_{\rm res}(\ell)
    =
    2\frac{k_0}{M}\ell,
\end{equation}
so that the propagator is
\begin{equation}
    iG(\omega,\ell)
    =
    \frac{i}{
        \omega
        -2\frac{k_0}{M}\ell
        +i0
    }.
\end{equation}

The interaction vertices follow directly from Eq.~\eqref{EFT_Lagrangian}. In particular, the leading contact operator gives the vertex
\begin{equation}
    -i\frac{k_0^2}{\pi}C_0.
\end{equation}
The factor $k_0^2/\pi$ originates from the radial $s$-wave normalization factor derived above. Accordingly, the normalization of the one-dimensional momentum-surface states differs from that of the usual $s$-wave two-particle states. Close to
the surface,
\begin{equation}
    \lvert k\rangle
    \mapsto
    \frac{\sqrt{\pi}}{k_0}
    \lvert \delta k\rangle.
\end{equation}
The normalization factor $\sqrt{\pi}/k_0$ on each external state compensates the factor $k_0^2/\pi$ in the interaction. For example, the leading contact operator gives
\begin{equation}
    \left(
        \frac{\sqrt{\pi}}{k_0}
    \right)
    \left(
        -i\frac{k_0^2}{\pi}C_0
    \right)
    \left(
        \frac{\sqrt{\pi}}{k_0}
    \right)
    =
    -iC_0,
\end{equation}
reproducing the normalization of the original two-particle amplitude.

More generally, evaluating the derivative operators between on-shell external states with $k=k'=k_0+\delta k$ gives the tree-level momentum-surface EFT amplitude
\begin{equation}
\label{tree:lebel:EFT}
    i\mathcal A_{\rm EFT}^{\rm tree}(k)
    =
    -i
    \sum_{n=0}^{\infty}
    C_{n}\left(2k_0\delta k\right)^n .
\end{equation}
Here $C_{n}$ is understood collectively to denote all Wilson coefficients multiplying independent operator structures that contribute at order $(\delta k)^n$ in the residual-momentum expansion.

\newpage
\bibliographystyle{JHEP}
\bibliography{main}

\end{document}